\documentclass[usenatbib]{mn2e}
\usepackage{times}
\usepackage{graphicx}
\usepackage{fixltx2e}  
\usepackage{amsmath}  
\usepackage{amssymb}
\usepackage{longtable}  
\usepackage{environ}  
\usepackage{threeparttable}  
\usepackage{threeparttablex}  
\usepackage{epstopdf}
\usepackage{aas_macros}
\usepackage{color}
\title[A Council of Giants]{A Council of Giants}
\author[M. L. McCall]
{
Marshall L. McCall\thanks{E-mail:  mccall@yorku.ca} \\
Department of Physics and Astronomy, York University, Toronto, Ontario, Canada  L3T 3R1
}
\date{Accepted 2014 January 28. \\
Received 2013 December 31. \\
In original form 2013 April 30.}                                           

\begin{document}

\maketitle


\maketitle
\begin{abstract}


Distances and near-infrared luminosities of the brightest galaxies in the Local Volume have been re-evaluated in order to gain a fully homogeneous collection of data for elucidating the framework of the Local Sheet and its relevance to Local Group evolution.
It is demonstrated that the Local Sheet is both geometrically and dynamically distinct from the Local Supercluster and that the evolution of the Sheet and Local Group were probably interconnected.  The Sheet is inclined by 
$8^\circ$  
with respect to the Local Supercluster, and the dispersion of giant members about the mid-plane is only 
$230 \, \rm kpc$.  
A ``Council of Giants'' with a radius of
$3.75 \, \rm Mpc$  
encompasses the Local Group, demarcating a clear upper limit to the realm of influence of the Local Group.  The only two giant elliptical galaxies in the Sheet sit on opposite sides {\color{black} of the Council}, raising the possibility that they have somehow shepherded the evolution of the Local Group.
{\color{black}
The position vector of the Andromeda galaxy with respect to the Milky Way deviates only 
$11^\circ$  
from the Sheet plane and only 
$11^\circ$  
from the projected axis of the ellipticals.
The Local Group appears to be moving away from a ridge in the potential surface of the Council  
on a path 
parallel to the elliptical axis.  
}
Spin directions of the giants in the Council are distributed over the sky in a pattern which is very different from that of giants beyond, possibly in reaction to the central mass asymmetry that developed into the Local Group.  
By matching matter densities of Group and Council giants, 
the edge of the volume of space {\color{black} most likely to} have contributed to the development of the Local Group is shown to be {\color{black} very close} to where gravitational forces from the Local Group and the Council balance.  
The boundary specification reveals that the Local Sheet formed out of a density perturbation of very low amplitude
($\sim 10\%$),  
but that normal matter was incorporated into galaxies with relatively high efficiency 
($\sim 40\%$).  
It appears that the development of the giants of the Local Sheet was guided by a pre-existing flattened framework of matter.

\end{abstract}

\begin{keywords}
large-scale structure of Universe -- Local Group -- galaxies:  kinematics and dynamics -- galaxies:  formation --  galaxies:  evolution -- galaxies:  distances and redshifts.
\end{keywords}

\section{
Introduction
}


Galaxies are organized into an expanding cosmic web of filamentary and sheet-like structures bounding volumes which are largely devoid of matter.   
However, very little is known observationally about the structure of structures and its linkages to galaxy evolution because it is difficult to constrain accurate relative positions of constituents from a distant vantage point.  The Local Sheet, a structure of which we are a part, offers an opportunity for advancement owing to our perspective from within and the proximity to measure reliable distances to members directly.  

Any study of local structure must start with a volume-limited sample of galaxies.  Efforts to construct such a sample began with the definition of the Local Volume \citep{kra79a,huc86a,sch92a},  
which in the rendition initiating this work (the Local Volume Catalog, or LVC -- \citealt{kar04a})
contains all known galaxies either with distances less than 
$10 \, \rm Mpc$  
or with radial velocities less than 
$550 \, \rm km \, s^{-1}$ 
with respect to the Local Group (a Hubble flow distance of 
$7.7 \, \rm Mpc$). 
Within the Local Volume, the Milky Way, Andromeda, and the smaller companions which comprise the Local Group reside in a layer of galaxies, mostly dwarfs, which has an apparent thickness of about 
$1.5 \, \rm Mpc$ \citep{sch92b,pee93a,pee01a,kar04a,kar05a,tul08a,fin12a}.  
At various times, the layer has been referred to as the 
``Local Cloud'' \citep{vau75a},
the ``Coma-Sculptor Cloud'' \citep{tul87a}, 
the ``local plane'' \citep{pee93a},  
the ``local filament'' \citep{kly03a},  
the ``Local pancake'' \citep{kar04a},  
and 
the ``Local Sheet'' \citep{pee01a,tul08a,pee10a}.  
At a certain level, it is the proximate manifestation of the Local Supercluster, whose existence was in fact established in part using the most luminous members of the layer \citep{vau53a}.  
However, models of the local velocity field seem to require that the Local Group be housed in a flattened body of galaxies distinct from the Local 
Supercluster \citep{kly03a}.  
Indeed, it has been argued that the supergalactic arrangement of nearby groups in the plane of the sky is evidence for such a body \citep{vau75a}.   
Also, peculiar velocities of galaxies show a 
sharp discontinuity at a distance of about
$7 \, \rm Mpc$ \citep{tul08a}.  
Studies of local structure are traditionally anchored to the supergalactic coordinate system, but whether or not this is the appropriate framework to adopt has not been examined thoroughly.  

Any local flattened structure distinct from the Local Supercluster ought  to be traced most reliably by its most {\color{black} luminous} members,
because they pinpoint the location of the largest concentrations {\color{black} of dark matter}.  Consequently, to isolate such a structure {\color{black} and elucidate its character}, this paper focuses on carefully mapping the distribution and properties of luminous spiral and elliptical galaxies in the Local Volume.  {\color{black} The framework is vital for guiding studies of the dwarf population locally} \citep{fin12a}, results from which will be presented separately.

\section{
Sample
}

A sample of luminous galaxies in the Local Volume was constructed primarily from the LVC.  For a galaxy to be included in the sample, it was required that the tabulated absolute magnitude in $B$ be equal to or brighter than 
$-18.0$.  
The adopted luminosity cutoff
is fully 
$3 \, \rm mag$  
brighter 
than the median absolute magnitude of galaxies near the edge of the LVC (8 to $10 \, \rm Mpc$).   
This means that LVC selection criteria, not survey detection thresholds, determine the representation of galaxies.    
Furthermore, the cutoff is comparable to the brightness of the largest dwarfs
(e.g., the Large Magellanic Cloud).
Thus, it is faint enough to ensure
that true giants, as defined by stellar mass, are completely sampled, 
even accounting for possible errors in LVC distances or biasing of blue luminosities by star formation.

A total of 56 galaxies in the LVC satisfied the brightness criterion.
Added to the sample were the M33-like spiral 
NGC 300,  
which was listed in the LVC as being slightly fainter
than the absolute magnitude cutoff, and
NGC 1023, NGC 4631, and NGC 5023,  
which modern distance determinations seemed to place within 
$10 \, \rm Mpc$.  
Thus, the final sample was comprised of 
60  
galaxies.
After re-evaluating all distances and brightnesses, 
7  
galaxies 
proved to be fainter than the absolute magnitude cutoff, and 
6  
lay beyond the nominal distance limit.

Sample galaxies are listed with their properties in Table~\ref{tbl_sample},
{\color{black} and the sources of the observations underlying the tabulated parameters are identified in Table~\ref{tbl_sources}.  All distance-dependent quantities are anchored to the nuclear maser distance for M106 \citep{hum13a}.}
Details about the origins and usage of {\color{black} tabulations are given in the sections to follow}.
Listed uncertainties, which are standard deviations, account for all random sources of error, but they
exclude the error in the distance zero-point (where relevant) because it is systematic.  Thus, the errors reflect how uncertain a property of
one galaxy is with respect to that of any other.  

\section{Method}

\subsection{Motions}
To study the kinematics of the Local Group with respect to neighbouring galaxies, 
the heliocentric line-of-sight velocity of each galaxy was {\color{black} adopted, whenever possible, to be a published measurement of the systemic velocity derived by fitting} a map of the internal velocity field.  Measurements {\color{black} made this way} were preferred to those from integrated spectral line profiles because they are less susceptible to perturbation by asymmetries in the spatial distribution of matter, particularly in the case of neutral hydrogen.

For galaxies beyond the Local Group, the heliocentric velocity was corrected for local expansion using a value for the Hubble constant founded upon infrared observations of Cepheids 
{\color{black} and a period-luminosity relation anchored to the distance of M106 \citep{rie11a,rie12a}.  
After accounting for the recent revision to the maser distance to M106 \citep{hum13a}, the following value for the Hubble constant was adopted:}
$H_0 = 71.6 \pm 2.9 \, \rm km \, s^{-1} \, Mpc^{-1}$.  
To place a velocity in the frame of reference of the Local Group,
the reflex motion of the Sun with respect to the
the luminosity-weighted centroid (in $K_s$) of the Local Group was removed using modern determinations of the motion of the Sun with respect to the Local Standard of Rest \citep{sch10a}, the orbital velocity of the Local Standard of Rest about the Milky Way \citep{mcm11a,mar12a}, and the orbital motion of 
Andromeda with respect to the Sun
\citep{mar12a}  
after appropriately correcting the tangential component for the revision to the distance to Andromeda presented in Table~\ref{tbl_sample}.
{\color{black} For kinematic studies of galaxies beyond the Local Group,} each interacting pair (Maffei 1 and 2; M81 and M82) was regarded as a single unit moving at the luminosity-weighted
mean velocity of its constituents.

{\color{black}
Internal ordered motions were characterized by the rotational velocity in the plateau of the rotation curve.  When possible, this was judged from a published fit to the internal velocity field.  However, for NGC~5068 and E274-G001, the rotational velocity had to be gauged from the HI line width at 20\% of the peak flux, and for M74 (NGC~628), which is almost face-on, the rotational velocity was estimated from the absolute magnitude in $K_s$ using the Tully-Fisher relation.
}
 
\subsection{Orientations}
The orientation of each sample galaxy, i.e.,  the tilt of the spin axis relative to the line of sight and the position angle of the {\color{black} line of nodes}, had to be constrained to correct magnitudes for internal extinction (in the case of a disk galaxy), to correct the apparent rotational velocity for projection, and to evaluate the direction of the angular momentum vector {\color{black} of the optical disk.  In this paper, the position angle is measured east from north to the nearest limb, and thereby takes on values between $0^\circ$ and $180^\circ$.}

For a disk galaxy, one gauge of tilt is the ratio of the semi-minor to the semi-major axis of the disk, better known as the axis ratio.  It was established where possible from the outermost isophotes of the deepest optical maps of surface brightness in the reddest possible passbands, and otherwise from the compilation of HyperLeda \citep{pat03a}.  
The corresponding tilt was derived assuming that disks are oblate spheroids with an edge-on axis ratio {\color{black} $q_0$ dependent on the Revised Hubble 
Type $T$ as follows:
\begin{eqnarray}
q_0 & = 0.20  & \quad -3.5 < T < 3.5 \\
& = 0.13 & \quad \phantom{-} 3.5 \le T < 9.5  \nonumber \\
& = 0.57 & \quad   \phantom{-} 9.5 \le T \nonumber
\end{eqnarray}  
(\citealt{sak00a,sta92a}; see also \citealt{ver97a}).   
}
Also, where possible, independent estimates of tilt were acquired from extant fits to maps of velocity fields.  A critical assessment of photometric and kinematic determinations was made for each galaxy, and an appropriate value of the tilt of the optically visible extent was adopted after taking into account such factors as {\color{black} passband, spatial coverage, obscuration, morphology, distortions, interactions,} and the tilt itself.  Generally, an isophotal tilt was preferred for galaxies viewed close to edge-on 
(tilt $> 80^\circ$ {\color{black} -- see \citealt{ver01a}}),  
and a kinematic value was preferred for galaxies close to face-on 
(tilt $< 30^\circ$).   
In between, {\color{black} usually} an average was adopted, in which case the uncertainty was taken to be half of the difference between the preferred photometric and kinematic values.  Major axis position angles were similarly estimated from optical photometry and velocity fields,  {\color{black} but with no restrictions on kinematic measurements.  Photometric and kinematic estimates of orientations are summarized in Table~\ref{tbl_orientations}, and adopted values are summarized in both Tables~\ref{tbl_sample} and \ref{tbl_orientations}.  Sources of observations are provided in Table~\ref{tbl_sources}.} 

Depending upon tilt, there are two or four possible orientations of the angular momentum vector for any given axis ratio and position angle.  
For each disk galaxy, the ambiguity was broken by identifying 
{\color{black} the direction of rotation of the limb specified by the position angle of the line of nodes as well as the top or near side
from the handedness of the} winding of the spiral arms or, for a few highly inclined systems, 
the pattern of 
obscuration \citep{kap83a}.  
For a few predominantly spheroidal systems, the orientation of the angular momentum vector for the main body could be established from the kinematics of old stellar components
(planetary nebulae in the case of Centaurus A: \citealt{wil86a,hui95a,woo07a}).  {\color{black} Codes conveying the direction of rotation and the handedness of the spiral pattern (or the dustiest side of the disk) are provided in Table~\ref{tbl_sample} next to the position angle and tilt, respectively.

The directions of the derived angular momentum vectors are given in Table~\ref{tbl_sample} in a coordinate system anchored to the Sheet.  With respect to positions mapped by \citet{kap83a} for
20  
galaxies in the sample (excluding NGC 247), the corresponding supergalactic positions differ on average by 
$11^\circ$. 
The largest discrepancies occur for galaxies which are heavily obscured and/or close to face-on.  Results presented in this paper ought to be preferred because they benefit from more modern constraints on tilts and position angles.  In the case of  NGC~247, it appears that \citet{kap83a} misidentified the receding side.

For some galaxies, tilted-ring models of HI velocity fields have enabled evaluation of disk orientations far beyond the extent of the stars.  Generally, results are comparable to those found for optically visible matter.  However, there are some galaxies for which the tilt appears to vary quite substantially with radius, indicative of a warp (e.g., the Circinus galaxy, for which the tilt drops from 
$66^\circ$  
to 
$47^\circ$  
over the radius range 
$10\arcmin$  
to
$25\arcmin$:  
\citealt{cur08a}).    
For such objects, it is conceivable that the spin axis derived for the optical disk is not aligned with the spin axis of the dark matter halo, although it is also possible that an interaction has distorted the outer velocity field.  Tilts and position angles presented in Table~\ref{tbl_sample} and the angular momentum vectors which follow from them are quite homogeneously conveying the orientation of baryonic matter within the visible extent of the sample galaxies, but are not necessarily conveying the orientation of dark matter beyond.  Note, however, that for galaxies for which the radial variation of tilt has been mapped, the rotational velocity in the plateau of the rotation curve ($V_\textit{flat}$ in Table~\ref{tbl_sample}) was determined using a tilt appropriate for that radial domain.
}

\subsection{Extinction}

Any structure of which we are a part spans the entire sky, so obscuration by dust in the Milky Way varies drastically across it. 
Without accommodating for the effective wavelength shifts afflicting broad-band photometry, significant systematic errors in distances and luminosities can arise for targets heavily-obscured by
dust \citep{mcc04a},  
be they galaxies at low Galactic latitudes or even Cepheid variables inside galaxies at high Galactic latitudes.  Corrections to the apparent colours and brightnesses of standard candles and of the galaxies themselves were evaluated using the York Extinction Solver (YES: \citealt{mcc04a}).  First, the optical depth at $1 \, \rm \mu m$ was derived for a spectral energy distribution (SED) characteristic of the probe of reddening.  Then, the extinction was evaluated using a SED characteristic of the target to be corrected.  Motions of the probe and target were accommodated by shifting SEDs in wavelength by amounts consistent with heliocentric velocities.
K-corrections for the targets were determined simultaneously with the extinction.
For tilted spiral galaxies, the extra extinction over face-on due to internal dust was estimated self-consistently using an algorithm constructed from observations of colours as a function of tilt \citep{mcc04a}.  

Except for the LMC {\color{black} and SMC (which were employed in the calibration of standard candles)}, all extinction analyses, Galactic and extragalactic, were founded upon a monochromatic reddening law generated from the algorithm of 
Fitzpatrick \citep{fit99a}.  
Through an appropriate choice of parameter, the law was tuned to deliver a ratio of total to selective extinction $A_\textit{V} / E(B-V)$ equal to 
3.07  
for the spectral energy distribution of Vega upon integration over $B$ and $V$ passbands \citep{mcc04a}.  
This value is appropriate for the diffuse component of the interstellar medium of the Milky Way \citep{mcc00a},  
of which most of the dust obscuring extragalactic targets should be a part, and similarly should be characteristic of internal extinction in other galaxies with disks like that of the Milky Way.  
{\color{black} For the LMC and SMC, corrections for obscuration by internal dust were accomplished using reddening laws directly measured for those environments\citep{gor03a}.}

For galaxies situated 
$10^\circ$  
or more away from the Galactic plane, 
optical depths due to dust in the Milky Way were derived from an all-sky map of the $B-V$ colour
excesses of elliptical galaxies \citep{sch98a}.  
Individual determinations of optical depth were made for 
the most heavily obscured galaxies
using HII regions (Maffei 2 -- \citealt{fin07a}; IC 342 -- \citealt{fin07a}; 
Circinus -- this paper, using \citealt{oli99a}),  
the $\hbox{Mg}_2$ index (Maffei 1 -- \citealt{fin03a,fin07a}),  
or colours of lightly obscured analogues (Dwingeloo~1 -- this paper).   
More comprehensive discussions of the extinction analyses and their impact on distances and luminosities are provided elsewhere \citep{mcc04a,fin07a}.

\subsection{Near-infrared Magnitudes}

{\color{black} The Tully-Fisher relation for spirals and the Fundamental Plane for ellipticals suggest that relative total masses of bright galaxies can be judged from stellar masses.
}
To be a reliable proxy for stellar mass, the luminosity of each galaxy 
was characterized in the infrared,
where it is not very sensitive to the star formation 
rate.  Corrections for obscuration by internal and external dust were minimized by focussing on $2.2 \, \rm \mu m$ ($K_s$), the reddest infrared passband for which data were readily available for most galaxies in the sample.  

For all but {\color{black} three galaxies}, apparent magnitudes in $K_s$ were derived from 2MASS
observations \citep{jar03a,skr06a}.   
{\color{black} Better measurements were available for 
NGC~1569 \citep{vad05a}  
and M82 \citep{ich95a}.  
}
The magnitude for the Milky Way was estimated indirectly from its rotational velocity using the Tully-Fisher relation in $K_s$ for galaxies in the Ursa Major cluster,
which was constructed self-consistently for this paper (\S\ref{sec_tfrelations}).   Apparent magnitudes in $B$ and $V$, which were needed to 
correct $K_s$ magnitudes for imperfections in 2MASS photometry, were preferentially adopted from modern digital imaging studies.  

\begin{figure}
\centering
\includegraphics[angle=0,width=7cm]{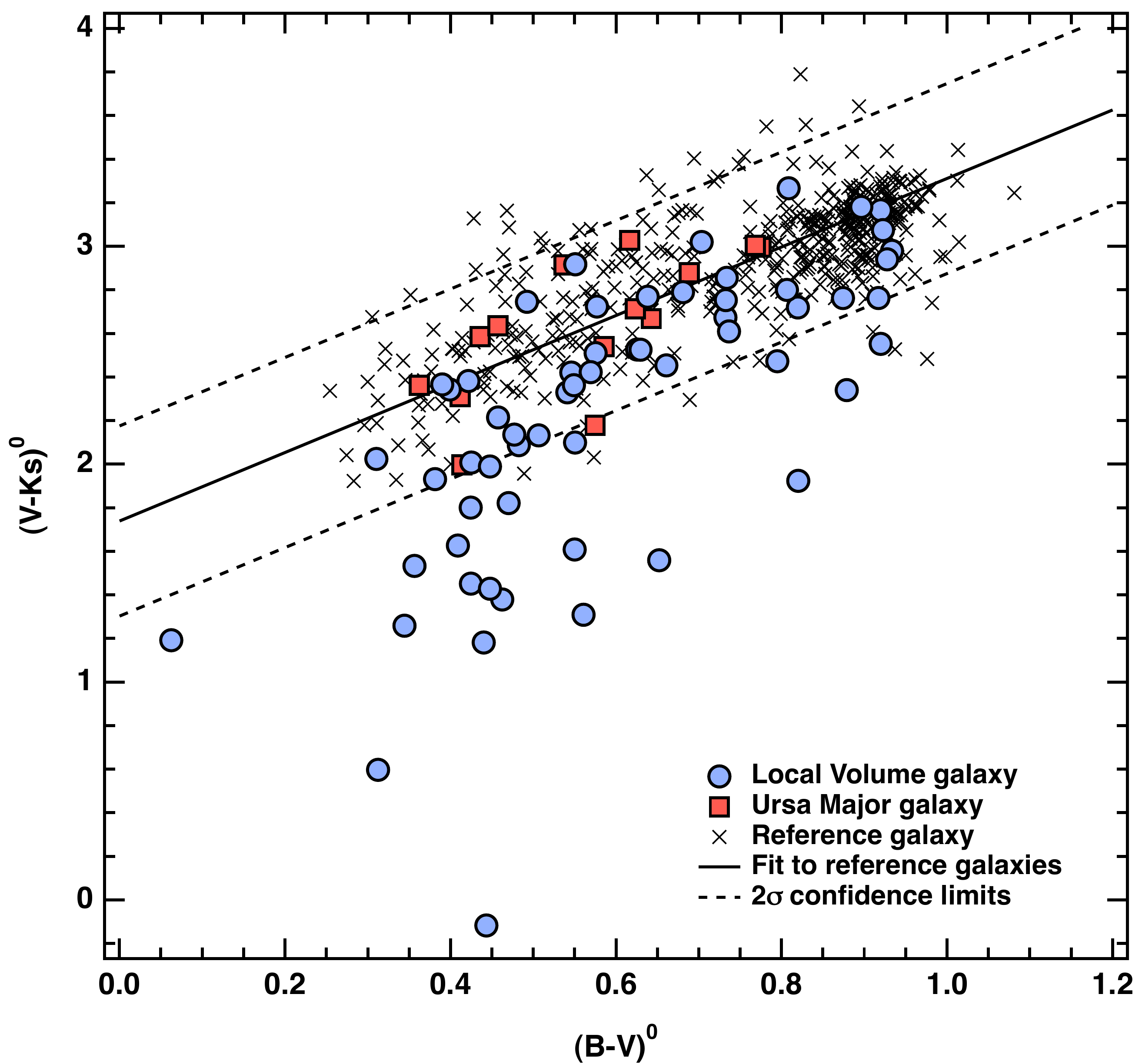}
\caption{
$(V-K_s)^0$ versus $(B-V)^0$ for galaxies across the Hubble sequence.  Colours are corrected for Galactic extinction, internal extinction, and redshift.  Galaxies in the reference sample, for which $K_s$ magnitudes were measured independently from 2MASS, are identified with crosses.  The fit and $2\sigma$ confidence limits are displayed as solid and dashed lines, respectively.  {\color{black} Highly-inclined galaxies in the Ursa Major cluster, which are marked by red squares, show precisely the same trend, verifying the integrity of internal extinction corrections.}  Galaxies in the Local Volume sample, for which $K_s$ magnitudes were measured by 2MASS, are marked with blue circles.  Near-infrared colours for Local Volume galaxies deviate blueward of the locus defined by the reference sample, showing that 2MASS magnitudes for them tend to be too faint.} 
\label{fig_ccdiagram}
\end{figure}

Magnitudes from 2MASS for galaxies with low surface brightnesses are known to be systematically too faint \citep{kir08a},  
and even measurements for bright galaxies are compromised by the finite extrapolation radii.  Furthermore, 2MASS magnitudes for objects spanning two or more survey strips {\color{black} or in crowded fields are} suspect.  This is evidenced by the large deviation of Andromeda from the Tully-Fisher relation in $K_s$.  To flag and correct deficiencies in 2MASS measurements, the relationship between the fully corrected colours $(V-K_s)^0$ and $(B-V)^0$ for the Local Volume sample was compared with that for a set of reference galaxies with {\color{black} impeccable} independent near-infrared photometry (Figure~\ref{fig_ccdiagram}).  Since total magnitudes in $B$ and $V$ are generally more robust than those in 2MASS $K_s$, it was expected that flaws in the $K_s$ magnitudes would be revealed as deviations in $(V-K_s)^0$ from the norm for $(B-V)^0$. 

The fiducial colour-colour relation was established using deep near-infrared and optical observations of a {\color{black} reference} sample of galaxies spanning the Hubble sequence with apparent sizes small compared to the imaging 
arrays \citep{jon94a,gav03a,eis07a}.  To minimize uncertainties arising from corrections for internal extinction, spirals were required to have tilts close to face-on 
{\color{black}
($b/a > 0.625$:  see \citealt{jon94a}).   
}
As shown in Figure~\ref{fig_ccdiagram},
$(V-K_s)^0$ is linearly correlated with $(B-V)^0$ across the Hubble sequence.
Allowing for uncertainties in both coordinates, the {\color{black} line which best fits the reference sample is given by}
\begin{equation} \label{ccfit}
(V-K_s)^0 = (1.739 \pm 0.054)  + (1.572 \pm 0.071) (B-V)^0
\end{equation}  
with the vertical dispersion being only 
$0.22 \, \rm mag$.  
A collection of highly-inclined spirals in the Ursa Major cluster \citep{ver01a}
displays precisely the same trend, verifying the reliability of the algorithm employed to correct disk galaxies for internal extinction.  

Values of $(V-K_s)^0$ for {\color{black} Local Volume} galaxies proved to be systematically shifted blueward of expectations by $0.20 \, \rm mag$.  
Thus, all 2MASS $K_s$ magnitudes were brightened accordingly.  Even then, some galaxies remained as blueward ``dropouts''
(by $2.3 \, \rm mag$  
in the case of Dwingeloo~1 -- see Figure~\ref{fig_ccdiagram}), showing that their $K_s$ magnitudes were still too faint.
For the 
15  
galaxies showing $(V-K_s)^0$ straying blueward of the fiducial relation by more than 
$2 \sigma$,  
$K_s$ magnitudes were additionally adjusted by an amount equal to the deviation. 
{\color{black}
In other words, the $K_s$ magnitude for each was derived from the $V$ magnitude by using Equation~\ref{ccfit} to evaluate the $V-K_s$ colour expected for its observed $B-V$ colour.
}

\subsection{Distances}

\subsubsection{Overview}
\label{sec_overview}

Catalogued distances for nearby galaxies 
are compromised by inhomogeneous approaches to analyses and imperfect alignment of the zero-points of different {\color{black} indices} \citep{fin07a}.
To come to definitive conclusions about local structure, it is imperative that these problems be eliminated.  Thus, distances (and, in turn, luminosities) for this paper were derived from first principles.  Required foundational observational parameters for the galaxies and their constituents were extracted or measured from published data.  

For distance determinations, preference was given to four techniques
founded upon stellar constituents:
(1) 
the period-luminosity (P-L) relations for Cepheid variables in $V$ and $I$;
(2) 
the luminosity cutoff for planetary nebulae (PNe) in the light of $\rm [O III]\lambda5007$;
(3)
the characteristic luminosity of fluctuations in surface brightness (SBF) in $I$;
and (4)
the luminosity of the tip of the red giant branch (TRGB) in $I$.  
For galaxies lacking measurements of stellar constituents, distances were determined from the Fundamental Plane or the Tully-Fisher relation.

First, distance scales for stellar constituents were unified with the Cepheid scale. This was accomplished through pairwise comparison of distances to galaxies for which more than one technique could be applied.  To this end, the Local Volume sample was augmented
with HST Key Project Cepheid 
calibrators \citep{fre01a}  
and several galaxies of 
low metallicity for which good distances were
derivable from multiple techniques 
(LMC, NGC~3109, SMC, WLM).  
Finally, all scales were shifted equally to bring the mean distance of M106 (NGC~4258) into coincidence with its distance of 
$7.60 \pm 0.23 \, \rm Mpc$  
derived from nuclear masers
\citep{hum13a}.   

Cepheid P-L relations in $V$ and $I$ were adopted to be those of
LMC Cepheids as defined by the OGLE~II project \citep{mac06a},  
but after appropriately correcting (via YES) the zero-points for 
a systematic error in the mean 
extinctions \citep{fin07a}.  
To avoid a bias in galaxy mass or metallicity, the evaluation
of the slope of the metallicity dependence of Cepheid distances
was based solely upon a comparison of {\color{black} uncorrected distances from Cepheids with distances from the tip of the red giant branch,
which was possible
for 18  
galaxies.} 

To bring zero-points for the other stellar indicators on to the Cepheid scale, and to solve for the dependence of the
PNe luminosity cutoff on metallicity, an analysis was made of
127  
independent distance estimates for 
34  
galaxies for which distances could be derived by more than one of the stellar methods.
The difference in distance moduli for every distance pairing possible for every galaxy (a total of 
73  
pairs) was computed, and the sum of the squares was minimized. 
Distance moduli flagged as leading to 
differences exceeding 
$0.3 \, \rm mag$ 
in absolute value
($2\sigma$  
for the final fit) 
were pinpointed and rejected.  

After aligning scales, the unweighted mean of the four distance moduli for 
M106 (NGC~4258) came out to be
$0.124 \pm 0.112 \, \rm mag$  
below  
the geometrical value derived from nuclear
masers \citep{hum13a}.  
{\color{black}
The uncertainty here accounts for the error in the mean stellar distance ($0.091 \, \rm mag$)  
and the maser distance 
($0.066 \, \rm mag$).}  
Zero-points for the stellar indicators were adjusted accordingly to deliver distance moduli on the maser scale.  
{\color{black} For the LMC, the resultant mean distance modulus was
$18.47 \pm 0.13$.  
}

Distances to 
52  
of the sample galaxies could be derived from one
or more of the stellar techniques.  
Distances to galaxies for which more than one technique could be applied were computed by averaging unrejected moduli with no weighting.  
The distance to the centre of the Milky Way
was adopted to be 
$8.29 \pm 0.16 \, \rm kpc$  
based upon a modern synthesis of extant measurements \citep{mar12a}.
For the remaining seven galaxies,
distances were derived from an updated version of
the Fundamental Plane for ellipticals in 
$I$ (Maffei~1)  
or 
new constructions of the Tully-Fisher relation in 
$V$ (NGC~3344, Circinus)  
or in 
$I$ (NGC~672, Maffei~2, Dwingeloo~1, and NGC~2903).   

\subsubsection{Cepheids}

The adopted P-L relations for Cepheids were
\begin{align}
M_\textit{V,Ceph} &= (-4.222 \pm 0.021) - (2.779 \pm 0.031) (\log P - 1) \\  
M_\textit{I,Ceph} &= (-4.923 \pm 0.014) - (2.979 \pm 0.021) (\log P - 1)
\end{align}
where $M_\textit{V,Ceph}$ and $M_\textit{I,Ceph}$ are the absolute magnitudes of the Cepheids in $V$ and $I$, respectively, and $P$ is the period of variability in days.  Zero-points have been adjusted to the maser scale on the basis of the outcome of the pair-wise distance analysis described in \S\ref{sec_overview}.
Zero-point errors do not include the uncertainty in the maser distance to M106, which is systematic.

For each galaxy with Cepheid observations, apparent distance moduli in $V$ and $I$ were combined to solve for the extinction and the extinction-free distance modulus simultaneously.  Extinction coefficients and K-corrections were derived using the spectral energy distribution of
a G0 supergiant.   
The extinction-free distance modulus, $\mu$, 
was presumed to depend upon metallicity as follows \citep{fre01a}:
\begin{equation}
\mu = \mu_\textit{true} - \gamma (8.5 - \zeta)  
\end{equation}
Here $\mu_\textit{true}$ is the true distance modulus, 
$\zeta$ is the mean logarithmic oxygen abundance by number 
(i.e., $12 + \log n(\hbox{O}) / n(\hbox{H})$)  
at the location of the Cepheids
as judged from HII regions (in a uniform way), and
$\gamma$ is a constant,.  
The comparison of Cepheid distances with those from the TRGB yielded  
$\gamma = -0.180 \pm 0.047$.   

\subsubsection{Planetary Nebulae}

The luminosity in $\rm [O~III]\lambda5007$ of the brightest planetary nebulae (PNe) is a constant until the metallicity drops below a threshold \citep{cia02a}.   
In the past, corrections for metallicity below the threshold have been
founded upon the mean oxygen abundance
adopted for Cepheids, which has been based upon observations
of HII regions.  Typically, observed planetary nebulae are distributed differently from Cepheids, and of course come from an older population, so the mean metallicity of Cepheids is not necessarily representative.
To determine distances from PNe, it is more 
sensible to use an index of the
mass of the host galaxy as a proxy for metallicity.
The index adopted here was the pseudo absolute magnitude in $K_s$, 
$M_\textit{Ks}^{\prime}$, which would be computed from the unextinguished apparent magnitude of the galaxy, $m_\textit{Ks}$, using a pseudo distance modulus, 
$\mu^\prime$, derived from a PNe luminosity limit equal to what would be observed in a high-metallicity system, where it appears to be a constant.  Defining $m^*$ to be the unextinguished apparent magnitude of the brightest PNe in $\rm [O~III]\lambda5007$, and $M_\textit{Ks}^\textit{ref}$ to be the galaxian absolute magnitude threshold where the limit becomes sensitive to metallicity, then the absolute magnitude $M^*$ of the PNe brightness limit was adopted to be
\begin{align}
M^* &= z_\textit{P} && \text{for} \  M_\textit{Ks}^{\prime} \le  M_\textit{Ks}^\textit{ref} \nonumber \\  
&= z_\textit{P} +  k_\textit{P}  \left[ M_\textit{Ks}^{\prime} - M_\textit{Ks}^\textit{ref} \right] && \text{for} \  M_\textit{Ks}^{\prime} > M_\textit{Ks}^\textit{ref}  
\end{align}
where
$M_\textit{Ks}^{\prime} = m_\textit{Ks} - \mu^\prime$,  
$\mu^\prime = m^* - z_\textit{P}$, 
and $z_\textit{P}$ and $k_\textit{P}$ are constants.  The pairwise analysis of distances described in \S\ref{sec_overview} yielded 
$M_\textit{Ks}^\textit{ref} = -23.0 \pm 0.5$,   
$k_\textit{P} = 0.106 \pm 0.049$,  
and 
$z_\textit{P} = -4.573 \pm 0.042$  
on the maser scale, where uncertainties are due to random errors only.

\subsubsection{Surface Brightness Fluctuations}

Absolute magnitudes in $I$ of surface brightness fluctuations, 
$\overline{M}_\textit{I}$, were derived from
\begin{equation}
\overline{M}_\textit{I} = z_\textit{S} + k_\textit{S} \left[ (V-I)_\textit{bkg} - 1.15) \right]  
\end{equation}
where $(V-I)_\textit{bkg}$ is the colour of the galaxy background where fluctuations
are measured \citep{ton01a},  
and $z_\textit{S}$ and $k_\textit{S}$ are constants.  
In correcting apparent fluctuation magnitudes for extinction,
spectral energy distributions were approximated to be
similar to that of an 
M4 giant.  
However, the spectral energy distribution of an elliptical galaxy was employed to correct the galaxy background colours.
The value of $k_\textit{S}$ was adopted to be 
$4.5\pm 0.25$ \citep{ton01a},  
from which
the pairwise distance analysis yielded 
$z_\textit{S} = -1.700 \pm 0.066$  
on the maser scale.

\subsubsection{Tip of the Red Giant Branch}

Absolute magnitudes in $I$ at the tip of the red giant branch, $M_\textit{I,TRGB}$, were derived from
\begin{equation}
M_\textit{I,TRGB} = z_\textit{T} + k_\textit{T}  \left[ (V-I)_\textit{TRGB} - 1.6 \right]
\end{equation}  
where $(V-I)_\textit{TRGB}$ is the colour of the tip of the 
red giant branch \citep{riz07a},
and $z_\textit{T}$ and $k_\textit{T}$ are constants.
In correcting apparent magnitudes 
for extinction, spectral energy distributions were 
approximated to be similar to that of an 
M0 giant.  
The value of $k_\textit{T}$ was adopted to be 
$0.217 \pm 0.020$ \citep{riz07a},  
from which the pairwise distance analysis yielded 
$z_\textit{T} = -4.053 \pm 0.028$  
on the maser scale.

\subsubsection{Fundamental Plane}

The $I$-band fundamental plane for dynamically hot systems was defined using galaxies in the Coma cluster \citep{fin03a}.   
The distance to the cluster
was anchored to the $I$-band fundamental planes defined by the Leo~I Group and
the Fornax and Virgo clusters
and to the $I$-band Tully-Fisher relation for
HST Key Project galaxies with Cepheid distances \citep{fre01a}.
Fundamental-Plane and Tully-Fisher distances
to the Coma cluster from the Key Project were updated differentially by determining the mean shift
in calibrator distances brought about by changes to the extinction,
the introduction of K-corrections, and revisions to the P-L relations for Cepheids,
and also taking into account revisions to extinction and K-corrections
for Coma galaxies.  
Revised distance moduli from the two methods
differed by only 
$0.003 \, \rm mag$.  
The average was
$34.753 \pm 0.089$  
on the maser scale, which was adopted to set the
zero-point of the Fundamental Plane.
On the maser scale, the metric length $R_e$ in kiloparsecs of the semimajor axis of the elliptical
isophote encompassing half of the total light in $I$ of a dynamically hot system was finalized to be
\begin{align}
\log R_e & = (-7.74 \pm 0.69) + (0.83 \pm 0.06) \langle \mu \rangle_e / 2.5  
\nonumber \\ & \qquad\quad + (0.87 \pm 0.19) \log \sigma_{e8}   
\end{align}
where $\langle \mu\rangle_e$ is the fully-corrected mean ``face-on'' surface brightness within the effective isophote in units of $\, \rm mag \, arcsec^{-2}$, and $\sigma_{e8}$ is the aperture-corrected velocity dispersion in units of $\rm km \, s^{-1}$.  
For the Coma cluster calibrators, the rms scatter in $\log R_e$ 
was 
$0.09 \, \rm dex$ \citep{fin03a}.  

\subsubsection{Tully-Fisher Relations}
\label{sec_tfrelations}

In constructing Tully-Fisher relations,
the amplitude of ordered motions was characterized by the tilt-corrected
rotational velocity in the plateau of the rotation curve ($V_\textit{flat}$), which is
a more reliable gauge of luminosity than the correspondingly corrected 21 cm line width \citep{ver01b}.  
{\color{black}
For galaxies whose velocity fields were modeled with tilted rings, the tilt was adopted to be that displayed at radii in the plateau of the rotation curve.}
Relations in $V$ and $I$ were constructed using disk galaxies for which distances were derived from stellar indicators
as described above.  
Calibrating galaxies were required to have
$\log 2 V_\textit{flat} \ge 2.1$,  
to be tilted by more than 
$40^\circ$,  
and to be extinguished by tilt by less than 
$0.75 \, \rm mag$  
in $I$.
The $V$-band relation was established using the same set of galaxies
as used for the $I$-band relation in order to reduce
the chance of any systematic error in $V$-band distances with respect
to $I$-band distances.  From 
29  
galaxies, the following relations
on the maser scale were derived :
\begin{align}
M_\textit{V,TF} & = (-20.793 \pm 0.044) \nonumber \\
& \qquad\quad - (8.539 \pm 0.284) \left[ \log(2 V_\textit{flat}) - 2.5 \right] \\  
M_\textit{I,TF} & = (-21.656 \pm 0.050) \nonumber \\
& \qquad\quad - (9.243 \pm 0.315) \left[ \log(2 V_\textit{flat}) - 2.5 \right]  
\end{align}
The rms scatter of the fits was 
$0.36$ and $0.34 \, \rm mag$  
in $V$ and $I$, respectively.

To judge the infrared luminosity of the Milky Way, galaxies
in the Ursa Major cluster were employed to define the Tully-Fisher relation
in $K_s$.  This sample was selected because of the availability of good near-infrared photometry deeper than that of 2MASS
carried out with arrays large compared to the galaxies 
\citep{tul96a,ver01a}.  
Local galaxies were not employed due to the greater uncertainty
in apparent magnitudes.  To set the zero-point,
the Key Project Tully-Fisher distance to the Ursa Major cluster \citep{fre01a}
was updated in the same way as that for the Coma cluster.  This led
to a distance modulus of
$31.570 \pm 0.121$  
on the maser scale.  
Selecting galaxies in the same way as for the Tully-Fisher relations in $V$ and $I$,
the following relation
on the maser scale was derived from 
18  
cluster members:
\begin{align}
M_\textit{Ks,TF} &= (-23.483 \pm 0.077) \nonumber \\
& \qquad\quad - (11.384 \pm 0.563) \left[ \log(2 V_\textit{flat}) - 2.5 \right]  
\end{align}
The rms scatter was 
$0.29 \, \rm mag$.  

\subsection{Luminosities}

Luminosities were computed for sample galaxies from the adopted distances and adjusted fully-corrected $K_s$ magnitudes.
The luminosity of the Milky Way was estimated 
from the Tully-Fisher relation in $K_s$ by adopting
$V_\textit{flat}$ to be
$226 \pm 11 \, \rm km \, s^{-1}$.  
This value was
based upon the recent upward revision of the rotation rate at the solar 
radius \citep{rei09a,mar12a}  
and a comparison of observed and predicted shapes of rotation curves for the Milky Way \citep{xue08a,sof09a,mcm11a},  
its look-alike NGC~891 \citep{san79a,sof97a},  
and Andromeda \citep{car06a}.   

\begin{figure}
\centering
\includegraphics[angle=0,width=7cm]{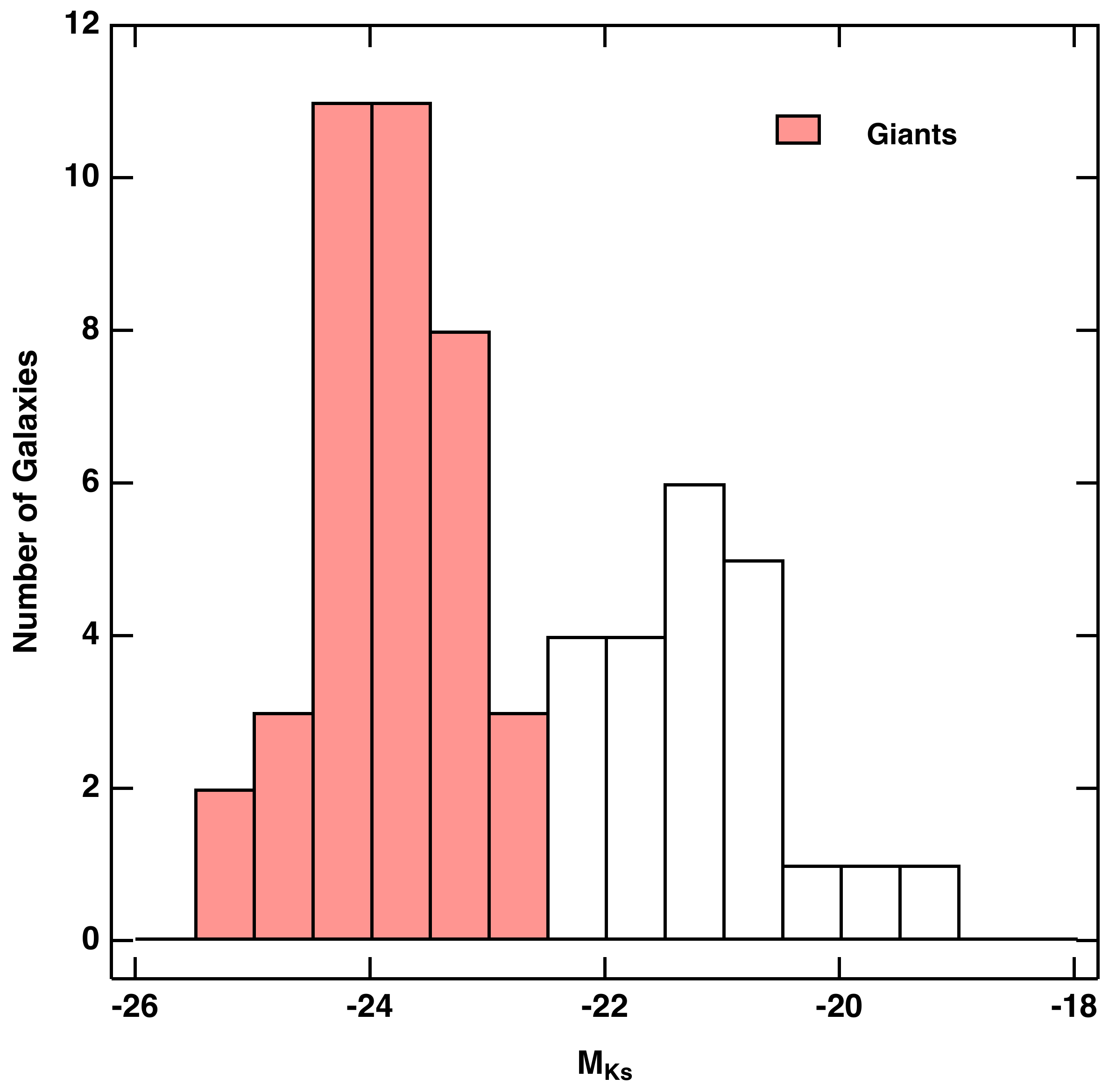}
\caption{
Luminosities of sample galaxies.  The index of luminosity is the absolute magnitude in $K_s$.  Galaxies designated as ``giants'' are highlighted in pink.
} 
\label{fig_luminosities}
\end{figure}

Figure~\ref{fig_luminosities} displays the distribution of absolute magnitudes in 
$K_s$ for sample galaxies.  
As seen 
elsewhere \citep{kar05b},  
there is a Gaussian-like peak centred around 
$-24.0$,  
with numbers declining for 
$1.5 \, \rm mag$  
faintward and then rising again until sampling becomes incomplete.
The behaviour suggests that there are two superimposed populations, one luminous and one faint \citep{bin88a}.   
In this paper, the focus will be on the peak, namely galaxies with 
$M_\textit{Ks} \le -22.5$.  
These galaxies will be referred to as ``giants''.  

\section{
Analysis
}

\subsection{
The Local Sheet and Council of Giants
}

Past discussions of local structure have been guided by a
coordinate system which is defined by the Local Supercluster.  However, a fit of a plane to the positions of the 
eight giants and three interacting pairs of giants  
(luminosity-weighted)
within 
$6 \, \rm Mpc$  
of the Milky Way
reveals an extremely flattened aggregate inclined to the supergalactic
plane by 
$7\fdg95 \pm 0\fdg12$   
with a north pole at supergalactic coordinates
$(L,B) = (241\fdg74 \pm 0\fdg74, 82\fdg05 \pm 0\fdg12)$.  
Errors here stem from the uncertainties in distances alone.
The Sun is perpendicularly offset 
northward  
of the mid-plane of the aggregate by
$129 \pm 4 \, \rm kpc$.  
{\color{black} Considering all giants individually, the standard deviation $\sigma_z$ about the mid-plane 
is only
$233 \, \rm kpc$.  
The apparent dispersion about the supergalactic plane is 
$357 \, \rm kpc$,  
which is
53\%  
higher.
The dispersions are negligibly amplified by distance errors.
}

The plane just defined will be regarded in this paper as the mid-plane of the structure to be called the ``Local Sheet''.  Discussions of the local organization of galaxies will be founded upon a coordinate system whose $x$--$y$ plane is coincident with the mid-plane {\color{black} and whose $x$-axis points along the intersection with the supergalactic plane.  This system will be referred to as "Sheet coordinates".}  Of the 
sample galaxies within 
$0.5 \, \rm Mpc$  
of the $x$--$y$ plane, 
87\%  
are less than 
$6 \, \rm Mpc$  
distant.
Of the sample galaxies nearer than
$6 \, \rm Mpc$,  
81\%  
lie within 
$0.5 \, \rm Mpc$  
of the $x$--$y$ plane.

\begin{figure*}
\centering
\includegraphics[keepaspectratio=true,angle=0,width=10cm]{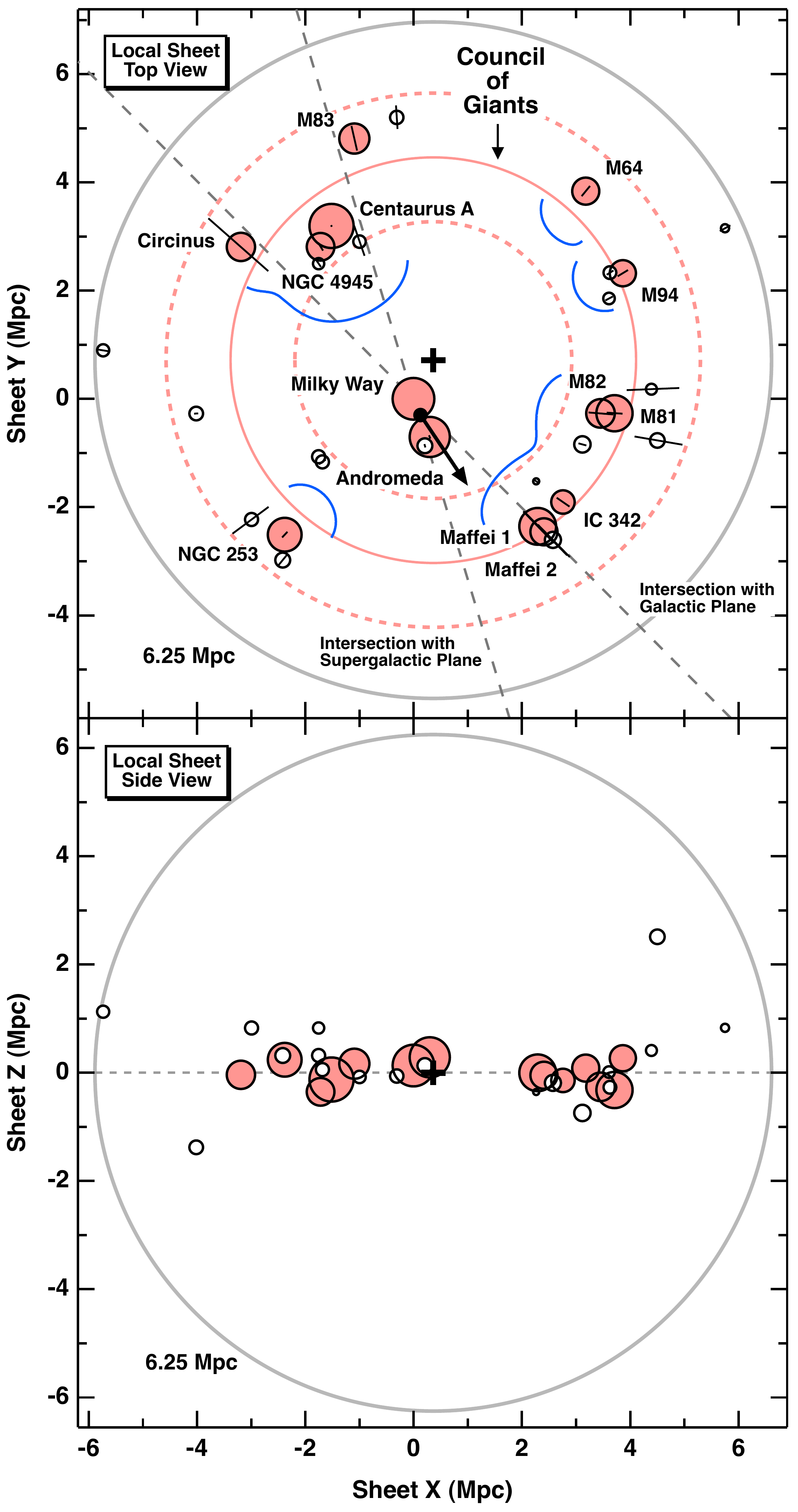}
\caption{
The spatial distribution of sample galaxies within 
$6.25 \, \rm Mpc$  
of the centre of the Council of Giants.  Shown are top and side views in a coordinate system with an $x$--$y$ plane coincident with the mid-plane of the Local Sheet, which is displayed as a dashed grey line in the lower panel.  To optimize clarity, the $x$-axis of Sheet coordinates has been rotated
$107^\circ$  
clockwise  
{\color{black} around the $z$-axis} from the direction of the line of intersection with the supergalactic plane.  In both panels, all sample galaxies within the spherical volume delineated by the large grey circles are displayed.  Galaxies are marked by circlets whose diameters are proportional to the cube root of the stellar mass.  Giants are highlighted in pink, and a bold cross marks the centre of the Council of Giants.   In the top view, black bars superimposed upon the galaxy markers convey the uncertainties in distance.  The luminosity-weighted centroid of the Local Group is noted with a small black disk, and the trajectory of the Local Group with respect to Council giants is conveyed by the attached arrow.  The {\color{black} solid pink} circle is the fit to the Council of Giants.  The inner dashed {\color{black} pink} circle marks the edge of the cylindrical realm of influence of the Local Group defined by density matching.  The outer dashed {\color{black} pink} circle correspondingly marks the outer edge of the density-matched volume of the Council.  Curves in blue are the loci of potential maxima as viewed {\color{black} today} from the centroid of the Local Group in directions parallel to the plane of the Sheet. {\color{black} Dashed grey lines mark the intersections of the Sheet with the Galactic and supergalactic planes.}    
}
\label{fig_localsheet}
\end{figure*}

Top and side views of the Local Sheet are presented in Figure~\ref{fig_localsheet}.  {\color{black} To expose salient features, displays are presented in "rotated Sheet coordinates", i.e., from the perspective generated by rotating the $x$-axis of Sheet coordinates by $107^\circ$ clockwise around the $z$-axis.}
By restricting attention to luminous galaxies, and by limiting the region displayed to that part of the Local Volume where giants are most tightly confined, the Figure shows more clearly than ever before the stark contrast between the Local Sheet and its surroundings.

\begin{figure}
\centering
\includegraphics[angle=0,width=7cm]{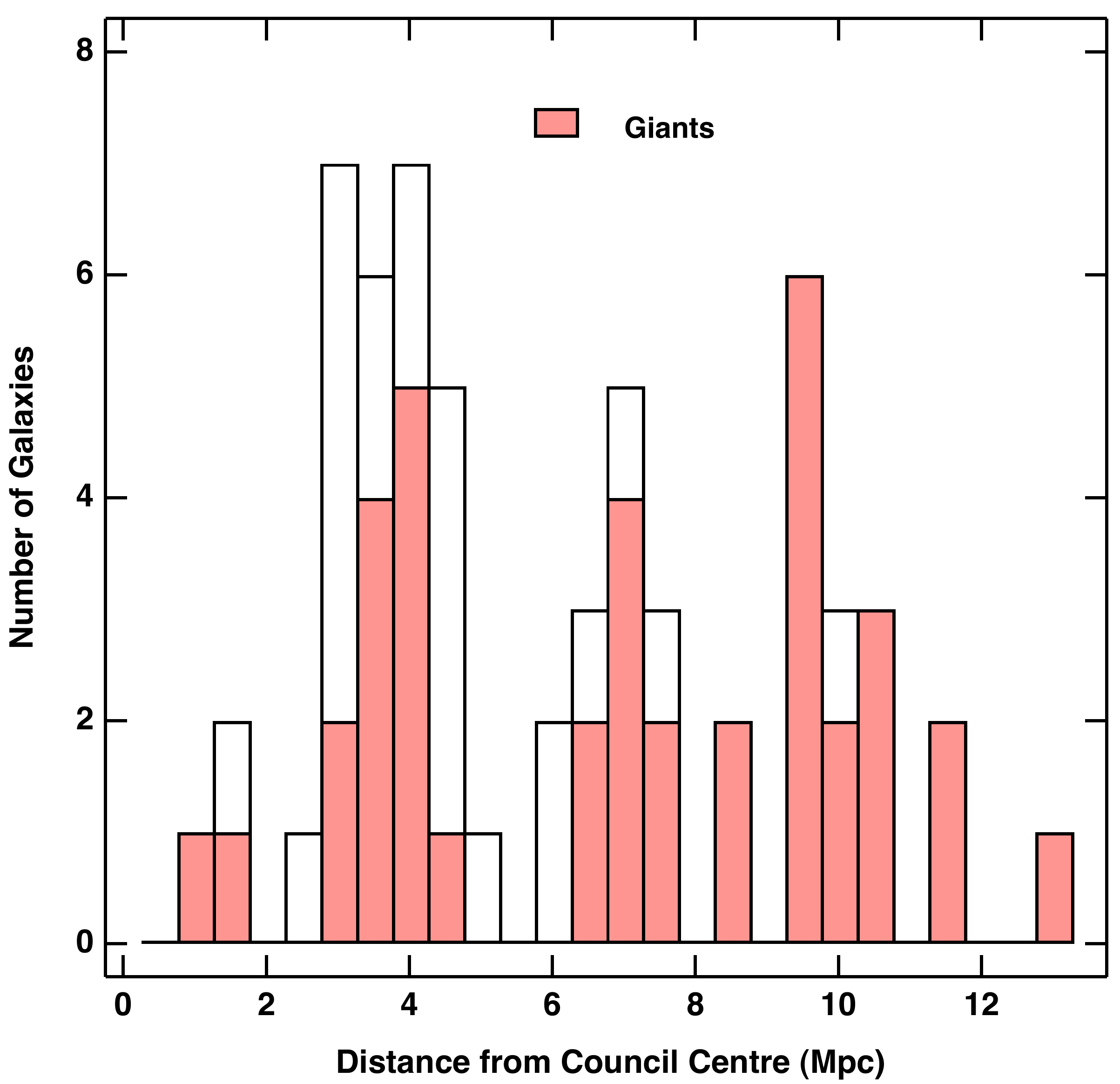}
\caption{
Distances of sample galaxies from the centre of the Council of Giants.  Giants are highlighted in pink.  The Council is evidenced by the peak centred at 
$3.75 \, \rm Mpc$.  
} 
\label{fig_distances}
\end{figure}

Beyond Andromeda, all sample giants (and most non-giants) within 
$6 \, \rm Mpc$  
of the Sun are confined to a narrow annulus encompassing the Local
Group.  This configuration will be referred to as the ``Council of Giants''.  The Council clearly stands out in Figure~\ref{fig_distances}, which shows how sample galaxies are distributed over distance from its centre (defined below).
Although it is by no means certain that the Council is anything more than an accidental arrangement of galaxies, it is worthy of its name by virtue of definitively limiting the extent of the Local Group's influence.  Also, as will be shown below, angular momenta of members expose a degree of dynamical unity.

What is displayed in Figure~\ref{fig_localsheet} are {\it all} galaxies within
$6.25 \, \rm Mpc$  
of the {\it Council centre}.
The circle best fitting the Council (handling binaries as before) has a radius of
$3.746 \pm 0.098 \, \rm  Mpc$,  
where the uncertainty is from a Monte Carlo simulation accounting for random errors in distances only (the systematic error owing to the uncertainty in the distance zero-point is 
$0.113 \, \rm Mpc$).  
{\color{black} The fit is marked by a solid pink circle in Figure~\ref{fig_localsheet}.}
The cosmic standard deviation of radial positions is estimated to be 
$0.48 \, \rm Mpc$  
after distance uncertainties are removed.
The centre of the Council (black cross) is
$0.81 \pm 0.13 \, \rm Mpc$  
from the Sun and offset along the Sheet from the centroid of the
Local Group (small solid disk) by 
$1.06 \, \rm Mpc$.  
The only giant elliptical galaxies,
Centaurus~A and Maffei~1, sit on opposite sides
of the Council, being separated in azimuth by 
$175^\circ$.  
Sample galaxies are tightly confined to the Local Sheet out to
$4 \, \rm Mpc$  
{\color{black} from the centre}, beyond which they start to become more widely dispersed vertically.

The $3 \sigma$ edge of the giant component of the Local Sheet is 
$5.2 \, \rm Mpc$  
from the centre of the Council, at which position the histogram of distances displays a clear gap (Figure~\ref{fig_distances}) .  From this perspective, the $1 \sigma$ thickness of the Sheet is only 
5\%  
of the extent.  Based upon a recent friends-of-friends analysis, dwarf irregular members of the Local Sheet are spread over an elliptical area whose
boundary ranges
$4.8$ to $7.0 \, \rm Mpc$  
from the Council centre \citep{fin12a}.  

Relative to Council galaxies and pairs, the velocity of the Local Group along the plane of the Sheet is
$11 \pm 12 \, \rm km \, s^{-1}$  
away from the Council centre toward
$-56^\circ \pm 70^\circ$  
with respect to the displayed $x$ direction of {\color{black}} rotated Sheet coordinates.
{\color{black} The vector is displayed as a thick black arrow in the top panel of Figure~\ref{fig_localsheet}}.
As viewed from the Council centre, the apex of the motion is
$47^\circ$ 
from the direction of the Local Group.  The Council appears to be in radial equilibrium  with respect to the Local Group, because after correction for the Group's translation, the mean of radially projected velocities is only
$-1 \, \rm km \, s^{-1}$,  
with a standard deviation of
$51 \, \rm km \, s^{-1}$.   
Uncertainties in heliocentric velocities and distances account for
$20 \, \rm km \, s^{-1}$  
of the spread, 
so the true velocity dispersion radial to the Council centre is
$47 \, \rm km \, s^{-1}$.  
None of the results above change significantly if the Hubble constant is varied within the range of its uncertainty.  As seen from the Sun, the velocity dispersion of isolated dwarf irregular galaxies with respect to the Local Group is
$35 \, \rm km \, s^{-1}$  \citep{fin12a},  
so motions of Council giants may be enhanced somewhat by the gravitational influence of neighbours.

Motions perpendicular to the Sheet are nearly tangential to the line of sight, so they {\color{black} cannot} be measured reliably.  However, cosmological simulations indicate that they may be {\color{black} just as cold as the radial motions \citep{fin12a}.  If so, then the time for a typical giant to pass through the Sheet (i.e., $2 \sigma_z$, or 
$465 \, \rm kpc$)  
would be
$9.6 \, \rm Gyr$,   
which is a significant fraction of the age of the Universe.  The crossing time for isolated dwarfs is even longer 
\citep{fin12a}.  
Thus, it appears that the galaxies in the Local Sheet have not had enough time to adjust dynamically to the gravitational environment in which they find themselves.}  

\begin{figure*}
\includegraphics[angle=0,width=15cm]{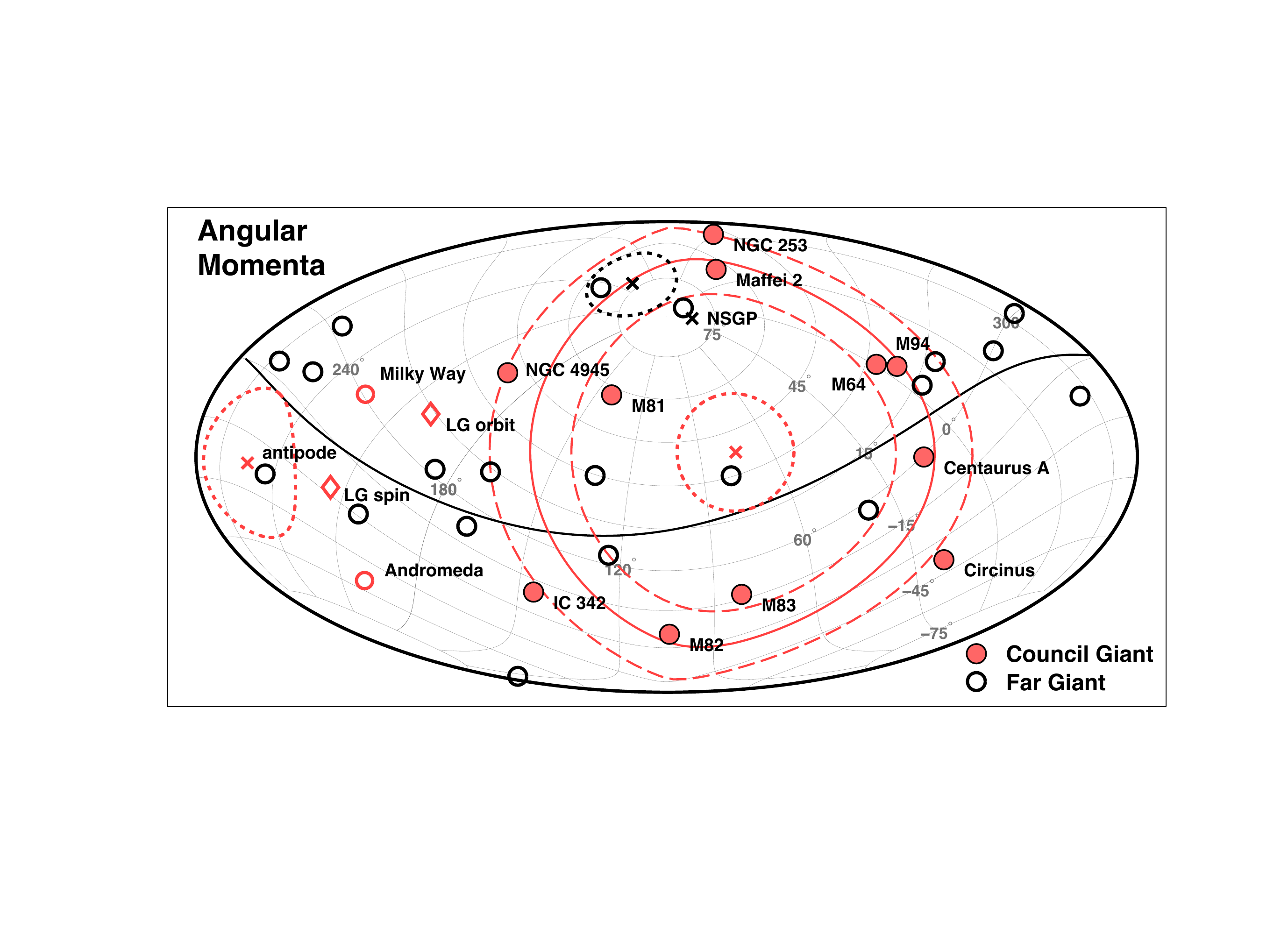}
\caption{
Angular momenta {\color{black} of galaxies} in the Local Volume.  {\color{black} Longitudes and latitudes} of angular momentum vectors {\color{black} in rotated Sheet coordinates} are conveyed via a Hammer projection.  The coordinate system is the same as that of Figure~\ref{fig_localsheet}.  
For reference, the north pole of the supergalactic coordinate system is marked by a black cross labeled ``NSGP''.   Pink markers and curves highlight features of giant galaxies in the Local Sheet, whereas black is reserved for giant galaxies beyond.    Solid pink circlets mark the directions of rotational angular momentum vectors for Council giants.  Open pink circlets show them for the Milky Way and Andromeda, and the open pink diamond labeled ``LG spin'' 
($215^\circ,-28^\circ$)  
marks the unweighted mean.  The open pink diamond labeled ``LG orbit'' 
($200^\circ,+14^\circ$)  
depicts the orbital angular momentum of Andromeda.  The solid pink curve is the small circle best fitting the locus of spin vectors of giants in the Council.  Concentric dashed pink circles outline the extent of the uncertainty in the cone angle. The pole of the small circle 
($74^\circ, +38^\circ$)  
and its uncertainty are marked by a pink cross and enclosing small dotted pink circle, respectively.  The opposite pole and error circle are marked with a pink cross labeled "antipode" and a surrounding dotted pink circle, respectively.  Open black circlets depict the directions of rotational angular momentum vectors for giants beyond the Local Sheet, and the heavy black curve is the great circle fitting all but the three close to the Sheet poles.  The north pole of the great circle 
($250^\circ,+73^\circ$)  
and its corresponding error circle are marked by a black cross and enclosing dotted black circle, respectively.
For most galaxies, the uncertainty in the direction of the spin vector is smaller than the symbol depicting it.
}
\label{fig_angmom}
\end{figure*}

It is possible to determine unambiguously the direction of the spin angular
momentum vector for most of the giants in the sample.  
Figure~\ref{fig_angmom} presents a Hammer projection displaying
the directions of the vectors for giants within the Sheet
(pink symbols)  
and beyond 
(black symbols)  
in {\color{black} rotated Sheet coordinates.} 
Spin angular momenta for Council giants (solid pink circlets) are aligned around a 
small circle with a 
radius of 
$71^\circ \pm 14^\circ$ 
and a pole
which is $38^\circ \pm 20^\circ$ 
above the plane of the Sheet (at supergalactic coordinates
$(L,B) = (125^\circ,+41^\circ)$). 
Rotational angular momenta
for the Milky Way and Andromeda (open pink circlets) point to the other side of the sky, the unweighted mean being $34^\circ$ 
away from the antipode.  
The orbital angular momentum of Andromeda (pink diamond), estimated from the most recent measurement of the proper motion of 
Andromeda \citep{mar12a} 
using the revision to the heliocentric distance presented here, {\color{black} is on the side of the antipode, too, deviating 
$14^\circ$  
from the plane of the Sheet}.  
However, because the orbit is nearly radial, this direction is extremely uncertain.
Angular momentum vectors for giants beyond the Council (open black circlets) largely follow a great circle tilted by only
$17^\circ \pm 13^\circ$ 
with respect to the Sheet plane and 
$23^\circ \pm 13^\circ$ 
with respect to the supergalactic plane. 
The three galaxies which deviate have angular momentum vectors pointing close to the poles.  Notably, the unweighted
mean spin vector for the Milky Way and Andromeda (open pink diamond labeled ``LG spin'') points only 
$13^\circ$ 
from the great circle.

\begin{figure}
\centering
\includegraphics[angle=0,width=7cm]{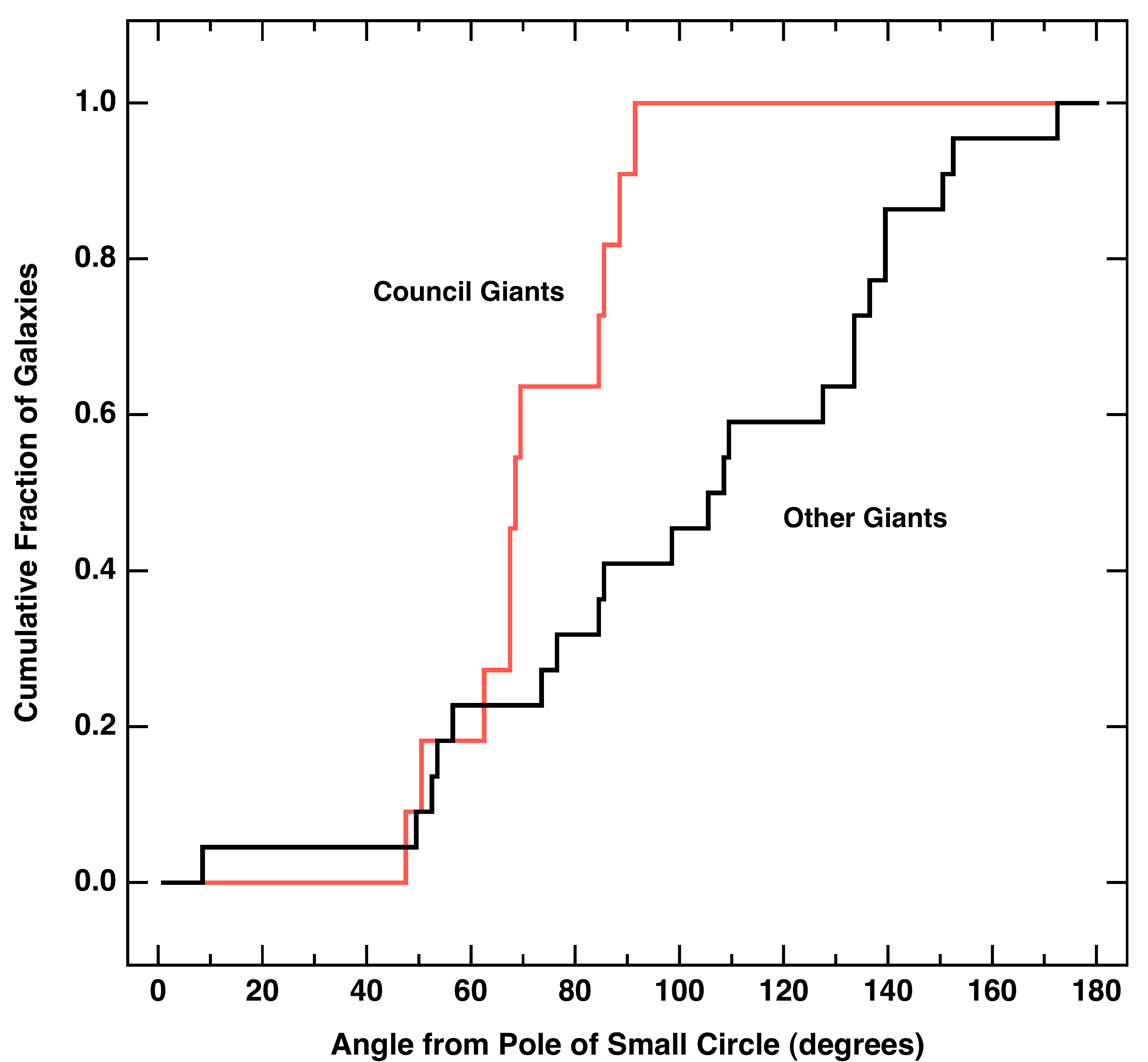}
\caption{
Cumulative histograms of spin directions relative to the small circle pole defined
by the angular momenta of Council giants.  The histogram for Council giants is shown in {\color{black} pink}.  The histogram for giants beyond the Council or in the Local Group is shown in black.
}
\label{fig_angmom_hist}
\end{figure}

The great circle's north pole, located at  
$(L,B) = (280^\circ,+67^\circ)$, 
is far from {\color{black} the small circle's pole}, lying 
$70^\circ \pm 24^\circ$ 
away.   Cumulative histograms of the deviations of angular momenta from the pole of the small circle further illustrate the differences between the two distributions (Figure~\ref{fig_angmom_hist}).
A Kuiper test reveals that the probability that the two samples are drawn from the same population is only 
$2.3$\%.   
In fact, this estimate should actually be regarded as an upper limit because the true nature of the distribution of spin orientations for giants not in the Council is smeared out in the cumulative histogram.

What is most important here 
(and missed previously by \citealt{kap83a})  
is that the angular momentum vectors of Council giants are ordered in a way which
is completely different from what is seen for giants beyond.  In this sense, the Local Sheet is dynamically distinct from the Local Supercluster.
Angular momentum vectors for dark matter haloes in a sheet are predicted to align with the plane of the sheet \citep{bai05a,lib12a},
so the observed spins of giants beyond the Council, and perhaps even the Milky Way and Andromeda, may reflect order associated with flattening on larger scales, such as that of the Local Supercluster.  There is a hint that even the Council giants once followed this pattern;  the pole of the small circle reflects the mean direction of their angular momenta, and it is located only 
$20^\circ \pm 24^\circ$  
from the great circle along which vectors for more distant giants are arranged.
It is unknown how the present arrangement of Council vectors developed, but one might speculate that it is somehow tied to torquing arising from the embedded asymmetry embodied by the mass distribution of the
Local Group \citep{lon08a}.  

\subsection{
Range and Overdensity of the Local Group
}

The Council of Giants defines a clear upper
bound to the extent of matter which contributed to the
formation of the Local Group.  In fact, its size can be used to judge
the ``realm of influence'' of the Group
and, in turn, the overdensity of the Local Sheet.
Suppose that the matter in the Local Group is spread above and below the Sheet in a cylinder around the Council centre having radius $R_\textit{LG}$. Suppose also that the matter of the Council is dispersed contiguously in a cylindrical shell with the same vertical dimension and extending inwards to $R_\textit{LG}$ from its observed radius $R_\textit{C}$ and outwards by an equal amount 
{\color{black} to $R_\textit{edge}$.  One might expect $R_\textit{edge} - R_\textit{C}$ to equal $R_\textit{C} - R_\textit{LG}$ if the driver of evolution is gravity.}
From the standpoint of continuity, there must be a value of $R_\textit{LG}$ within which the density of matter associated with the Local Group matches the density of matter associated with Council galaxies.  This radius defines the extent of the volume which could have contributed to the development of the Local Group, as well as the extent of the zone over which the Council prevailed. 
The ratio $R_\textit{LG} / R_\textit{C}$  is determined solely by the ratio
of the mass contained in Council galaxies, $\mathcal{M}_\textit{C}$, to
the mass contained in the Local Group, $\mathcal{M}_\textit{LG}$:
\begin{equation}
R_\textit{LG} / R_\textit{C} = 
2 \left[ \left( \mathcal{M}_\textit{C} / \mathcal{M}_\textit{LG} + 1 \right)^{1/2} + 1 \right]^{-1}
\end{equation}  

It is reasonable to presume that non-giants constitute comparable and relatively small fractions of the mass in the Local Group and the Council \citep{kar04a}, so giants alone can be used to judge $\mathcal{M}_\textit{C} / \mathcal{M}_\textit{LG}$ reliably.   
{\color{black}
However, to do so, the ratio of dark to baryonic mass in the Council relative to the Local Group must be constrained.

The most recent study of weak lensing \citep{vel14a} has quantified masses for galaxies within the virial radius, i.e., within the radius at which the density of matter exceeds the cosmic average by a factor of 200.   The ratio of this ``virial mass'' to the stellar mass for galaxies with spiral-like colour classes is constant within errors for stellar masses spanning the range 
$0.2 - 4 \times 10^{10} h_{70}^{-2} \, \rm \mathcal{M}_\odot$,  
i.e., over most of the range covered by giant spirals in the Local Sheet 
($h_{70} = H_0 / 70 \, \rm km \, s^{-1} \, Mpc^{-1}$
is a scaled value of the Hubble constant).
Thus, a fixed stellar mass fraction for spirals is reasonable.  There is evidence that the virial-to-stellar mass ratio for galaxies with elliptical-like colour classes does vary slowly with stellar mass over the range 
$2 - 40 \times 10^{10} h_{70}^{-2} \, \rm \mathcal{M}_\odot$,  
but in the mass range of interest here {\color{black} such galaxies} appear to have stellar mass fractions comparable to those of spirals.  Taken at face value, the results indicate that the ratio of the virial to the stellar mass for Maffei~1 and Cen~A should exceed that of the Milky Way by factors of
1.3 and 1.5,  
respectively.  Nevertheless, \citet{vel14a} caution that such a comparison is precarious because different mass ranges are probed by the two colour classes.  As a baseline for this paper, the total mass of every giant {\color{black} has been assumed} to be a fixed multiple of its stellar mass.  This approximation does not lead to a large error in $\mathcal{M}_\textit{C} / \mathcal{M}_\textit{LG}$ or in quantities derived from it because the two elliptical galaxies in the Council constitute only
36\%  
of the stellar mass there (see Table~\ref{tbl_sensitivities}).
Note that stellar masses are likely to be better indices of total masses than baryonic masses (i.e., stars plus gas).  Baryonic masses under-weight ellipticals relative to spirals because gas expelled by ellipticals is missed.
}

{\color{black} 
Stellar masses were determined from luminosities in $K_s$ utilizing 
mass-to-light ratios estimated from chemo-photometric models by \citet{por04a} founded upon the initial mass function (IMF) of \citet{kro98a}.  Syntheses by \citet{por04a} were chosen because of the careful attention given to asymptotic giants and the realistic construction of composite systems of stars.  The Kroupa IMF is superior to that of Salpeter \citep{sal55a} because it yields synthetic mass-to-light ratios in $B$, $V$, $I$, and $K$ which agree well with the values observed locally for the disk of the Milky Way \citep{por09a}.  Specifically,  stellar mass-to-light ratios were estimated from integrated 
$B-V$ colours  
using
\begin{equation}
\log {\mathcal M}_\textit{stars} / {\mathcal L}_\textit{Ks} = -0.298 + 0.73 \left[ (B-V) - 0.6 \right]
\end{equation}  
In converting absolute magnitudes to  luminosities, the absolute magnitude of the Sun in $K_s$ was adopted to be 
$3.315$ \citep{hol06a, fly06a}.}  
Fortunately, mass-to-light ratios in $K_s$ do not vary steeply with 
colour, so despite their uncertainty, relative stellar masses can be computed with some confidence. 

Stellar masses for the giant galaxies yield
$\mathcal{M}_\textit{C} / \mathcal{M}_\textit{LG} =  2.72 \pm 0.54$.  
Then, density matching leads to 
$R_\textit{LG} = 2.56 \pm 0.17 \, \rm Mpc$.  
Correspondingly, the outer boundary of the shell in which the smoothed density of Council giants matches that of the Local Group is at radius
$R_\textit{edge} = 4.94 \pm 0.27 \, \rm Mpc$.  
{\color{black}
Note that errors here do not include the uncertainty in the distance zero-point.  In Figure~\ref{fig_localsheet}, $R_\textit{LG}$ and $R_\textit{edge}$ are marked by dashed pink circles.
}

{\color{black}
Table~\ref{tbl_sensitivities} shows how {\color{black} $\mathcal{M}_\textit{C} / \mathcal{M}_\textit{LG}$, $R_\textit{LG}$, and $R_\textit{edge}$} depend upon input assumptions.  Changes are within errors if the stellar mass-to-light ratios are held fixed or if total-to-stellar mass ratios follow the trends suggested by weak lensing.  The {\color{black} value of $R_\textit{LG}$}  increases by
15\%  
if a spherical geometry is adopted.  If the specification of $R_\textit{edge}$ is modified to require that the mass of the Council be distributed equally between a shell with width $R_\textit{edge} - R_\textit{C}$ and a shell with width $R_\textit{C} - R_\textit{LG}$, i.e., to require the Council to have accumulated just as much mass from beyond $R_\textit{C}$ as from within, then $R_\textit{LG}$ decreases slightly because the density overall must rise in response to the reduced volume of the outer shell.  Again, the change is within the uncertainty.
}

It is possible to estimate the overdensity, $\Delta$, of the Local Sheet, defined here as the ratio of the observed matter density to the mean for the Universe, by comparing the mass of the Local Group with the mass of matter expected within its realm of influence at the background density.  To this end, the cylindrical volume through which matter is dispersed is approximated to have height equal to $2 R_{LG}$.  The matter density parameter derived by the Planck consortium
\citep{ade13a}   
is $\Omega_m h^2 = 0.1426 \pm 0.0025$  
(``Planck $+$ WP''), where
$h = H_0 / 100 \, \rm km \, s^{-1} \, Mpc^{-1}$.  
Thus, within the cylinder,
$\mathcal{M}_\textit{LG} / \Delta = (4.16 \pm 0.85) \times 10^{12} \, \rm \mathcal{M}_\odot$,  
independent of $h$.
This is close to the mass of the Local Group as judged from timing \citep{mar12a}, which is 
$(4.34 \pm 0.54) \times 10^{12} \, \rm \mathcal{M}_\odot$  
after accounting for revisions to the heliocentric distance and total heliocentric velocity of Andromeda (this work) and to the age of the Universe 
\citep{ade13a}.   
The corresponding overdensity is 
$1.04 \pm 0.25$.   
The overdensity rises to 
$1.21 \pm 0.47$  
if cosmological simulations are employed to virialize the timing mass \citep{mar12a}.

{\color{black} 
The estimate for the overdensity does not depend strongly on input parameters (see Table~\ref{tbl_sensitivities}).  Notably, it is insensitive to the choice of geometry}.
Of course, it would be higher if there were significant amounts of matter beyond the Local Group not incorporated in galaxies (see \citealt{fin12a}).
{\color{black} However,} the result fits expectations from models of the velocity field of galaxies within 
$40 h^{-1} \, \rm Mpc$ \citep{kly03a},  
which predict an overdensity of about unity within 
$5 h^{-1} \, \rm Mpc$. 

Most of the light of the Local Group comes from its two giants, so the total mass-to-light ratio in $K_s$ gleaned from density matching is 
$(8.9 \pm 1.5) \Delta$   
in solar units.  For Sheet giants overall, it is somewhat greater,
$(10.1 \pm 1.7) \Delta$,  
because of the enhanced stellar mass-to-light ratios of the ellipticals.
{\color{black}
The timing mass for the Local Group implies a total-to-stellar mass ratio for spiral galaxies of 
$18.5 \pm 4.1$,  
which is significantly lower than the value of 
$43 \pm 9$  
suggested by weak lensing.
}

Comparing the fraction of the mass of Sheet giants in stellar form with the cosmic ratio of baryons to matter, {\color{black} density matching yields an}
efficiency
of galaxy formation  \citep{ost12a} equal to
$(0.365 \pm 0.060) / \Delta$.   
{\color{black} Utilizing the timing mass for the Local Group, the efficiency works out to}
$0.349 \pm 0.077$.  
The estimates must be
regarded as lower limits to the true efficiency because gas incorporated in galaxies has not been accommodated.
{\color{black}
How the estimate for the efficiency derived from density matching depends upon input parameters is summarized in Table~\ref{tbl_sensitivities}.
}

If all matter within $R_\textit{edge}$ were dispersed evenly across the plane of the Sheet, the surface mass density would be 
$(0.202 \pm 0.014) \Delta \, \rm \mathcal{M}_\odot \, pc^{-2}$. 
This corresponds to 
$0.092 \pm 0.027$ Milky Ways  
per square Megaparsec, independent of $\Delta$.

\subsection{
Evolution of the Local Group
}

The reservoir of matter available to contribute to the development of the Local Group must have been limited by {\color{black} the gravitation of} surrounding material.  In the upper panel of Figure~\ref{fig_localsheet}, blue curves trace maxima in the potential surface described by the gravitational fields of the 
14  
giants in the Local Sheet as viewed {\color{black} today} from the luminosity-weighted centroid of the Local Group.  {\color{black} Relative masses were gauged from stellar masses assuming a fixed total-to-stellar mass ratio.}  Council galaxies, especially the ellipticals, clearly restrict the domain of the Local Group.  
If mass equivalent to the Local Group giants is placed at the centre of the Council, and mass equivalent to Council giants is uniformly spread around the Council at radius $R_\textit{C}$, then the potential in the plane of the Sheet peaks at a radius of 
$2.6 \, \rm Mpc$.  
{\color{black} This is identical to the radius of the realm of influence of the Local Group derived from density matching.
Neither result depends on the mass scale, and sensitivities to relative masses are extremely weak.}

{\color{black} 
Because of the unique arrangement of galaxies in the Council, the two elliptical galaxies would have gravitationally confined any mass concentration in the Council in two orthogonal directions.
It is conceivable that galaxies in the Canes Venatici I group (NGC 4736 and NGC 4826) and the Sculptor group (NGC 253) were particularly vulnerable, given that the ellipticals are roughly equidistant.
}

There is additional, albeit circumstantial, evidence that the development of the Local Group was influenced by local structure.  The position vector of Andromeda with respect to the Milky Way is inclined by only 
$11^\circ$  
with respect to the Local Sheet.  Projected on to the Sheet, it deviates by only
$11^\circ$  
from the axis of the ellipticals.  Furthermore,
accounting for tangential 
motion \citep{mar12a},  
the current trajectory of Andromeda is {\color{black} is at an angle of only 
$3^\circ$  
with respect to the plane of the Sheet in a direction 
$13^\circ$  
from the elliptical axis.   These observations hint that binarity may have been connected somehow to the existence of the elliptical dipole.}

Studies of the spatial anisotropy of dwarfs in the Local Group also suggest a connection with the organization of matter beyond the Local Group \citep{pas07a,lee08a}.  The axis of least dispersion is
$29^\circ$  
from the pole of the Local Sheet, and the potential field inferred from the axis of greatest dispersion suggests that tidal forces are maximized in a direction only 
$8^\circ$ to $15^\circ$  
away from Maffei~1.  
The recently-discovered extended array of dwarf galaxies in orbit around the Andromeda galaxy
\citep{iba13a,con13a}  
occupies a plane inclined by only 
$18^\circ$  
to the Local Sheet.  Such a close alignment would be expected if the system were an outcome of {\color{black} an interaction of bodies confined to moving} within a pre-existing flattened framework of matter. 

Along the line joining the two elliptical galaxies, the potential 
from Council galaxies (i.e.,
excluding the influence of the Milky Way and Andromeda) peaks 
$0.3 \, \rm Mpc$ 
from the Council centre 
($0.8 \, \rm Mpc$ 
from the centroid of the Local Group).  There is a broad shallow minimum in the perpendicular direction. Thus, the centre represents a point of instability.  If the Local Group started there, it would likely have moved in the general direction of an elliptical.  Indeed,
projected on to the Local Sheet, the Local Group is moving away from the Council centre in the direction of Maffei~1 on a trajectory which is only 
$1^\circ$  
from parallel to the axis of the ellipticals.
However, at the current position of the Local Group, the potential arising from Council giants as {\color{black} they are configured today} is
3\%  
higher than at the centre 
{\color{black} (a barrier of $30 \, \rm km \, s^{-1}$).}   
Also, even after augmenting the translational motion by $3\sigma$, the Local Group would be able to move only 
60\%  
of the way from the centre to its present position in a time less than the age of the Universe.  Consequently, in a relative sense, it is likely that the Local Group developed at a place offset from the Council centre.  

\section{
Discussion
}

Being giant elliptical galaxies, it is likely that both Maffei 1 and Centaurus A
developed strong winds during their evolution as a result of heating by massive stars and supernovae \citep{mat71a,lar74a,mat02a,pip04a,pin10a,cot12a}.  
During the Sedov-Taylor phase of their expansion, the winds could have shepherded gas located between the
two galaxies, possibly contributing to the growth of disks in the Local Group.  The winds might also have had a bearing on confining gas and triggering star formation in nearby galaxies.  Stellar velocity dispersions offer a means of probing the salient details \citep{mcc98a,mcc03a}.  

{\color{black} An energy-driven} wind would have developed once the gas temperature rose to the virial temperature, which is set by the potential.  That temperature can be gauged today from the kinetic energy per unit mass of stars, $u_s$, assuming that dark matter dominates the mass.  The velocity dispersions for Maffei~1 and Cen~A are
$180 \pm 7 \, \rm km \, s^{-1}$  
and
$126 \pm 9 \, \rm km \, s^{-1}$,  
respectively,
after correction to a radius equal to one-eighth of the effective radius \citep{fin03a,sil05a,duf79a} .  
Thus, the virial temperatures for Maffei~1 and Cen~A are 
$(2.4 \pm 0.2) \times 10^6 \, \rm K$  
and 
$(1.2 \pm 0.2) \times 10^6 \, \rm \rm K$,  
respectively.  
{\color{black}
The escape velocity is supersonic at the virial temperature, and
the terminal velocity of ejected gas could have been as high as
three  
times the speed of sound,
}
i.e., 
$(700 \pm 30) \, \rm km \, s^{-1}$  
in the case of Maffei~1 and 
$(490 \pm 30) \, \rm km \, s^{-1}$  
in the case of Cen~A
\citep{kha53a, pac53a, che85a, mur05a, opp06a}.
At the time of the wind, the internal energy of gas heated by stellar processes would have been linked to the total mass of stars formed, $\mathcal{M}_s$.  Consequently, the virial condition for mass loss requires that the mass of gas that was ejected be proportional to 
$\mathcal{M}_s / u_s$.   
Because the stellar mass of Cen~A is
$1.7 \pm 0.7$  
times greater than that of Maffei~1, Cen~A would have ejected 
$3.5 \pm 1.6$  
times more
gas, and the Sedov-Taylor radius would have been 
$1.5 \pm 0.2$  
times larger \citep{mck95a}.  
If the overdensity of intergalactic baryons were comparable to or below that estimated for all matter, then the ejected gas could have expanded freely all the way to the Local Group.  For example,
if gas left Maffei~1 at a redshift of 
2.5  
or Cen~A at a redshift of 5.0,  
it would have reached the Local Group by a redshift of 
1.5,  
the epoch of peak star 
formation
9.5  
billion years ago \citep{soi08a}. 
Thus, it is feasible that gas expelled from the Council ellipticals could have influenced the baryonic evolution of disk galaxies in the Local Sheet.

How well the Local Sheet retains intergalactic baryons, be they primordial or from galactic winds, depends upon its virial temperature.   
{\color{black} The virial temperature $T_\textit{vir}$ near the mid-plane of a sheet with radius $R$ and surface mass density $\Sigma$ is given by
\begin{equation}
T_\textit{vir} = \frac{2 \pi G}{3 k} A m_p \Sigma R
\end{equation}
where $G$ is the gravitational constant, $k$ is Boltzmann's constant, $A$ is the mean molecular weight, and $m_p$ is the mass of the proton.
Adopting a radius of
$5.2 \, \rm Mpc$  
for the giant component,}
the estimated mean surface mass density of matter implies a virial temperature of 
$7 \times 10^5 \, \rm K$  
and an escape velocity of 
$240 \, \rm km \, s^{-1}$.  
\citet{fin12a} has suggested that there may be as much as a factor of two more mass in the Sheet than is found in galaxies, so the actual virial temperature could easily be over a million degrees.  Therefore, warm gas may permeate the Sheet, but gas ejected by the ellipticals in directions at large angles to the Sheet would have escaped.

\section{
Conclusions
}

{\color{black} 
Properties of the Local Sheet and its Council of Giants are summarized in Table~\ref{tbl_summary}.
}
This study suggests that a structure with the geometry of the Sheet was instrumental in guiding the formation and evolution of constituent galaxies.  It also suggests that a binary, or the precursor of it, can influence the angular momentum acquired by neighbouring galaxies.  It is unlikely that a randomly dispersed collection of galaxies could have agglomerated into a structure as cold as the Sheet in a way which could generate an interacting pair of galaxies {\color{black} near the middle of a ring of galaxies with opposing ellipticals and ordered spins}.  Indeed, modern cosmological simulations reveal galaxies developing from dark cores fed by flows of gas along pre-existing filaments of dark matter \citep{dan12a}.  

Unfortunately, there is only one Local Sheet.  Further insights into the formation and evolution of the Local Sheet, and particularly guidance on the interplay between the Local Group and the Local Sheet, will require the identification of like systems in the greater Universe. 

\section*{Acknowledgements}
The author thanks R. Fingerhut for many stimulating conversations about the Local Sheet as she progressed in her investigations of the dwarf component, and for her comments on the first draft of this paper.   {\color{black} The author is grateful to S. Sakai for her help in understanding the foundations of the Key Project Tully-Fisher relations and to K. Herrmann for her assistance in interpreting her measurements of luminosity functions of planetary nebulae.} Thanks are conveyed also to 
R.~M. Stesky of Pangaea Scientific for support with Spheristat, a geophysical software package seconded to study the organization of angular momentum vectors, and to G. Conidis for statistical insights.  Particular gratitude is expressed to S. McCall,  A. Boudakian, M. Doherty, C. Law, J. Marshall, M. Ng, and F. Shariff, whose attention to the author's health allowed this work to be completed, and to N. McCall for thought-provoking discussions and for his strength during times of adversity.  Finally, the author is grateful to the Natural Sciences and Engineering Research Council of Canada for its continuing support.  

\bibliographystyle{mn2e_from_williams}
\bibliography{mccall_references,mccall_references_tab2}

\clearpage
\onecolumn
\setcounter{table}{0}

\begin{center}
\begin{ThreePartTable}

\begin{TableNotes}

\item[]
(1) 
Name of galaxy, in order of right ascension;
Longitude in rotated Sheet coordinates, in degrees;
Latitude in rotated Sheet coordinates, in degrees.
The natural Sheet coordinate system has its north pole (the direction of positive $z$) at supergalactic coordinates
$(L,B) = (241\fdg74 \pm 0\fdg74, 82\fdg05 \pm 0\fdg12)$  
with the $x$-$y$ plane perpendicularly offset from the Sun 
southward  
of the supergalactic plane by 
$129 \pm 4 \, \rm kpc$.  
The positive $x$-axis points parallel to the line of nodes towards 
$L = 151\fdg74$  
and the positive $y$-axis points towards the supergalactic longitude of the north pole.
In the rotated system, the $x$-axis has been rotated by 
$106\fdg74$ clockwise,  
so longitudes of galaxies are concomitantly greater. 
(2) 
Numerical index of the morphological stage in the Revised Hubble System;
Heliocentric radial velocity, in $\rm km \, s^{-1}$;
Radial velocity, corrected for the Hubble flow if the heliocentric distance is beyond $1 \, \rm Mpc$), in the frame of reference of the luminosity-weighted centroid of the Local Group, in $\rm km \, s^{-1}$.  {\color{black} The value adopted for the Hubble constant was 
$71.6 \, \rm km \, s^{-1} \, Mpc^{-1}$.}  
(3) 
Optical depth of interstellar dust {\color{black} in the Milky Way} at $1 \, \rm \mu m$;
Heliocentric distance modulus on the maser scale \citep{hum13a}, in $\rm mag$;
Method used to determine the distance modulus ($\mbox{C}$ = Cepheid variables in $V$ and $I$; \mbox{FP} = fundamental plane; $\mbox{P}$ = planetary nebulae; $\mbox{S}$ = surface brightness fluctuations in $I$; $\mbox{T}$ = tip of the red giant branch in $I$; \mbox{TF} = Tully-Fisher relation).  The distance to the centre of the Milky Way is from \citet{mar12a}.  For any distance determined using more than one method, the tabulated error is the standard deviation of the estimates.  Otherwise, the error comes from propagating uncertainties in observational parameters.  The uncertainty in the zero-point of the distance scale is not included.  Even though M106 (NGC~4258) sets the zero-point via its masers, the uncertainty recorded for its distance is based upon the dispersion of its stellar indicators.
(4) 
Rotated Cartesian Sheet coordinates, in $\rm Mpc$.  The origin is the projection of the Sun onto the plane of the Local Sheet.  In this system, the luminosity-weighted centroid of the Local Group is at 
$(X,Y,Z) = (0.124, -0.297, 0.200)$,  
and the centre of the Council of Giants is at 
$(X,Y,Z) = (0.362, 0.718, 0.000)$.  
(5) 
Absolute magnitude in $K_s$, in $\rm mag$;
Absolute magnitude in $V$, in $\rm mag$;
Fully-corrected integrated $B-V$ colour, in $\rm mag$.
(6)
Logarithm of the stellar mass, in solar units, {\color{black} based upon an absolute magnitude for the Sun of 
$3.315 \, \rm mag$ in $K_s$  
(the uncertainty accounts only for the error in the luminosity)};
Inclination to the plane of the sky, in degrees; 
Position angle of the line of nodes, measured in degrees eastward from north to the first limb (epoch 1950 assumed).  
Symbols in parentheses next to the inclination and position angle resolve the ambiguity in the orientation of the spin vector.  In the case of the inclination, the symbol $+$ ($-$) signifies arms open counter-clockwise (clockwise) or that the near side is at the position angle of the receding limb plus (minus) $90^\circ$. In the case of the position angle, the symbol signifies whether the specified limb is receding ($+$) or approaching ($-$).  The symbol $0$ signifies indeterminate.
(7)
Tilt-corrected rotational velocity in the plateau of the rotation curve, in $\rm km \, s^{-1}$;
Longitude of the angular momentum vector in rotated Sheet coordinates, in degrees;
Latitude of the angular momentum vector in rotated Sheet coordinates, in degrees.


\end{TableNotes}

\setlength{\jot}{-0.0pt}

\begin{longtable}{ccccccc}

\caption{Galaxies in the Sample\label{tbl_sample}}  
\\

\hline
\noalign{\smallskip}

\begin{array}[t]{c} \mbox{Galaxy} \\  L_\textit{sheet} \\ B_\textit{sheet} \end{array} &
\begin{array}[t]{c} 	T \\  V_\odot \\ V_\textit{LG} \end{array}	&
\begin{array}[t]{c} 	 \tau_{\scriptscriptstyle 1} \\ \mathit{DM} \\ \mbox{Method} \end{array}	&
\begin{array}[t]{c} 	X_\textit{sheet} \\ Y_\textit{sheet} \\ Z_\textit{sheet}	\end{array}	&
\begin{array}[t]{c}	M_\textit{Ks}	\\ M_\textit{V} \\ B-V \end{array}	&
\begin{array}[t]{c} 	\log {\mathcal M}_\textit{stars} \\ i \\ \mathit{PA} 	\end{array}	&
\begin{array}[t]{c} 	V_\textit{flat} \\  L_\textit{sheet} (\mathit{AM}) \\ B_\textit{sheet}(\mathit{AM})	\end{array}
\\ 

\noalign{\smallskip}

(1) & 
(2) & 
(3) & 
(4) & 
(5) & 
(6) &
(7)
\\

\noalign{\smallskip}
\hline
\noalign{\smallskip}
\endfirsthead

\caption{cont'd.} \\

\hline
\noalign{\smallskip}

\begin{array}[t]{c} \mbox{Galaxy} \\  L_\textit{sheet} \\ B_\textit{sheet} \end{array} &
\begin{array}[t]{c} 	T \\  V_\odot \\ V_\textit{LG} \end{array}	&
\begin{array}[t]{c} 	 \tau_{\scriptscriptstyle 1} \\ \mathit{DM} \\ \mbox{Method} \end{array}	&
\begin{array}[t]{c} 	X_\textit{sheet} \\ Y_\textit{sheet} \\ Z_\textit{sheet}	\end{array}	&
\begin{array}[t]{c}	M_\textit{Ks}	\\ M_\textit{V} \\ B-V \end{array}	&
\begin{array}[t]{c} 	\log {\mathcal M}_\textit{stars} \\ i \\ \mathit{PA} 	\end{array}	&
\begin{array}[t]{c} 	V_\textit{flat} \\  L_\textit{sheet} (\mathit{AM}) \\ B_\textit{sheet}(\mathit{AM})	\end{array}
\\ 

\noalign{\smallskip}
 
(1) & 
(2) & 
(3) & 
(4) & 
(5) & 
(6) &
(7)
\\

\noalign{\smallskip}
\hline
\endhead

\noalign{\smallskip}
\hline

\endfoot
\hline
\noalign{\smallskip}
\insertTableNotes

\endlastfoot


$\begin{aligned}[t]	\mbox{\bf NGC 55\phantom{oo 1}}	&	\\	211.30	&	\\	8.87	&	\end{aligned}$ &	$\begin{aligned}[t]	\phantom{-11}	8.7	&	\\		120.5	& \pm	3.0	\phantom{1}	\\		-51.6		& \pm	6.3	\phantom{1}	\end{aligned}$ &	$\begin{aligned}[t]	0.015	& \pm	0.002	\\	26.573	& \pm	0.050	\\	&	\mbox{C,T}	\end{aligned}$ &	$\begin{aligned}[t]		-1.756	& \pm	0.040	\\	-1.067	& \pm	0.025	\\	0.321	& \pm	0.004	\end{aligned}$ &	$\begin{aligned}[t]	-21.36	& \pm	0.23	\\	-18.96	& \pm	0.07	\\	0.425	& \pm	0.050	\end{aligned}$ &	$\begin{aligned}[t]	\phantom{1}	9.445	& \pm	0.092	\\	81.2	& \pm	1.6	\; (	-	)	\phantom{1}	\\	105.0	& \pm	4.0	\; (	+	)	\end{aligned}$ &	$\begin{aligned}[t]	\phantom{111}	84	& \pm	2	\phantom{1111}	\\		112.7	& \pm	1.6	\phantom{1}	\\	\phantom{1}	-2.2	& \pm	4.0	\phantom{1}	\end{aligned}$	\\	\noalign{\smallskip}
$\begin{aligned}[t]	\mbox{\bf Andromeda\phantom{1}}	&	\\	292.90	&	\\	20.72	&	\end{aligned}$ &	$\begin{aligned}[t]	\phantom{-11}	3.0	&	\\		-301.0	& \pm	1.0	\phantom{1}	\\		-62.5	& \pm	9.8	\phantom{1}	\end{aligned}$ &	$\begin{aligned}[t]	0.070	& \pm	0.011	\\	24.453	& \pm	0.093	\\	&	\mbox{C,P,S,T}	\end{aligned}$ &	$\begin{aligned}[t]		0.296	& \pm	0.013	\\	-0.701	& \pm	0.030	\\	0.288	& \pm	0.007	\end{aligned}$ &	$\begin{aligned}[t]	-24.94	& \pm	0.25	\\	-22.17	& \pm	0.13	\\	0.652	& \pm	0.041	\end{aligned}$ &	$\begin{aligned}[t]		11.042	& \pm	0.101	\\	78.0	& \pm	0.5	\; (	-	)	\phantom{1}	\\	37.9	& \pm	0.5	\; (	+	)	\end{aligned}$ &	$\begin{aligned}[t]	\phantom{11}	226	& \pm	5	\phantom{111}	\\		198.8	& \pm	0.9	\phantom{1}	\\		-54.8	& \pm	0.5	\phantom{1}	\end{aligned}$	\\	\noalign{\smallskip}
$\begin{aligned}[t]	\mbox{\bf NGC 247\phantom{o 1}}	&	\\	230.89	&	\\	4.76	&	\end{aligned}$ &	$\begin{aligned}[t]	\phantom{-11}	6.9	&	\\		161.0	& \pm	7.0	\phantom{1}	\\		-72.5		& \pm	12.7		\end{aligned}$ &	$\begin{aligned}[t]	0.020	& \pm	0.003	\\	27.922	& \pm	0.078	\\	&	\mbox{T}	\end{aligned}$ &	$\begin{aligned}[t]		-2.420	& \pm	0.087	\\	-2.976	& \pm	0.107	\\	0.320	& \pm	0.007	\end{aligned}$ &	$\begin{aligned}[t]	-21.84	& \pm	0.26	\\	-19.41	& \pm	0.14	\\	0.440	& \pm	0.050	\end{aligned}$ &	$\begin{aligned}[t]	\phantom{1}	9.648	& \pm	0.103	\\	74.8	& \pm	0.8	\; (	+	)	\phantom{1}	\\	170.6	& \pm	0.5	\; (	+	)	\end{aligned}$ &	$\begin{aligned}[t]	\phantom{11}	105	& \pm	7	\phantom{111}	\\		173.9	& \pm	1.8	\phantom{1}	\\	\phantom{-}	66.5	& \pm	0.6	\phantom{1}	\end{aligned}$	\\	\noalign{\smallskip}
$\begin{aligned}[t]	\mbox{\bf NGC 253\phantom{o 1}}	&	\\	226.46	&	\\	4.02	&	\end{aligned}$ &	$\begin{aligned}[t]	\phantom{-11}	5.1	&	\\		236.0	& \pm	1.0	\phantom{1}	\\		10.0	& \pm	6.6	\phantom{1}	\end{aligned}$ &	$\begin{aligned}[t]	0.022	& \pm	0.003	\\	27.695	& \pm	0.045	\\	&	\mbox{P,T}	\end{aligned}$ &	$\begin{aligned}[t]		-2.382	& \pm	0.049	\\	-2.506	& \pm	0.052	\\	0.243	& \pm	0.002	\end{aligned}$ &	$\begin{aligned}[t]	-24.37	& \pm	0.05	\\	-21.40	& \pm	0.07	\\	0.638	& \pm	0.050	\end{aligned}$ &	$\begin{aligned}[t]		10.805	& \pm	0.019	\\	75.9	& \pm	0.9	\; (	+	)	\phantom{1}	\\	51.1	& \pm	0.5	\; (	-	)	\end{aligned}$ &	$\begin{aligned}[t]	\phantom{11}	217	& \pm	4	\phantom{111}	\\		297.2	& \pm	1.4	\phantom{1}	\\	\phantom{-}	48.6	& \pm	0.5	\phantom{1}	\end{aligned}$	\\	\noalign{\smallskip}
$\begin{aligned}[t]	\mbox{\bf NGC 300\phantom{o 1}}	&	\\	214.56	&	\\	1.66	&	\end{aligned}$ &	$\begin{aligned}[t]	\phantom{-11}	6.9	&	\\		144.5	& \pm	3.0	\phantom{1}	\\		-37.8	& \pm	7.3	\phantom{1}	\end{aligned}$ &	$\begin{aligned}[t]	0.015	& \pm	0.002	\\	26.558	& \pm	0.080	\\	&	\mbox{C,P,T}	\end{aligned}$ &	$\begin{aligned}[t]		-1.687	& \pm	0.062	\\	-1.162	& \pm	0.043	\\	0.059	& \pm	0.003	\end{aligned}$ &	$\begin{aligned}[t]	-21.19	& \pm	0.26	\\	-18.59	& \pm	0.14	\\	0.550	& \pm	0.050	\end{aligned}$ &	$\begin{aligned}[t]	\phantom{1}	9.468	& \pm	0.103	\\	46.4	& \pm	3.6	\; (	-	)	\phantom{1}	\\	106.8	& \pm	1.2	\; (	-	)	\end{aligned}$ &	$\begin{aligned}[t]	\phantom{111}	94	& \pm	8	\phantom{1111}	\\		348.5	& \pm	3.6	\phantom{1}	\\	\phantom{1}	-3.6	& \pm	0.9	\phantom{1}	\end{aligned}$	\\	\noalign{\smallskip}
$\begin{aligned}[t]	\mbox{\bf M33\phantom{ngeloo 1}}	&	\\	283.49	&	\\	8.60	&	\end{aligned}$ &	$\begin{aligned}[t]	\phantom{-11}	6.0	&	\\		-180.0	& \pm	1.0	\phantom{1}	\\		11.8	& \pm	8.9	\phantom{1}	\end{aligned}$ &	$\begin{aligned}[t]	0.048	& \pm	0.008	\\	24.741	& \pm	0.078	\\	&	\mbox{C,P,T}	\end{aligned}$ &	$\begin{aligned}[t]		0.207	& \pm	0.007	\\	-0.863	& \pm	0.031	\\	0.134	& \pm	0.000	\end{aligned}$ &	$\begin{aligned}[t]	-21.77	& \pm	0.24	\\	-19.30	& \pm	0.10	\\	0.462	& \pm	0.021	\end{aligned}$ &	$\begin{aligned}[t]	\phantom{1}	9.636	& \pm	0.096	\\	54.0	& \pm	0.5	\; (	+	)	\phantom{1}	\\	22.5	& \pm	0.5	\; (	-	)	\end{aligned}$ &	$\begin{aligned}[t]	\phantom{11}	106	& \pm	4	\phantom{111}	\\		299.8	& \pm	0.7	\phantom{1}	\\	\phantom{-}	52.6	& \pm	0.5	\phantom{1}	\end{aligned}$	\\	\noalign{\smallskip}
$\begin{aligned}[t]	\mbox{\bf M74\phantom{ngeloo 1}}	&	\\	268.97	&	\\	-2.17	&	\end{aligned}$ &	$\begin{aligned}[t]	\phantom{-11}	5.2	&	\\	\phantom{-}	655.5	& \pm	1.5		\\	\phantom{-}	162.8	& \pm	15.8		\end{aligned}$ &	$\begin{aligned}[t]	0.080	& \pm	0.013	\\	29.759	& \pm	0.067	\\	&	\mbox{P}	\end{aligned}$ &	$\begin{aligned}[t]		-0.160	& \pm	0.005	\\	-8.936	& \pm	0.276	\\	-0.339	& \pm	0.014	\end{aligned}$ &	$\begin{aligned}[t]	-23.14	& \pm	0.09	\\	-20.85	& \pm	0.27	\\	0.482	& \pm	0.015	\end{aligned}$ &	$\begin{aligned}[t]		10.196	& \pm	0.034	\\	9.3	& \pm	0.9	\; (	+	)	\phantom{1}	\\	25.0	& \pm	5.0	\; (	+	)	\end{aligned}$ &	$\begin{aligned}[t]	\phantom{11}	147	& \pm	16	\phantom{11}	\\		266.6	& \pm	0.8	\phantom{1}	\\		-11.9	& \pm	0.8	\phantom{1}	\end{aligned}$	\\	\noalign{\smallskip}
$\begin{aligned}[t]	\mbox{\bf NGC 672\phantom{o 1}}	&	\\	280.92	&	\\	-2.01	&	\end{aligned}$ &	$\begin{aligned}[t]	\phantom{-11}	6.0	&	\\	\phantom{-}	422.0	& \pm	2.0	\phantom{1}	\\	\phantom{-}	182.8	& \pm	50.9		\end{aligned}$ &	$\begin{aligned}[t]	0.089	& \pm	0.014	\\	28.810	& \pm	0.348	\\	&	\mbox{TF in I}	\end{aligned}$ &	$\begin{aligned}[t]		1.094	& \pm	0.175	\\	-5.666	& \pm	0.909	\\	-0.202	& \pm	0.053	\end{aligned}$ &	$\begin{aligned}[t]	-20.54	& \pm	0.35	\\	-18.12	& \pm	0.37	\\	0.457	& \pm	0.052	\end{aligned}$ &	$\begin{aligned}[t]	\phantom{1}	9.140	& \pm	0.141	\\	64.3	& \pm	0.7	\; (	+	)	\phantom{1}	\\	67.5	& \pm	3.5	\; (	+	)	\end{aligned}$ &	$\begin{aligned}[t]	\phantom{111}	78	& \pm	5	\phantom{111}	\\		219.5	& \pm	1.2	\phantom{1}	\\		-32.3	& \pm	3.1	\phantom{1}	\end{aligned}$	\\	\noalign{\smallskip}
$\begin{aligned}[t]	\mbox{\bf NGC 891\phantom{o 1}}	&	\\	297.21	&	\\	-5.52	&	\end{aligned}$ &	$\begin{aligned}[t]	\phantom{-11}	3.0	&	\\		528.0	& \pm	2.0	\phantom{1}	\\	\phantom{1}	-6.8	& \pm	31.4		\end{aligned}$ &	$\begin{aligned}[t]	0.074	& \pm	0.012	\\	30.022	& \pm	0.098	\\	&	\mbox{P,T}	\end{aligned}$ &	$\begin{aligned}[t]		4.592	& \pm	0.207	\\	-8.931	& \pm	0.403	\\	-0.971	& \pm	0.050	\end{aligned}$ &	$\begin{aligned}[t]	-24.66	& \pm	0.10	\\	-21.54	& \pm	0.21	\\	0.551	& \pm	0.051	\end{aligned}$ &	$\begin{aligned}[t]		10.856	& \pm	0.040	\\	88.3	& \pm	1.5	\; (	-	)	\phantom{1}	\\	22.5	& \pm	0.5	\; (	-	)	\end{aligned}$ &	$\begin{aligned}[t]	\phantom{11}	227	& \pm	5	\phantom{111}	\\		340.3	& \pm	10.2		\\	\phantom{-}	83.8	& \pm	1.1	\phantom{1}	\end{aligned}$	\\	\noalign{\smallskip}
$\begin{aligned}[t]	\mbox{\bf NGC 925\phantom{o 1}}	&	\\	289.10	&	\\	-9.13	&	\end{aligned}$ &	$\begin{aligned}[t]	\phantom{-11}	6.9	&	\\	\phantom{-}	546.3	& \pm	3.9	\phantom{1}	\\	\phantom{-}	27.0	& \pm	15.8		\end{aligned}$ &	$\begin{aligned}[t]	0.086	& \pm	0.014	\\	29.902	& \pm	0.048	\\	&	\mbox{C}	\end{aligned}$ &	$\begin{aligned}[t]		3.081	& \pm	0.068	\\	-8.899	& \pm	0.197	\\	-1.513	& \pm	0.036	\end{aligned}$ &	$\begin{aligned}[t]	-22.30	& \pm	0.07	\\	-20.30	& \pm	0.14	\\	0.424	& \pm	0.052	\end{aligned}$ &	$\begin{aligned}[t]	\phantom{1}	9.819	& \pm	0.029	\\	61.0	& \pm	5.0	\; (	+	)	\phantom{1}	\\	109.3	& \pm	2.7	\; (	-	)	\end{aligned}$ &	$\begin{aligned}[t]	\phantom{11}	112	& \pm	6	\phantom{1111}	\\		350.6	& \pm	5.0	\phantom{1}	\\	\phantom{1}	-6.0	& \pm	2.4	\phantom{1}	\end{aligned}$	\\	\noalign{\smallskip}
$\begin{aligned}[t]	\mbox{\bf NGC 1023\phantom{ 1}}	&	\\	295.25	&	\\	-9.87	&	\end{aligned}$ &	$\begin{aligned}[t]	\phantom{1}	-2.7	&	\\	\phantom{-}	617.0	& \pm	1.0	\phantom{1}	\\	\phantom{-11}	9.2	& \pm	54.7		\end{aligned}$ &	$\begin{aligned}[t]	0.069	& \pm	0.011	\\	30.181	& \pm	0.170	\\	&	\mbox{P,S}	\end{aligned}$ &	$\begin{aligned}[t]		4.558	& \pm	0.357	\\	-9.665	& \pm	0.757	\\	-1.860	& \pm	0.156	\end{aligned}$ &	$\begin{aligned}[t]	-24.16	& \pm	0.17	\\	-20.98	& \pm	0.21	\\	0.933	& \pm	0.051	\end{aligned}$ &	$\begin{aligned}[t]		10.935	& \pm	0.069	\\	72.2	& \pm	1.3	\; (	0	)	\phantom{1}	\\	85.0	& \pm	1.0	\; (	+	)	\phantom{1}	\end{aligned}$ &	$\begin{aligned}[t]	\phantom{11}	237	& \pm	30	\phantom{11}	\\		& \cdots  	\\		& \cdots 	\end{aligned}$	\\	\noalign{\smallskip}
$\begin{aligned}[t]	\mbox{\bf Maffei 1\phantom{oo 1}}	&	\\	314.24	&	\\	0.01	&	\end{aligned}$ &	$\begin{aligned}[t]	\phantom{1}	-5.0	&	\\	\phantom{-1}	66.4	& \pm	5.0	\phantom{1}	\\	\phantom{-1}	41.1	& \pm	31.2		\end{aligned}$ &	$\begin{aligned}[t]	1.691	& \pm	0.066	\\	27.583	& \pm	0.269	\\	&	\mbox{FP in I}	\end{aligned}$ &	$\begin{aligned}[t]		2.290	& \pm	0.284	\\	-2.352	& \pm	0.291	\\	0.000	& \pm	0.016	\end{aligned}$ &	$\begin{aligned}[t]	-24.24	& \pm	0.40	\\	-21.12	& \pm	0.33	\\	0.879	& \pm	0.133	\end{aligned}$ &	$\begin{aligned}[t]		10.928	& \pm	0.159 	\\	& \cdots \phantom{11}	\; (	0	) 		\\	83.9	& \pm	0.7	\; (	0	) 	\phantom{1\;}	\end{aligned}$ &	$\begin{aligned}[t]	\phantom{1} &  \cdots  	\\		\phantom{1} &  \cdots  	 \\		 \phantom{1} &  \cdots 		\end{aligned}$	\\	\noalign{\smallskip}
$\begin{aligned}[t]	\mbox{\bf Maffei 2\phantom{oo 1}}	&	\\	314.45	&	\\	-0.74	&	\end{aligned}$ &	$\begin{aligned}[t]	\phantom{-11}	4.0	&	\\		-23.0	& \pm	1.0	\phantom{1}	\\		-61.5	& \pm	41.7		\end{aligned}$ &	$\begin{aligned}[t]	2.017	& \pm	0.211	\\	27.683	& \pm	0.356	\\	&	\mbox{TF in I}	\end{aligned}$ &	$\begin{aligned}[t]		2.406	& \pm	0.395	\\	-2.452	& \pm	0.402	\\	-0.044	& \pm	0.028	\end{aligned}$ &	$\begin{aligned}[t]	-23.90	& \pm	0.73	\\	-21.42	& \pm	0.69	\\	0.470	& \pm	0.371	\end{aligned}$ &	$\begin{aligned}[t]		10.493	& \pm	0.290	\\	67.0	& \pm	1.0	\; (	+	)	\phantom{1}	\\	24.5	& \pm	1.5	\; (	-	)	\end{aligned}$ &	$\begin{aligned}[t]	\phantom{11}	170	& \pm	4	\phantom{111}	\\		317.0	& \pm	3.2	\phantom{1}	\\	\phantom{-}	64.1	& \pm	1.0	\phantom{1}	\end{aligned}$	\\	\noalign{\smallskip}
$\begin{aligned}[t]	\mbox{\bf Dwingeloo 1}	&	\\	314.64	&	\\	-2.90	&	\end{aligned}$ &	$\begin{aligned}[t]	\phantom{-11}	6.0	&	\\	\phantom{-}	107.9	& \pm	0.4	\phantom{1}	\\	\phantom{-}	45.7	& \pm	31.7		\end{aligned}$ &	$\begin{aligned}[t]	1.710	& \pm	0.104	\\	27.825	& \pm	0.248	\\	&	\mbox{TF in I}	\end{aligned}$ &	$\begin{aligned}[t]		2.572	& \pm	0.294	\\	-2.604	& \pm	0.298	\\	-0.185	& \pm	0.036	\end{aligned}$ &	$\begin{aligned}[t]	-22.15	& \pm	0.46	\\	-19.71	& \pm	0.40	\\	0.443	& \pm	0.096	\end{aligned}$ &	$\begin{aligned}[t]	\phantom{1}	9.773	& \pm	0.184	\\	51.0	& \pm	2.0	\; (	-	)	\phantom{1}	\\	111.4	& \pm	0.6	\; (	+	)	\end{aligned}$ &	$\begin{aligned}[t]	\phantom{11}	113	& \pm	4	\phantom{1111}	\\		185.4	& \pm	2.0	\phantom{1}	\\	\phantom{1}	-0.9	& \pm	0.6	\phantom{1}	\end{aligned}$	\\	\noalign{\smallskip}
$\begin{aligned}[t]	\mbox{\bf NGC 1313\phantom{ 1}}	&	\\	183.81	&	\\	-18.87	&	\end{aligned}$ &	$\begin{aligned}[t]	\phantom{-11}	7.0	&	\\		480.0	& \pm	2.0	\phantom{1}	\\		-19.2	& \pm	7.1	\phantom{1}	\end{aligned}$ &	$\begin{aligned}[t]	0.124	& \pm	0.020	\\	28.167	& \pm	0.020	\\	&	\mbox{T}	\end{aligned}$ &	$\begin{aligned}[t]		-4.019	& \pm	0.037	\\	-0.268	& \pm	0.002	\\	-1.376	& \pm	0.014	\end{aligned}$ &	$\begin{aligned}[t]	-21.67	& \pm	0.24	\\	-19.39	& \pm	0.10	\\	0.344	& \pm	0.053	\end{aligned}$ &	$\begin{aligned}[t]	\phantom{1}	9.511	& \pm	0.096	\\	45.4	& \pm	2.6	\; (	+	)	\phantom{1}	\\	2.5	& \pm	1.5	\; (	+	)	\end{aligned}$ &	$\begin{aligned}[t]	\phantom{11}	104	& \pm	6	\phantom{111}	\\		146.0	& \pm	4.0	\phantom{1}	\\		-56.2	& \pm	1.8	\phantom{1}	\end{aligned}$	\\	\noalign{\smallskip}
$\begin{aligned}[t]	\mbox{\bf IC 342\phantom{loo 1}}	&	\\	325.37	&	\\	-2.41	&	\end{aligned}$ &	$\begin{aligned}[t]	\phantom{-11}	5.9	&	\\	\phantom{-}	25.0	& \pm	3.0	\phantom{1}	\\		-19.4	& \pm	14.3		\end{aligned}$ &	$\begin{aligned}[t]	0.677	& \pm	0.056	\\	27.633	& \pm	0.092	\\	&	\mbox{C,P}	\end{aligned}$ &	$\begin{aligned}[t]		2.758	& \pm	0.117	\\	-1.904	& \pm	0.081	\\	-0.141	& \pm	0.011	\end{aligned}$ &	$\begin{aligned}[t]	-23.51	& \pm	0.10	\\	-21.30	& \pm	0.19	\\	0.425	& \pm	0.065	\end{aligned}$ &	$\begin{aligned}[t]		10.302	& \pm	0.041	\\	25.0	& \pm	3.0	\; (	-	)	\phantom{1}	\\	39.0	& \pm	3.0	\; (	+	)	\end{aligned}$ &	$\begin{aligned}[t]	\phantom{11}	192	& \pm	22	\phantom{11}	\\		145.7	& \pm	1.4	\phantom{1}	\\		-20.4	& \pm	3.0	\phantom{1}	\end{aligned}$	\\	\noalign{\smallskip}
$\begin{aligned}[t]	\mbox{\bf NGC 1569\phantom{ 1}}	&	\\	326.11	&	\\	-7.37	&	\end{aligned}$ &	$\begin{aligned}[t]	\phantom{-11}	9.6	&	\\	\	-81.6	&	 \pm	4.4	\phantom{1}	\\		-103.9	& \pm	11.0		\end{aligned}$ &	$\begin{aligned}[t]	0.804	& \pm	0.129	\\	27.212	& \pm	0.032	\\	&	\mbox{T}	\end{aligned}$ &	$\begin{aligned}[t]		2.264	& \pm	0.033	\\	-1.521	& \pm	0.022	\\	-0.353	& \pm	0.007	\end{aligned}$ &	$\begin{aligned}[t]	-19.03	& \pm	0.11	\\	-18.43	& \pm	0.37	\\	0.062	& \pm	0.126	\end{aligned}$ &	$\begin{aligned}[t]	\phantom{1}	8.246	& \pm	0.045	\\	90.0	& \pm	1.0	\; (	-	)	\phantom{1}	\\	119.3	& \pm	1.1	\; (	+	)	\end{aligned}$ &	$\begin{aligned}[t]	\phantom{111}	35	& \pm	4	\phantom{111}	\\		232.6	& \pm	0.6	\phantom{1}	\\		-19.0	& \pm	1.1	\phantom{1}	\end{aligned}$	\\	\noalign{\smallskip}
$\begin{aligned}[t]	\mbox{\bf NGC 2403\phantom{ 1}}	&	\\	344.94	&	\\	-12.94	&	\end{aligned}$ &	$\begin{aligned}[t]	\phantom{-11}	6.0	&	\\	\phantom{-}	133.2	& \pm	2.2	\phantom{1}	\\	\phantom{-1}	22.8	& \pm	9.9	\phantom{1}	\end{aligned}$ &	$\begin{aligned}[t]	0.045	& \pm	0.007	\\	27.620	& \pm	0.050	\\	&	\mbox{C,P}	\end{aligned}$ &	$\begin{aligned}[t]		3.116	& \pm	0.072	\\	-0.838	& \pm	0.019	\\	-0.742	& \pm	0.020	\end{aligned}$ &	$\begin{aligned}[t]	-22.26	& \pm	0.24	\\	-19.87	& \pm	0.09	\\	0.409	& \pm	0.050	\end{aligned}$ &	$\begin{aligned}[t]	\phantom{1}	9.791	& \pm	0.094	\\	60.5	& \pm	2.5	\; (	+	)	\phantom{1}	\\	125.3	& \pm	0.8	\; (	+	)	\end{aligned}$ &	$\begin{aligned}[t]	\phantom{11}	134	& \pm	2	\phantom{111}	\\		287.8	& \pm	4.4	\phantom{1}	\\		-58.9	& \pm	1.1	\phantom{1}	\end{aligned}$	\\	\noalign{\smallskip}
$\begin{aligned}[t]	\mbox{\bf NGC 2683\phantom{ 1}}	&	\\	10.21	&	\\	-40.59	&	\end{aligned}$ &	$\begin{aligned}[t]	\phantom{-11}	3.1	&	\\	\phantom{-}	415.0	& \pm	1.0	\phantom{1}	\\		-169.3	& \pm	55.1		\end{aligned}$ &	$\begin{aligned}[t]	0.037	& \pm	0.006	\\	29.425	& \pm	0.357	\\	&	\mbox{S}	\end{aligned}$ &	$\begin{aligned}[t]		5.672	& \pm	0.933	\\	1.022	& \pm	0.168	\\	-4.938	& \pm	0.833	\end{aligned}$ &	$\begin{aligned}[t]	-23.49	& \pm	0.36	\\	-20.62	& \pm	0.36	\\	0.732	& \pm	0.050	\end{aligned}$ &	$\begin{aligned}[t]		10.521	& \pm	0.143	\\	79.3	& \pm	2.1	\; (	+	)	\phantom{1}	\\	42.8	& \pm	1.3	\; (	-	)	\end{aligned}$ &	$\begin{aligned}[t]	\phantom{11}	156	& \pm	8	\phantom{111}	\\		294.5	& \pm	1.8	\phantom{1}	\\	\phantom{-1}	0.0	& \pm	1.7	\phantom{1}	\end{aligned}$	\\	\noalign{\smallskip}
$\begin{aligned}[t]	\mbox{\bf NGC 2784\phantom{ 1}}	&	\\	104.04	&	\\	-57.59	&	\end{aligned}$ &	$\begin{aligned}[t]	\phantom{1}	-2.1	&	\\	\phantom{-}	708.0	& \pm	10.0		\\		-295.2	& \pm	65.6		\end{aligned}$ &	$\begin{aligned}[t]	0.244	& \pm	0.039	\\	30.087	& \pm	0.259	\\	&	\mbox{S}	\end{aligned}$ &	$\begin{aligned}[t]		-1.339	& \pm	0.160	\\	5.356	& \pm	0.639	\\	-8.696	& \pm	1.053	\end{aligned}$ &	$\begin{aligned}[t]	-24.04	& \pm	0.26	\\	-20.67	& \pm	0.30	\\	0.920	& \pm	0.061	\end{aligned}$ &	$\begin{aligned}[t]		10.877	& \pm	0.104	\\	66.4	& \pm	1.1	\; (	0	)	\phantom{1}	\\	73.0	& \pm	1.0	\; (	+	)	\end{aligned}$ &	$\begin{aligned}[t]	\phantom{11}	203	& \pm	10	\phantom{11}	\\		& \cdots	\\		& \cdots		\end{aligned}$	\\	\noalign{\smallskip}
$\begin{aligned}[t]	\mbox{\bf NGC 2787\phantom{ 1}}	&	\\	353.25	&	\\	-7.67	&	\end{aligned}$ &	$\begin{aligned}[t]	\phantom{1}	-1.1	&	\\	\phantom{-}	723.0	& \pm	10.0		\\	\phantom{-}	309.0	& \pm	67.2		\end{aligned}$ &	$\begin{aligned}[t]	0.149	& \pm	0.024	\\	29.422	& \pm	0.362	\\	&	\mbox{S}	\end{aligned}$ &	$\begin{aligned}[t]		7.524	& \pm	1.254	\\	-0.891	& \pm	0.149	\\	-1.021	& \pm	0.192	\end{aligned}$ &	$\begin{aligned}[t]	-22.40	& \pm	0.36	\\	-19.13	& \pm	0.37	\\	0.923	& \pm	0.054	\end{aligned}$ &	$\begin{aligned}[t]		10.224	& \pm	0.145	\\	55.5	& \pm	1.6	\; (	-	)	\phantom{1}	\\	109.0	& \pm	1.0	\; (	-	)	\end{aligned}$ &	$\begin{aligned}[t]	\phantom{11}	210	& \pm	15	\phantom{11}	\\		194.3	& \pm	2.2	\phantom{1}	\\	\phantom{-}	61.9	& \pm	1.5	\phantom{1}	\end{aligned}$	\\	\noalign{\smallskip}
$\begin{aligned}[t]	\mbox{\bf NGC 2903\phantom{ 1}}	&	\\	30.00	&	\\	-43.60	&	\end{aligned}$ &	$\begin{aligned}[t]	\phantom{-11}	4.0	&	\\	\phantom{-}	555.6	& \pm	1.3	\phantom{1}	\\		-173.0	& \pm	38.4		\end{aligned}$ &	$\begin{aligned}[t]	0.035	& \pm	0.006	\\	29.733	& \pm	0.188	\\	&	\mbox{TF in I}	\end{aligned}$ &	$\begin{aligned}[t]		5.492	& \pm	0.476	\\	3.170	& \pm	0.275	\\	-6.038	& \pm	0.535	\end{aligned}$ &	$\begin{aligned}[t]	-23.99	& \pm	0.19	\\	-21.37	& \pm	0.20	\\	0.546	& \pm	0.050	\end{aligned}$ &	$\begin{aligned}[t]		10.586	& \pm	0.076	\\	61.2	& \pm	0.5	\; (	-	)	\phantom{1}	\\	23.0	& \pm	1.0	\; (	-	)	\end{aligned}$ &	$\begin{aligned}[t]	\phantom{11}	188	& \pm	4	\phantom{111}	\\		260.5	& \pm	0.7	\phantom{1}	\\	\phantom{-1}	2.2	& \pm	0.7	\phantom{1}	\end{aligned}$	\\	\noalign{\smallskip}
$\begin{aligned}[t]	\mbox{\bf M81\phantom{ngeloo 1}}	&	\\	355.97	&	\\	-4.89	&	\end{aligned}$ &	$\begin{aligned}[t]	\phantom{-11}	2.4	&	\\		-39.4 	& \pm	2.8	\phantom{1}	\\		-173.3	& \pm	11.2		\end{aligned}$ &	$\begin{aligned}[t]	0.091	& \pm	0.015	\\	27.867	& \pm	0.086	\\	&	\mbox{C,P,S,T}	\end{aligned}$ &	$\begin{aligned}[t]		3.709	& \pm	0.147	\\	-0.262	& \pm	0.010	\\	-0.318	& \pm	0.018	\end{aligned}$ &	$\begin{aligned}[t]	-24.34	& \pm	0.09	\\	-21.66	& \pm	0.12	\\	0.795	& \pm	0.052	\end{aligned}$ &	$\begin{aligned}[t]		10.905	& \pm	0.035	\\	57.2	& \pm	1.8	\; (	-	)	\phantom{1}	\\	151.3	& \pm	1.1	\; (	-	)	\end{aligned}$ &	$\begin{aligned}[t]	\phantom{11}	199	& \pm	11	\phantom{11}	\\		141.3	& \pm	2.5	\phantom{1}	\\	\phantom{-}	57.3	& \pm	1.4	\phantom{1}	\end{aligned}$	\\	\noalign{\smallskip}
$\begin{aligned}[t]	\mbox{\bf M82\phantom{ngeloo 1}}	&	\\	355.59	&	\\	-4.25	&	\end{aligned}$ &	$\begin{aligned}[t]	\phantom{-11}	3.0	&	\\	\phantom{-}	199.0	& \pm	7.0	\phantom{1}	\\	\phantom{-1}	86.6	& \pm	15.2		\end{aligned}$ &	$\begin{aligned}[t]	0.181	& \pm	0.029	\\	27.709	& \pm	0.144	\\	&	\mbox{T}	\end{aligned}$ &	$\begin{aligned}[t]		3.450	& \pm	0.229	\\	-0.266	& \pm	0.018	\\	-0.257	& \pm	0.026	\end{aligned}$ &	$\begin{aligned}[t]	-23.82	& \pm	0.25	\\	-20.67	& \pm	0.25	\\	0.625	& \pm	0.057	\end{aligned}$ &	$\begin{aligned}[t]		10.573	& \pm	0.099	\\	76.0	& \pm	1.8	\; (	-	)	\phantom{1}	\\	67.0	& \pm	3.0	\; (	+	)	\end{aligned}$ &	$\begin{aligned}[t]	\phantom{11}	110	& \pm	5	\phantom{111}	\\		104.1	& \pm	2.0	\phantom{1}	\\		-24.4	& \pm	2.9	\phantom{1}	\end{aligned}$	\\	\noalign{\smallskip}
$\begin{aligned}[t]	\mbox{\bf NGC 3115\phantom{ 1}}	&	\\	73.88	&	\\	-47.06	&	\end{aligned}$ &	$\begin{aligned}[t]	\phantom{1}	-2.9	&	\\	\phantom{-}	663.0	& \pm	5.0	\phantom{1}	\\		-287.4	& \pm	25.4		\end{aligned}$ &	$\begin{aligned}[t]	0.053	& \pm	0.009	\\	30.064	& \pm	0.111	\\	&	\mbox{P,S,T}	\end{aligned}$ &	$\begin{aligned}[t]		1.930	& \pm	0.099	\\	6.678	& \pm	0.341	\\	-7.471	& \pm	0.388	\end{aligned}$ &	$\begin{aligned}[t]	-24.39	& \pm	0.11	\\	-21.43	& \pm	0.12	\\	0.917	& \pm	0.051	\end{aligned}$ &	$\begin{aligned}[t]		11.016	& \pm	0.045	\\	86.0	& \pm	0.5	\; (	-	)	\phantom{1}	\\	43.5	& \pm	1.0	\; (	+	)	\end{aligned}$ &	$\begin{aligned}[t]	\phantom{11}	262	& \pm	9	\phantom{111}	\\		133.6	& \pm	1.0	\phantom{1}	\\	\phantom{-}	29.7	& \pm	0.7	\phantom{1}	\end{aligned}$	\\	\noalign{\smallskip}
$\begin{aligned}[t]	\mbox{\bf NGC 3344\phantom{ 1}}	&	\\	37.43	&	\\	-28.07	&	\end{aligned}$ &	$\begin{aligned}[t]	\phantom{-11}	4.0	&	\\	\phantom{-}	586.8	& \pm	0.4	\phantom{1}	\\		-450.9	& \pm	59.5		\end{aligned}$ &	$\begin{aligned}[t]	0.037	& \pm	0.006	\\	30.650	& \pm	0.155	\\	&	\mbox{TF in V}	\end{aligned}$ &	$\begin{aligned}[t]		9.407	& \pm	0.672	\\	7.200	& \pm	0.514	\\	-6.318	& \pm	0.460	\end{aligned}$ &	$\begin{aligned}[t]	-23.43	& \pm	0.16	\\	-20.90	& \pm	0.18	\\	0.541	& \pm	0.050	\end{aligned}$ &	$\begin{aligned}[t]		10.356	& \pm	0.065	\\	25.5	& \pm	0.5	\; (	+	)	\phantom{1}	\\	156.1	& \pm	0.7	\; (	-	)	\end{aligned}$ &	$\begin{aligned}[t]	\phantom{11}	163	& \pm	4	\phantom{1111}	\\	\phantom{1}	38.8	& \pm	0.3	\phantom{1}	\\	\phantom{1}	-3.1	& \pm	0.5	\phantom{1}	\end{aligned}$	\\	\noalign{\smallskip}
$\begin{aligned}[t]	\mbox{\bf M95\phantom{ngeloo 1}}	&	\\	51.68	&	\\	-33.11	&	\end{aligned}$ &	$\begin{aligned}[t]	\phantom{-11}	3.0	&	\\	\phantom{-}	772.0	& \pm	3.0	\phantom{1}	\\		-104.8	& \pm	13.2		\end{aligned}$ &	$\begin{aligned}[t]	0.032	& \pm	0.005	\\	30.086	& \pm	0.044	\\	&	\mbox{C,P,T}	\end{aligned}$ &	$\begin{aligned}[t]		5.368	& \pm	0.109	\\	6.791	& \pm	0.138	\\	-5.644	& \pm	0.117	\end{aligned}$ &	$\begin{aligned}[t]	-23.66	& \pm	0.06	\\	-20.71	& \pm	0.06	\\	0.733	& \pm	0.100	\end{aligned}$ &	$\begin{aligned}[t]		10.589	& \pm	0.023	\\	45.0	& \pm	2.0	\; (	-	)	\phantom{1}	\\	6.0	& \pm	7.0	\; (	-	)	\end{aligned}$ &	$\begin{aligned}[t]	\phantom{11}	197	& \pm	9	\phantom{1111}	\\		249.6	& \pm	4.7	\phantom{1}	\\	\phantom{1}	-7.9	& \pm	2.6	\phantom{1}	\end{aligned}$	\\	\noalign{\smallskip}
$\begin{aligned}[t]	\mbox{\bf M96\phantom{ngeloo 1}}	&	\\	51.82	&	\\	-32.40	&	\end{aligned}$ &	$\begin{aligned}[t]	\phantom{-11}	1.8	&	\\	\phantom{-}	910.0	& \pm	9.0	\phantom{1}	\\	\phantom{-1}	49.6	& \pm	38.8		\end{aligned}$ &	$\begin{aligned}[t]	0.028	& \pm	0.005	\\	30.041	& \pm	0.137	\\	&	\mbox{C,P,S}	\end{aligned}$ &	$\begin{aligned}[t]		5.283	& \pm	0.333	\\	6.717	& \pm	0.424	\\	-5.424	& \pm	0.350	\end{aligned}$ &	$\begin{aligned}[t]	-23.97	& \pm	0.14	\\	-21.16	& \pm	0.15	\\	0.736	& \pm	0.070	\end{aligned}$ &	$\begin{aligned}[t]		10.717	& \pm	0.056	\\	49.5	& \pm	1.6	\; (	+	)	\phantom{1}	\\	135.0	& \pm	5.0	\; (	+	)	\end{aligned}$ &	$\begin{aligned}[t]	\phantom{11}	224	& \pm	41	\phantom{11}	\\		114.3	& \pm	4.7	\phantom{1}	\\		-68.2	& \pm	3.7	\phantom{1}	\end{aligned}$	\\	\noalign{\smallskip}
$\begin{aligned}[t]	\mbox{\bf NGC 3377\phantom{ 1}}	&	\\	49.49	&	\\	-31.47	&	\end{aligned}$ &	$\begin{aligned}[t]	\phantom{1}	-4.8	&	\\	\phantom{-}	690.0	& \pm	5.0	\phantom{1}	\\		-198.6	& \pm	38.9		\end{aligned}$ &	$\begin{aligned}[t]	0.039	& \pm	0.006	\\	30.151	& \pm	0.130	\\	&	\mbox{P,S}	\end{aligned}$ &	$\begin{aligned}[t]		5.901	& \pm	0.353	\\	6.908	& \pm	0.414	\\	-5.562	& \pm	0.341	\end{aligned}$ &	$\begin{aligned}[t]	-22.92	& \pm	0.13	\\	-20.00	& \pm	0.22	\\	0.820	& \pm	0.050	\end{aligned}$ &	$\begin{aligned}[t]		10.355	& \pm	0.053	\\	90.0	& \pm	10.0	\; (	0	)		\\	43.7	& \pm	2.4	\; (	-	) \phantom{1}	\end{aligned}$ &	$\begin{aligned}[t]	\phantom{111}	88	& \pm	7	\phantom{111}	\\		305.1	& \pm	9.0	\phantom{1}	\\		-21.6	& \pm	6.0	\phantom{1}	\end{aligned}$	\\	\noalign{\smallskip}
$\begin{aligned}[t]	\mbox{\bf M105\phantom{geloo 1}}	&	\\	51.07	&	\\	-31.91	&	\end{aligned}$ &	$\begin{aligned}[t]	\phantom{1}	-4.8	&	\\	\phantom{-}	916.0	& \pm	5.0	\phantom{1}	\\	\phantom{-1}	53.8	& \pm	21.3		\end{aligned}$ &	$\begin{aligned}[t]	0.027	& \pm	0.004	\\	30.057	& \pm	0.073	\\	&	\mbox{P,S}	\end{aligned}$ &	$\begin{aligned}[t]		5.440	& \pm	0.183	\\	6.734	& \pm	0.226	\\	-5.390	& \pm	0.186	\end{aligned}$ &	$\begin{aligned}[t]	-23.99	& \pm	0.08	\\	-20.85	& \pm	0.10	\\	0.928	& \pm	0.050	\end{aligned}$ &	$\begin{aligned}[t]		10.862	& \pm	0.030	\\	90.0	& \pm	10.0	\; (	0	)		\\	70.0	& \pm	2.1	\; (	-	)	\phantom{1} \end{aligned}$ &	$\begin{aligned}[t]	\phantom{111}	56	& \pm	2	\phantom{1111}	\\		320.9	& \pm	8.5	\phantom{1}	\\	\phantom{1}	-0.3	& \pm	5.6	\phantom{1}	\end{aligned}$	\\	\noalign{\smallskip}
$\begin{aligned}[t]	\mbox{\bf NGC 3384\phantom{ 1}}	&	\\	51.06	&	\\	-31.83	&	\end{aligned}$ &	$\begin{aligned}[t]	\phantom{1}	-2.7	&	\\	\phantom{-}	737.0	& \pm	5.0	\phantom{1}	\\		-181.5	& \pm	37.1		\end{aligned}$ &	$\begin{aligned}[t]	0.031	& \pm	0.005	\\	30.218	& \pm	0.121	\\	&	\mbox{P,S,T}	\end{aligned}$ &	$\begin{aligned}[t]		5.868	& \pm	0.327	\\	7.261	& \pm	0.405	\\	-5.795	& \pm	0.330	\end{aligned}$ &	$\begin{aligned}[t]	-23.67	& \pm	0.12	\\	-20.29	& \pm	0.14	\\	0.896	& \pm	0.050	\end{aligned}$ &	$\begin{aligned}[t]		10.712	& \pm	0.049	\\	62.8	& \pm	1.6	\; (	0	)	\phantom{1}	\\	50.5	& \pm	2.5	\; (	-	)	\phantom{1}	\end{aligned}$ &	$\begin{aligned}[t]	\phantom{11}	108	& \pm	28	\phantom{11}	\\		& \cdots	\\		& \cdots		\end{aligned}$	\\	\noalign{\smallskip}
$\begin{aligned}[t]	\mbox{\bf NGC 3412\phantom{ 1}}	&	\\	50.44	&	\\	-30.97	&	\end{aligned}$ &	$\begin{aligned}[t]	\phantom{1}	-2.0	&	\\	\phantom{-}	850.0	& \pm	2.0	\phantom{1}	\\	-79.1		& \pm	40.5		\end{aligned}$ &	$\begin{aligned}[t]	0.032	& \pm	0.005	\\	30.258	& \pm	0.129	\\	&	\mbox{S}	\end{aligned}$ &	$\begin{aligned}[t]		6.113	& \pm	0.363	\\	7.400	& \pm	0.440	\\	-5.761	& \pm	0.350	\end{aligned}$ &	$\begin{aligned}[t]	-22.79	& \pm	0.13	\\	-19.83	& \pm	0.20	\\	0.874	& \pm	0.050	\end{aligned}$ &	$\begin{aligned}[t]		10.344	& \pm	0.052	\\	57.0	& \pm	1.6	\; (	0	)	\phantom{1}	\\	151.0	& \pm	0.9	\; (	+	)	\phantom{1}	\end{aligned}$ &	$\begin{aligned}[t]	\phantom{11}	121	& \pm	11	\phantom{11}	\\		& \cdots	\\		& \cdots		\end{aligned}$	\\	\noalign{\smallskip}
$\begin{aligned}[t]	\mbox{\bf NGC 3489\phantom{ 1}}	&	\\	50.79	&	\\	-28.67	&	\end{aligned}$ &	$\begin{aligned}[t]	\phantom{1}	-1.3	&	\\	\phantom{-}	702.0	& \pm	5.0	\phantom{1}	\\		-275.0	& \pm	49.5		\end{aligned}$ &	$\begin{aligned}[t]	0.019	& \pm	0.003	\\	30.395	& \pm	0.144	\\	&	\mbox{S}	\end{aligned}$ &	$\begin{aligned}[t]		6.619	& \pm	0.439	\\	8.112	& \pm	0.538	\\	-5.726	& \pm	0.388	\end{aligned}$ &	$\begin{aligned}[t]	-23.22	& \pm	0.14	\\	-20.22	& \pm	0.18	\\	0.807	& \pm	0.050	\end{aligned}$ &	$\begin{aligned}[t]		10.468	& \pm	0.058	\\	58.5	& \pm	3.0	\; (	-	)	\phantom{1}	\\	72.1	& \pm	0.9	\; (	+	)	\end{aligned}$ &	$\begin{aligned}[t]	\phantom{111}	97	& \pm	12	\phantom{11}	\\		169.4	& \pm	2.8	\phantom{1}	\\	\phantom{-}	13.8	& \pm	1.5	\phantom{1}	\end{aligned}$	\\	\noalign{\smallskip}
$\begin{aligned}[t]	\mbox{\bf NGC 3621\phantom{ 1}}	&	\\	105.02	&	\\	-28.20	&	\end{aligned}$ &	$\begin{aligned}[t]	\phantom{-11}	6.9	&	\\		728.5	& \pm	2.7		\\		-44.8		& \pm	12.8		\end{aligned}$ &	$\begin{aligned}[t]	0.091	& \pm	0.015	\\	29.283	& \pm	0.068	\\	&	\mbox{C,T}	\end{aligned}$ &	$\begin{aligned}[t]		-1.628	& \pm	0.051	\\	6.066	& \pm	0.190	\\	-3.368	& \pm	0.110	\end{aligned}$ &	$\begin{aligned}[t]	-22.99	& \pm	0.08	\\	-20.41	& \pm	0.08	\\	0.422	& \pm	0.052	\end{aligned}$ &	$\begin{aligned}[t]		10.093	& \pm	0.031	\\	64.4	& \pm	0.5	\; (	+	)	\phantom{1}	\\	163.3	& \pm	2.1	\; (	-	)	\end{aligned}$ &	$\begin{aligned}[t]	\phantom{11}	140	& \pm	4	\phantom{111}	\\	\phantom{1}	79.1	& \pm	2.0	\phantom{1}	\\	\phantom{-}	30.3	& \pm	0.9	\phantom{1}	\end{aligned}$	\\	\noalign{\smallskip}
$\begin{aligned}[t]	\mbox{\bf M66\phantom{ngeloo 1}}	&	\\	53.43	&	\\	-24.23	&	\end{aligned}$ &	$\begin{aligned}[t]	\phantom{-11}	3.0	&	\\	\phantom{-}	708.2	& \pm	1.1	\phantom{1}	\\		-148.4	& \pm	4.5	\phantom{1}	\end{aligned}$ &	$\begin{aligned}[t]	0.036	& \pm	0.006	\\	30.074	& \pm	0.004	\\	&	\mbox{C,P}	\end{aligned}$ &	$\begin{aligned}[t]		5.593	& \pm	0.010	\\	7.539	& \pm	0.014	\\	-4.224	& \pm	0.008	\end{aligned}$ &	$\begin{aligned}[t]	-24.50	& \pm	0.02	\\	-21.79	& \pm	0.04	\\	0.575	& \pm	0.050	\end{aligned}$ &	$\begin{aligned}[t]		10.809	& \pm	0.007	\\	63.4	& \pm	1.6	\; (	-	)	\phantom{1}	\\	173.0	& \pm	0.5	\; (	+	)	\end{aligned}$ &	$\begin{aligned}[t]	\phantom{11}	199	& \pm	11	\phantom{11}	\\		244.5	& \pm	0.7	\phantom{1}	\\		-37.7	& \pm	1.6	\phantom{1}	\end{aligned}$	\\	\noalign{\smallskip}
$\begin{aligned}[t]	\mbox{\bf NGC 4144\phantom{ 1}}	&	\\	24.07	&	\\	-3.06	&	\end{aligned}$ &	$\begin{aligned}[t]	\phantom{-11}	6.0	&	\\	\phantom{-}	265.0	& \pm	1.0	\phantom{1}	\\		-210.1	& \pm	8.2	\phantom{1}	\end{aligned}$ &	$\begin{aligned}[t]	0.017	& \pm	0.003	\\	29.346	& \pm	0.029	\\	&	\mbox{T}	\end{aligned}$ &	$\begin{aligned}[t]		6.739	& \pm	0.090	\\	3.011	& \pm	0.040	\\	-0.395	& \pm	0.007	\end{aligned}$ &	$\begin{aligned}[t]	-20.19	& \pm	0.05	\\	-18.00	& \pm	0.18	\\	0.447	& \pm	0.100	\end{aligned}$ &	$\begin{aligned}[t]	\phantom{1}	8.991	& \pm	0.022	\\	79.0	& \pm	1.4	\; (	-	)	\phantom{1}	\\	102.0	& \pm	0.5	\; (	-	)	\end{aligned}$ &	$\begin{aligned}[t]	\phantom{111}	74	& \pm	5	\phantom{111}	\\		283.7	& \pm	1.6	\phantom{1}	\\	\phantom{-}	30.1	& \pm	0.5	\phantom{1}	\end{aligned}$	\\	\noalign{\smallskip}
$\begin{aligned}[t]	\mbox{\bf NGC 4236\phantom{ 1}}	&	\\	2.38	&	\\	5.35	&	\end{aligned}$ &	$\begin{aligned}[t]	\phantom{-11}	7.9	&	\\		-10.0	 & \pm	5.0		\\		-174.4	& \pm	25.2		\end{aligned}$ &	$\begin{aligned}[t]	0.017	& \pm	0.003	\\	28.220	& \pm	0.240	\\	&	\mbox{T}	\end{aligned}$ &	$\begin{aligned}[t]	\phantom{-}	4.393	& \pm	0.486	\\	0.182	& \pm	0.020	\\	0.412	& \pm	0.031	\end{aligned}$ &	$\begin{aligned}[t]	-20.91	& \pm	0.33	\\	-18.68	& \pm	0.25	\\	0.312	& \pm	0.050	\end{aligned}$ &	$\begin{aligned}[t]	\phantom{1}	9.182	& \pm	0.133	\\	76.2	& \pm	0.5	\; (	-	)	\phantom{1}	\\	158.1	& \pm	2.0	\; (	+	)	\end{aligned}$ &	$\begin{aligned}[t]	\phantom{111}	85	& \pm	5	\phantom{111}	\\		154.3	& \pm	8.4	\phantom{1}	\\		-78.6	& \pm	1.0	\phantom{1}	\end{aligned}$	\\	\noalign{\smallskip}
$\begin{aligned}[t]	\mbox{\bf NGC 4244\phantom{ 1}}	&	\\	32.78	&	\\	-3.53	&	\end{aligned}$ &	$\begin{aligned}[t]	\phantom{-11}	6.1	&	\\		244.0	& \pm	2.0	\phantom{1}	\\		-47.2		& \pm	7.1	\phantom{1}	\end{aligned}$ &	$\begin{aligned}[t]	0.024	& \pm	0.004	\\	28.183	& \pm	0.037	\\	&	\mbox{T}	\end{aligned}$ &	$\begin{aligned}[t]		3.626	& \pm	0.062	\\	2.335	& \pm	0.040	\\	-0.266	& \pm	0.007	\end{aligned}$ &	$\begin{aligned}[t]	-21.40	& \pm	0.48	\\	-19.10	& \pm	0.42	\\	0.356	& \pm	0.050	\end{aligned}$ &	$\begin{aligned}[t]	\phantom{1}	9.410	& \pm	0.191	\\	84.5	& \pm	0.5	\; (	+	)	\phantom{1}	\\	43.6	& \pm	1.4	\; (	-	)	\end{aligned}$ &	$\begin{aligned}[t]	\phantom{111}	99	& \pm	2	\phantom{111}	\\		306.1	& \pm	0.6	\phantom{1}	\\		-30.8	& \pm	1.4	\phantom{1}	\end{aligned}$	\\	\noalign{\smallskip}
$\begin{aligned}[t]	\mbox{\bf M106\phantom{geloo 1}}	&	\\	23.71	&	\\	-1.37		&	\end{aligned}$ &	$\begin{aligned}[t]	\phantom{-11}	4.0	&	\\		450.0	& \pm	1.0	\phantom{1}	\\		-32.6		& \pm	20.6		\end{aligned}$ &	$\begin{aligned}[t]	0.018	& \pm	0.003	\\	29.404	& \pm	0.091	\\	&	\mbox{C,P,S,T}	\end{aligned}$ &	$\begin{aligned}[t]		6.953	& \pm	0.291	\\	3.054	& \pm	0.128	\\	-0.182	& \pm	0.013	\end{aligned}$ &	$\begin{aligned}[t]	-24.28	& \pm	0.09	\\	-21.71	& \pm	0.10	\\	0.549	& \pm	0.050	\end{aligned}$ &	$\begin{aligned}[t]		10.701	& \pm	0.037	\\	66.9	& \pm	0.9	\; (	-	)	\phantom{1}	\\	150.0	& \pm	0.5	\; (	-	)	\end{aligned}$ &	$\begin{aligned}[t]	\phantom{11}	205	& \pm	9	\phantom{111}	\\		235.4	& \pm	1.5	\phantom{1}	\\	\phantom{-}	65.3	& \pm	0.7	\phantom{1}	\end{aligned}$	\\	\noalign{\smallskip}
$\begin{aligned}[t]	\mbox{\bf NGC 4449\phantom{ 1}}	&	\\	27.24	&	\\	0.18	&	\end{aligned}$ &	$\begin{aligned}[t]	\phantom{-11}	9.8	&	\\		214.0	& \pm	6.0	\phantom{1}	\\		-25.6	& \pm	8.9	\phantom{1}	\end{aligned}$ &	$\begin{aligned}[t]	0.022	& \pm	0.003	\\	28.043	& \pm	0.042	\\	&	\mbox{T}	\end{aligned}$ &	$\begin{aligned}[t]	\phantom{-}	3.609	& \pm	0.070	\\	1.858	& \pm	0.036	\\	0.013	& \pm	0.002	\end{aligned}$ &	$\begin{aligned}[t]	-21.00	& \pm	0.06	\\	-18.45	& \pm	0.19	\\	0.399	& \pm	0.100	\end{aligned}$ &	$\begin{aligned}[t]	\phantom{1}	9.280	& \pm	0.023	\\	56.3	& \pm	2.6	\; (	-	)	\phantom{1}	\\	57.0	& \pm	7.0	\; (	-	)	\end{aligned}$ &	$\begin{aligned}[t]	\phantom{111}	75	& \pm	9	\phantom{111}	\\		261.7	& \pm	2.9	\phantom{1}	\\		-14.4	& \pm	5.7	\phantom{1}	\end{aligned}$	\\	\noalign{\smallskip}
$\begin{aligned}[t]	\mbox{\bf M104\phantom{geloo 1}}	&	\\	82.75	&	\\	-9.20	&	\end{aligned}$ &	$\begin{aligned}[t]	\phantom{-11}	1.1	&	\\		1100.0	& \pm	3.0	\phantom{1}	\\	\phantom{-}	260.0	& \pm	34.5		\end{aligned}$ &	$\begin{aligned}[t]	0.058	& \pm	0.009	\\	29.852	& \pm	0.151	\\	&	\mbox{P,S}	\end{aligned}$ &	$\begin{aligned}[t]		1.161	& \pm	0.081	\\	9.126	& \pm	0.635	\\	-1.491	& \pm	0.113	\end{aligned}$ &	$\begin{aligned}[t]	-25.10	& \pm	0.15	\\	-22.35	& \pm	0.19	\\	0.920	& \pm	0.051	\end{aligned}$ &	$\begin{aligned}[t]		11.303	& \pm	0.061	\\	84.8	& \pm	0.6	\; (	+	)	\phantom{1}	\\	89.9	& \pm	0.3	\; (	+	)	\end{aligned}$ &	$\begin{aligned}[t]	\phantom{11}	353	& \pm	10	\phantom{111}	\\		168.9	& \pm	0.6	\phantom{1}	\\	\phantom{1}	-8.0	& \pm	0.3	\phantom{1}	\end{aligned}$	\\	\noalign{\smallskip}
$\begin{aligned}[t]	\mbox{\bf NGC 4631\phantom{ 1}}	&	\\	39.11	&	\\	-0.63	&	\end{aligned}$ &	$\begin{aligned}[t]	\phantom{-11}	6.6	&	\\	\phantom{-}	617.0	& \pm	10.0		\\	\phantom{-1}	85.3	& \pm	14.0		\end{aligned}$ &	$\begin{aligned}[t]	0.019	& \pm	0.003	\\	29.381	& \pm	0.037	\\	&	\mbox{T}	\end{aligned}$ &	$\begin{aligned}[t]		5.833	& \pm	0.099	\\	4.742	& \pm	0.081	\\	-0.082	& \pm	0.004	\end{aligned}$ &	$\begin{aligned}[t]	-23.43	& \pm	0.04	\\	-21.21	& \pm	0.05	\\	0.310	& \pm	0.030	\end{aligned}$ &	$\begin{aligned}[t]		10.190	& \pm	0.018	\\	85.1	& \pm	0.5	\; (	+	)	\phantom{1}	\\	74.6	& \pm	11.4	\; (	+	)	\end{aligned}$ &	$\begin{aligned}[t]	\phantom{11}	145	& \pm	5	\phantom{111}	\\		124.1	& \pm	0.6	\phantom{1}	\\	\phantom{-1}	3.3	& \pm	11.3		\end{aligned}$	\\	\noalign{\smallskip}
$\begin{aligned}[t]	\mbox{\bf M94\phantom{ngeloo 1}}	&	\\	31.04	&	\\	3.43	&	\end{aligned}$ &	$\begin{aligned}[t]	\phantom{-11}	2.4	&	\\	\phantom{-}	306.7	& \pm	3.7	\phantom{1}	\\	\phantom{-1}	31.9	& \pm	9.4	\phantom{1}	\end{aligned}$ &	$\begin{aligned}[t]	0.020	& \pm	0.003	\\	28.271	& \pm	0.054	\\	&	\mbox{P,T}	\end{aligned}$ &	$\begin{aligned}[t]	\phantom{-}	3.862	& \pm	0.096	\\	2.325	& \pm	0.058	\\	0.270	& \pm	0.004	\end{aligned}$ &	$\begin{aligned}[t]	-23.39	& \pm	0.06	\\	-20.17	& \pm	0.07	\\	0.703	& \pm	0.050	\end{aligned}$ &	$\begin{aligned}[t]		10.458	& \pm	0.023	\\	40.5	& \pm	0.9	\; (	+	)	\phantom{1}	\\	111.3	& \pm	4.8	\; (	-	)	\end{aligned}$ &	$\begin{aligned}[t]	\phantom{11}	135	& \pm	21	\phantom{11}	\\		354.6	& \pm	1.6	\phantom{1}	\\	\phantom{-}	21.5	& \pm	2.9	\phantom{1}	\end{aligned}$	\\	\noalign{\smallskip}
$\begin{aligned}[t]	\mbox{\bf M64\phantom{ngeloo 1}}	&	\\	50.39	&	\\	0.99	&	\end{aligned}$ &	$\begin{aligned}[t]	\phantom{-11}	2.4	&	\\	\phantom{-}	407.4	& \pm	7.0	\phantom{1}	\\	\phantom{-1}	18.1	& \pm	11.7		\end{aligned}$ &	$\begin{aligned}[t]	0.047	& \pm	0.007	\\	28.489	& \pm	0.053	\\	&	\mbox{T}	\end{aligned}$ &	$\begin{aligned}[t]	\phantom{-}	3.179	& \pm	0.078	\\	3.842	& \pm	0.094	\\	0.086	& \pm	0.001	\end{aligned}$ &	$\begin{aligned}[t]	-23.43	& \pm	0.06	\\	-20.37	& \pm	0.08	\\	0.734	& \pm	0.050	\end{aligned}$ &	$\begin{aligned}[t]		10.496	& \pm	0.022	\\	56.4	& \pm	1.4	\; (	+	)	\phantom{1}	\\	113.5	& \pm	0.5	\; (	-	)	\end{aligned}$ &	$\begin{aligned}[t]	\phantom{11}	164	& \pm	7	\phantom{111}	\\		359.2	& \pm	1.4	\phantom{1}	\\	\phantom{-}	27.1	& \pm	0.6	\phantom{1}	\end{aligned}$	\\	\noalign{\smallskip}
$\begin{aligned}[t]	\mbox{\bf NGC 4945\phantom{ 1}}	&	\\	121.43	&	\\	-6.08	&	\end{aligned}$ &	$\begin{aligned}[t]	\phantom{-11}	6.1	&	\\	\phantom{-}	561.0	& \pm	3.0	\phantom{1}	\\	\phantom{-1}	84.4	& \pm	11.5		\end{aligned}$ &	$\begin{aligned}[t]	0.201	& \pm	0.032	\\	27.614	& \pm	0.057	\\	&	\mbox{T}	\end{aligned}$ &	$\begin{aligned}[t]		-1.720	& \pm	0.045	\\	2.814	& \pm	0.074	\\	-0.351	& \pm	0.013	\end{aligned}$ &	$\begin{aligned}[t]	-23.70	& \pm	0.06	\\	-20.97	& \pm	0.22	\\	0.629	& \pm	0.032	\end{aligned}$ &	$\begin{aligned}[t]		10.528	& \pm	0.024	\\	81.7	& \pm	2.2	\; (	+	)	\phantom{1}	\\	43.5	& \pm	1.5	\; (	+	)	\end{aligned}$ &	$\begin{aligned}[t]	\phantom{11}	174	& \pm	7	\phantom{111}	\\		193.5	& \pm	2.8	\phantom{1}	\\	\phantom{-}	39.5	& \pm	1.6	\phantom{1}	\end{aligned}$	\\	\noalign{\smallskip}
$\begin{aligned}[t]	\mbox{\bf NGC 5023\phantom{ 1}}	&	\\	28.75	&	\\	7.23	&	\end{aligned}$ &	$\begin{aligned}[t]	\phantom{-11}	5.9	&	\\	\phantom{-}	406.0	& \pm	1.0	\phantom{1}	\\	\phantom{-11}	3.7	& \pm	8.6	\phantom{1}	\end{aligned}$ &	$\begin{aligned}[t]	0.020	& \pm	0.003	\\	29.097	& \pm	0.037	\\	&	\mbox{T}	\end{aligned}$ &	$\begin{aligned}[t]	\phantom{-}	5.752	& \pm	0.098	\\	3.156	& \pm	0.054	\\	0.832	& \pm	0.012	\end{aligned}$ &	$\begin{aligned}[t]	-19.67	& \pm	0.07	\\	-17.54	& \pm	0.13	\\	0.381	& \pm	0.050	\end{aligned}$ &	$\begin{aligned}[t]	\phantom{1}	8.736	& \pm	0.027	\\	87.0	& \pm	2.0	\; (	0	)	\phantom{1}	\\	28.0	& \pm	0.5	\; (	+	) \phantom{1}	\end{aligned}$ &	$\begin{aligned}[t]	\phantom{111}	83	& \pm	2	\phantom{111}	\\		& \cdots	\\		& \cdots		\end{aligned}$	\\	\noalign{\smallskip}
$\begin{aligned}[t]	\mbox{\bf NGC 5068\phantom{ 1}}	&	\\	93.44	&	\\	-0.64	&	\end{aligned}$ &	$\begin{aligned}[t]	\phantom{-11}	6.0	&	\\	\phantom{-}	668.0	& \pm	3.0	\phantom{1}	\\	\phantom{-}	118.6	& \pm	13.9		\end{aligned}$ &	$\begin{aligned}[t]	0.116	& \pm	0.019	\\	28.586	& \pm	0.091	\\	&	\mbox{P}	\end{aligned}$ &	$\begin{aligned}[t]		-0.312	& \pm	0.013	\\	5.202	& \pm	0.218	\\	-0.058	& \pm	0.008	\end{aligned}$ &	$\begin{aligned}[t]	-21.27	& \pm	0.10	\\	-18.97	& \pm	0.20	\\	0.550	& \pm	0.053	\end{aligned}$ &	$\begin{aligned}[t]	\phantom{1}	9.501	& \pm	0.041	\\	28.6	& \pm	1.2	\; (	-	)	\phantom{1}	\\	104.0	& \pm	1.0	\; (	0	)	\end{aligned}$ &	$\begin{aligned}[t]	\phantom{111}	94	& \pm	9	\phantom{111}	\\		& \cdots	\\		& \cdots		\end{aligned}$	\\	\noalign{\smallskip}
$\begin{aligned}[t]	\mbox{\bf NGC 5102\phantom{ 1}}	&	\\	108.96	&	\\	-1.40	&	\end{aligned}$ &	$\begin{aligned}[t]	\phantom{1}	-3.0	&	\\	\phantom{-}	470.0	& \pm	5.0	\phantom{1}	\\	\phantom{-1}	37.7	& \pm	19.9		\end{aligned}$ &	$\begin{aligned}[t]	0.062	& \pm	0.010	\\	27.445	& \pm	0.206	\\	&	\mbox{P,T}	\end{aligned}$ &	$\begin{aligned}[t]		-1.000	& \pm	0.095	\\	2.910	& \pm	0.276	\\	-0.075	& \pm	0.019	\end{aligned}$ &	$\begin{aligned}[t]	-20.74	& \pm	0.21	\\	-18.09	& \pm	0.29	\\	0.661	& \pm	0.051	\end{aligned}$ &	$\begin{aligned}[t]	\phantom{1}	9.370	& \pm	0.084	\\	70.0	& \pm	2.0	\; (	+	)	\phantom{1}	\\	46.1	& \pm	3.1	\; (	+	)	\end{aligned}$ &	$\begin{aligned}[t]	\phantom{111}	95	& \pm	2	\phantom{111}	\\		172.7	& \pm	2.5	\phantom{1}	\\	\phantom{-}	31.4	& \pm	2.9	\phantom{1}	\end{aligned}$	\\	\noalign{\smallskip}
$\begin{aligned}[t]	\mbox{\bf Centaurus A}	&	\\	115.40	&	\\	-2.02	&	\end{aligned}$ &	$\begin{aligned}[t]	\phantom{1}	-5.0	&	\\	\phantom{-}	541.0	& \pm	7.0	\phantom{1}	\\	\phantom{-1}	66.1	& \pm	11.4		\end{aligned}$ &	$\begin{aligned}[t]	0.131	& \pm	0.021	\\	27.750	& \pm	0.010	\\	&	\mbox{C,P,T}	\end{aligned}$ &	$\begin{aligned}[t]		-1.518	& \pm	0.007	\\	3.197	& \pm	0.015	\\	-0.125	& \pm	0.001	\end{aligned}$ &	$\begin{aligned}[t]	-24.95	& \pm	0.25	\\	-21.92	& \pm	0.12	\\	0.820	& \pm	0.021	\end{aligned}$ &	$\begin{aligned}[t]		11.169	& \pm	0.099	\\	79.0	& \pm	10.0	\; (	-	)		\\	80.0	& \pm	5.0	\; (	-	)	\end{aligned}$ &	$\begin{aligned}[t]	\phantom{111}	77	& \pm	7	\phantom{1111}	\\	\phantom{1}	14.3	& \pm	10.0		\\	\phantom{1}	-0.6	& \pm	5.0	\phantom{1}	\end{aligned}$	\\	\noalign{\smallskip}
$\begin{aligned}[t]	\mbox{\bf M51\phantom{ngeloo 1}}	&	\\	25.86	&	\\	10.38	&	\end{aligned}$ &	$\begin{aligned}[t]	\phantom{-11}	4.0	&	\\		464.0	& \pm	3.0	\phantom{1}	\\		-20.2	& \pm	23.9		\end{aligned}$ &	$\begin{aligned}[t]	0.040	& \pm	0.006	\\	29.524	& \pm	0.106	\\	&	\mbox{P}	\end{aligned}$ &	$\begin{aligned}[t]	\phantom{-}	7.129	& \pm	0.348	\\	3.455	& \pm	0.169	\\	1.451	& \pm	0.065	\end{aligned}$ &	$\begin{aligned}[t]	-24.24	& \pm	0.11	\\	-21.30	& \pm	0.12	\\	0.492	& \pm	0.012	\end{aligned}$ &	$\begin{aligned}[t]		10.647	& \pm	0.044	\\	20.0	& \pm	5.0	\; (	-	)	\phantom{1}	\\	169.0	& \pm	4.2	\; (	+	)	\end{aligned}$ &	$\begin{aligned}[t]	\phantom{11}	224	& \pm	58	\phantom{11}	\\		202.8	& \pm	1.9	\phantom{1}	\\		-29.3	& \pm	4.9	\phantom{1}	\end{aligned}$	\\	\noalign{\smallskip}
$\begin{aligned}[t]	\mbox{\bf NGC 5195\phantom{ 1}}	&	\\	25.79	&	\\	10.46	&	\end{aligned}$ &	$\begin{aligned}[t]	\phantom{1}	-1.0	&	\\	\phantom{-}	601.0	& \pm	8.5	\phantom{1}	\\	\phantom{-}	148.3	& \pm	55.3		\end{aligned}$ &	$\begin{aligned}[t]	0.041	& \pm	0.007	\\	29.403	& \pm	0.264	\\	&	\mbox{S}	\end{aligned}$ &	$\begin{aligned}[t]	\phantom{-}	6.746	& \pm	0.820	\\	3.259	& \pm	0.396	\\	1.383	& \pm	0.152	\end{aligned}$ &	$\begin{aligned}[t]	-23.36	& \pm	0.27	\\	-19.89	& \pm	0.27	\\	0.809	& \pm	0.012	\end{aligned}$ &	$\begin{aligned}[t]		10.524	& \pm	0.106	\\	42.5	& \pm	1.8	\; (	+	)	\phantom{1}	\\	94.5	& \pm	3.5	\; (	-	)	\end{aligned}$ &	$\begin{aligned}[t]	\phantom{11}	105	& \pm	11	\phantom{11}	\\		342.6	& \pm	1.9	\phantom{1}	\\	\phantom{-}	12.1	& \pm	2.4	\phantom{1}	\end{aligned}$	\\	\noalign{\smallskip}
$\begin{aligned}[t]	\mbox{\bf M83\phantom{ngeloo 1}}	&	\\	102.83	&	\\	1.95	&	\end{aligned}$ &	$\begin{aligned}[t]	\phantom{-11}	5.0	&	\\		515.0	& \pm	1.0	\phantom{1}	\\		-26.5	& \pm	15.7		\end{aligned}$ &	$\begin{aligned}[t]	0.075	& \pm	0.012	\\	28.467	& \pm	0.103	\\	&	\mbox{C,P,T}	\end{aligned}$ &	$\begin{aligned}[t]		-1.096	& \pm	0.052	\\	4.813	& \pm	0.228	\\	0.168	& \pm	0.002	\end{aligned}$ &	$\begin{aligned}[t]	-24.08	& \pm	0.11	\\	-21.15	& \pm	0.12	\\	0.577	& \pm	0.032	\end{aligned}$ &	$\begin{aligned}[t]		10.642	& \pm	0.042	\\	25.0	& \pm	5.0	\; (	+	)	\phantom{1}	\\	46.0	& \pm	1.0	\; (	-	)	\end{aligned}$ &	$\begin{aligned}[t]	\phantom{11}	172	& \pm	32	\phantom{11}	\\	\phantom{1}	81.6	& \pm	4.4	\phantom{1}	\\		-13.0	& \pm	2.6	\phantom{1}	\end{aligned}$	\\	\noalign{\smallskip}
$\begin{aligned}[t]	\mbox{\bf M101\phantom{geloo 1}}	&	\\	18.50	&	\\	15.63	&	\end{aligned}$ &	$\begin{aligned}[t]	\phantom{-11}	5.9	&	\\	\phantom{-}	244.0	& \pm	1.0	\phantom{1}	\\		-156.2	& \pm	23.6		\end{aligned}$ &	$\begin{aligned}[t]	0.010	& \pm	0.002	\\	29.353	& \pm	0.130	\\	&	\mbox{C,P,T}	\end{aligned}$ &	$\begin{aligned}[t]	\phantom{-}	6.811	& \pm	0.408	\\	2.278	& \pm	0.136	\\	2.009	& \pm	0.113	\end{aligned}$ &	$\begin{aligned}[t]	-24.05	& \pm	0.14	\\	-21.48	& \pm	0.20	\\	0.390	& \pm	0.010	\end{aligned}$ &	$\begin{aligned}[t]		10.494	& \pm	0.056	\\	21.0	& \pm	3.0	\; (	-	)	\phantom{1}	\\	42.0	& \pm	2.0	\; (	+	)	\end{aligned}$ &	$\begin{aligned}[t]	\phantom{11}	202	& \pm	33	\phantom{11}	\\		186.0	& \pm	1.8	\phantom{1}	\\	\phantom{-1}	2.3	& \pm	2.5	\phantom{1}	\end{aligned}$	\\	\noalign{\smallskip}
$\begin{aligned}[t]	\mbox{\bf Circinus\phantom{oo 1}}	&	\\	138.65	&	\\	-0.51	&	\end{aligned}$ &	$\begin{aligned}[t]	\phantom{-11}	3.3	&	\\		441.5	& \pm	1.5	\phantom{1}	\\		-86.6	& \pm	52.8		\end{aligned}$ &	$\begin{aligned}[t]	0.649	& \pm	0.050	\\	28.145	& \pm	0.375	\\	&	\mbox{TF in V}	\end{aligned}$ &	$\begin{aligned}[t]		-3.193	& \pm	0.551	\\	2.809	& \pm	0.485	\\	-0.038	& \pm	0.029	\end{aligned}$ &	$\begin{aligned}[t]	-23.68	& \pm	0.38	\\	-20.69	& \pm	0.41	\\	0.681	& \pm	0.110	\end{aligned}$ &	$\begin{aligned}[t]		10.559	& \pm	0.150	\\	69.9	& \pm	2.7	\; (	-	)	\phantom{1}	\\	30.1	& \pm	6.1	\; (	-	)	\end{aligned}$ &	$\begin{aligned}[t]	\phantom{11}	154	& \pm	13	\phantom{11}	\\	\phantom{1}	20.5	& \pm	4.2	\phantom{1}	\\		-38.8	& \pm	5.4	\phantom{1}	\end{aligned}$	\\	\noalign{\smallskip}
$\begin{aligned}[t]	\mbox{\bf E274-G001\phantom{ 1}}	&	\\	125.07	&	\\	15.17	&	\end{aligned}$ &	$\begin{aligned}[t]	\phantom{-11}	6.6	&	\\	\phantom{-}	522.0	& \pm	2.0	\phantom{1}	\\	\phantom{-}	128.1	& \pm	11.2		\end{aligned}$ &	$\begin{aligned}[t]	0.293	& \pm	0.047	\\	27.482	& \pm	0.052	\\	&	\mbox{T}	\end{aligned}$ &	$\begin{aligned}[t]		-1.757	& \pm	0.042	\\	2.502	& \pm	0.060	\\	0.829	& \pm	0.017	\end{aligned}$ &	$\begin{aligned}[t]	-20.59	& \pm	0.29	\\	-17.97	& \pm	0.19	\\	0.561	& \pm	0.066	\end{aligned}$ &	$\begin{aligned}[t]	\phantom{1}	9.235	& \pm	0.116	\\	83.9	& \pm	2.6	\; (	+	)	\phantom{1}	\\	43.1	& \pm	1.0	\; (	+	)	\end{aligned}$ &	$\begin{aligned}[t]	\phantom{111}	77	& \pm	3	\phantom{111}	\\		213.0	& \pm	2.8	\phantom{1}	\\	\phantom{-}	19.2	& \pm	1.1	\phantom{1}	\end{aligned}$	\\	\noalign{\smallskip}
$\begin{aligned}[t]	\mbox{\bf Milky Way\phantom{ 1}}	&	\\	134.09	&	\\	87.57	&	\end{aligned}$ &	$\begin{aligned}[t]	\phantom{-11}	3.0	&	\\		-11.1		& \pm	1.2	\phantom{1}	\\		-46.5		& \pm	9.0	\phantom{1}	\end{aligned}$ &	$\begin{aligned}[t]	0.000	& \pm	0.000	\\	14.593	& \pm	0.042	\\	&	\cdots	\end{aligned}$ &	$\begin{aligned}[t]		-0.004	& \pm	0.000	\\	0.004	& \pm	0.000	\\	0.135	& \pm	0.000	\end{aligned}$ &	$\begin{aligned}[t]	-25.26	& \pm	0.29	\\	-22.12	& \pm	0.20	\\	0.551	& \pm	0.051	\end{aligned}$ &	$\begin{aligned}[t]		11.095	& \pm	0.118	\\	90.0	& \pm	0.0	\; (	+	)	\phantom{1}	\\	31.7	& \pm	0.0	\; (	+	)	\end{aligned}$ &	$\begin{aligned}[t]	\phantom{11}	226	& \pm	11	\phantom{11}	\\		224.8	& \pm	0.0	\phantom{1}	\\	\phantom{-1}	0.7	& \pm	0.0	\phantom{1}	\end{aligned}$	\\	\noalign{\smallskip}
$\begin{aligned}[t]	\mbox{\bf NGC 6503\phantom{ 1}}	&	\\	350.36	&	\\	28.83	&	\end{aligned}$ &	$\begin{aligned}[t]	\phantom{-11}	5.9	&	\\	\phantom{-1}	25.5	& \pm	0.4	\phantom{1}	\\		-106.4	& \pm	23.3		\end{aligned}$ &	$\begin{aligned}[t]	0.036	& \pm	0.006	\\	28.559	& \pm	0.213	\\	&	\mbox{T}	\end{aligned}$ &	$\begin{aligned}[t]		4.500	& \pm	0.441	\\	-0.765	& \pm	0.075	\\	2.513	& \pm	0.234	\end{aligned}$ &	$\begin{aligned}[t]	-21.57	& \pm	0.21	\\	-18.95	& \pm	0.22	\\	0.569	& \pm	0.030	\end{aligned}$ &	$\begin{aligned}[t]	\phantom{1}	9.635	& \pm	0.086	\\	72.7	& \pm	1.1	\; (	-	)	\phantom{1}	\\	120.8	& \pm	0.5	\; (	-	)	\end{aligned}$ &	$\begin{aligned}[t]	\phantom{11}	116	& \pm	1	\phantom{111}	\\		260.2	& \pm	1.4	\phantom{1}	\\		-39.8	& \pm	0.6	\phantom{1}	\end{aligned}$	\\	\noalign{\smallskip}
$\begin{aligned}[t]	\mbox{\bf NGC 6946\phantom{ 1}}	&	\\	329.99	&	\\	37.75	&	\end{aligned}$ &	$\begin{aligned}[t]	\phantom{-11}	5.9	&	\\	\phantom{-1}	43.7	& \pm	3.3	\phantom{1}	\\		-121.2	& \pm	32.3		\end{aligned}$ &	$\begin{aligned}[t]	0.390	& \pm	0.062	\\	28.947	& \pm	0.220	\\	&	\mbox{P}	\end{aligned}$ &	$\begin{aligned}[t]		4.269	& \pm	0.433	\\	-2.466	& \pm	0.250	\\	3.817	& \pm	0.374	\end{aligned}$ &	$\begin{aligned}[t]	-23.92	& \pm	0.22	\\	-21.59	& \pm	0.29	\\	0.506	& \pm	0.115	\end{aligned}$ &	$\begin{aligned}[t]		10.529	& \pm	0.089	\\	32.6	& \pm	1.0	\; (	+	)	\phantom{1}	\\	65.9	& \pm	3.2	\; (	-	)	\end{aligned}$ &	$\begin{aligned}[t]	\phantom{11}	200	& \pm	23	\phantom{11}	\\		355.0	& \pm	1.5	\phantom{1}	\\	\phantom{-}	13.1	& \pm	1.3	\phantom{1}	\end{aligned}$	\\	\noalign{\smallskip}
$\begin{aligned}[t]	\mbox{\bf IC 5052\phantom{oo 1}}	&	\\	171.09	&	\\	11.00	&	\end{aligned}$ &	$\begin{aligned}[t]	\phantom{-11}	7.1	&	\\	\phantom{-}	584.0	& \pm	3.0	\phantom{1}	\\	\phantom{-1}	26.1	& \pm	10.2		\end{aligned}$ &	$\begin{aligned}[t]	0.058	& \pm	0.009	\\	28.851	& \pm	0.037	\\	&	\mbox{T}	\end{aligned}$ &	$\begin{aligned}[t]		-5.736	& \pm	0.098	\\	0.900	& \pm	0.015	\\	1.128	& \pm	0.017	\end{aligned}$ &	$\begin{aligned}[t]	-21.12	& \pm	0.30	\\	-18.68	& \pm	0.21	\\	0.448	& \pm	0.051	\end{aligned}$ &	$\begin{aligned}[t]	\phantom{1}	9.366	& \pm	0.120	\\	85.7	& \pm	3.2	\; (	+	)	\phantom{1}	\\	140.0	& \pm	1.0	\; (	+	)	\end{aligned}$ &	$\begin{aligned}[t]	\phantom{111}	87	& \pm	4	\phantom{111}	\\	\phantom{1}	88.3	& \pm	3.3	\phantom{1}	\\		-14.9	& \pm	1.2	\phantom{1}	\end{aligned}$	\\	\noalign{\smallskip}
$\begin{aligned}[t]	\mbox{\bf NGC 7793\phantom{ 1}}	&	\\	216.62	&	\\	12.50	&	\end{aligned}$ &	$\begin{aligned}[t]	\phantom{-11}	7.4	&	\\		226.2	& \pm	1.2	\phantom{1}	\\		-33.1	& \pm	29.8		\end{aligned}$ &	$\begin{aligned}[t]	0.022	& \pm	0.003	\\	27.897	& \pm	0.241	\\	&	\mbox{T}	\end{aligned}$ &	$\begin{aligned}[t]		-2.995	& \pm	0.332	\\	-2.226	& \pm	0.247	\\	0.827	& \pm	0.078	\end{aligned}$ &	$\begin{aligned}[t]	-21.27	& \pm	0.25	\\	-18.94	& \pm	0.26	\\	0.477	& \pm	0.020	\end{aligned}$ &	$\begin{aligned}[t]	\phantom{1}	9.448	& \pm	0.099	\\	51.7	& \pm	2.1	\; (	-	)	\phantom{1}	\\	104.7	& \pm	5.4	\; (	-	)	\end{aligned}$ &	$\begin{aligned}[t]	\phantom{11}	112	& \pm	8	\phantom{1111}	\\		344.9	& \pm	2.2	\phantom{1}	\\	\phantom{1}	-3.8	& \pm	4.2	\phantom{1}	\end{aligned}$	\\	\noalign{\smallskip}


\end{longtable}

\end{ThreePartTable}

\end{center}
\clearpage

\begin{center}
\begin{ThreePartTable}

\begin{TableNotes}

\item[]
(1) 
Name of galaxy, in order of right ascension.
(2) 
Origin of heliocentric velocity.
(3)
Origin of axis ratio, tilt, and position angle.
(4)
Origin of velocity field.
(5) 
Origin of integrated photometry ($B-V$, $V$, and $K_s$).
(6)
Origin of observations required to determine distance.
\smallskip
\item[]
Translations: \\
 abl71	\citep{abl71a};
 ach95	\citep{ach95a};
 adl96	\citep{adl96a};
 agu03	\citep{agu03a};
 alb80	\citep{alb80a};
 ann08	\citep{ann08a};
 baj84	\citep{baj84a};
 bar75	\citep{bar75a};
 beg87	\citep{beg87a};
 ber95	\citep{ber95a};
 bla01	\citep{bla01a};
 blo08	\citep{blo08a};
 bot86	\citep{bot86a};
 bri01	\citep{bri01a};
 bur86	\citep{bur86a};
 bur96	\citep{bur96a};
 bus96	\citep{bus96a};
 but88	\citep{but88a};
 but99	\citep{but99a};
 but04	\citep{but04a};
 cao00	\citep{cao00a};
 cap87	\citep{cap87a};
 cap90	\citep{cap90a};
 cap93	\citep{cap93a};
 cap07	\citep{cap07a};
 car85	\citep{car85a};
 car90	\citep{car90a};
 car06	\citep{car06a};
 cas91	\citep{cas91a};
 cia89a	\citep{cia89a};
 cia89b	\citep{cia89b};
 cia91	\citep{cia91a};
 cia02	\citep{cia02a};
 cia04	\citep{cia04a};
 cop04	\citep{cop04a};
 cro02	\citep{cro02a};
 cur08	\citep{cur08a};
 dah93	\citep{dah93a};
 dah05	\citep{dah05a};
 dai06	\citep{dai06a};
 dre83	\citep{dre83a};
 duf79	\citep{duf79a};
 dur01	\citep{dur01a};
 els97	\citep{els97a};
 ems99	\citep{ems99a};
 ems04	\citep{ems04a};
 erw03	\citep{erw03a};
 fel97	\citep{fel97a};
 fer00	\citep{fer00a};
 fer07	\citep{fer07a};
 fin03	\citep{fin03a};
 fin07	\citep{fin07a};
 fis08	\citep{fis08a};
 fit90	\citep{fit90a};
 for96	\citep{for96a};
 fra02	\citep{fra02a};
 fre77	\citep{fre77a};
 fre01	\citep{fre01a};
 fry99	\citep{fry99a};
 gal04	\citep{gal04a};
 gar02	\citep{gar02a};
 gar03	\citep{gar03a};
 geb00	\citep{geb00a};
 gie08	\citep{gie08a};
 got70	\citep{got70a};
 gro08	\citep{gro08a};
 har99	\citep{har99a};
 hel04	\citep{hel04a};
 her96	\citep{her96a};
 her99	\citep{her99a};
 her05	\citep{her05a};
 her08	\citep{her08a};
 hla11	\citep{hla11a};
 hou61	\citep{hou61a};
 hui93	\citep{hui93a};
 hui95	\citep{hui95a};
 hum90	\citep{hum90a};
 hun98	\citep{hun98a};
 hun99	\citep{hun99a};
 hun06	\citep{hun06a};
 hur96	\citep{hur96a};
 ich95	\citep{ich95a};
 jac89	\citep{jac89a};
 jar03	\citep{jar03a};
 jon99	\citep{jon99a};
 kam92	\citep{kam92a};
 kar02a	\citep{kar02a};
 kar02b	\citep{kar02b};
 kar03a	\citep{kar03a};
 kar03b	\citep{kar03b};
 kar03c	\citep{kar03c};
 kar06	\citep{kar06a};
 kar07	\citep{kar07a};
 kim02	\citep{kim02a};
 kir08	\citep{kir08a};
 kis88	\citep{kis88a};
 kor04	\citep{kor04a};
 kun97	\citep{kun97a};
 mac00	\citep{mac00a};
 mac06	\citep{mac06a};
 mak99	\citep{mak99a};
 mar82	\citep{mar82a};
 mar01	\citep{mar01a};
 mar12	\citep{mar12a};
 may05	\citep{may05a};
 mcc04	\citep{mcc04b};   
 mcc05	\citep{mcc05a};
 mcm94	\citep{mcm94a};
 mcm11	\citep{mcm11a};
 moi04	\citep{moi04a};
 mol04	\citep{mol04a};
 mou05	\citep{mou05a};
 mou08	\citep{mou08a};
 mou09	\citep{mou09a};
 nei99	\citep{nei99a};
 new80a	\citep{new80a};
 new80b	\citep{new80b};
 noo08	\citep{noo08a};
 nor06	\citep{nor06a};
 oka76	\citep{oka76a};
 oka77	\citep{oka77a};
 oll96	\citep{oll96a};
 oos07	\citep{oos07a};
 ott01	\citep{ott01a};
 pat03	\citep{pat03a};
 pen80	\citep{pen80a};
 pen81	\citep{pen81a};
 pie92	\citep{pie92a};
 pis00	\citep{pis00a};
 puc90	\citep{puc90a};
 puc91	\citep{puc91a};
 ran94	\citep{ran94a};
 rei09	\citep{rei09a};
 rej05	\citep{rej05a};
 rek05	\citep{rek05a};
 rix95	\citep{rix95a};
 riz07	\citep{riz07a};
 rog79	\citep{rog79a};
 rub85	\citep{rub85a};
 ryd94	\citep{ryd94a};
 ryd95	\citep{ryd95a};
 sah02	\citep{sah02a};
 sak99	\citep{sak99a};
 sak00	\citep{sak00a};
 sak04	\citep{sak04a};
 san79	\citep{san79a};
 sch77	\citep{sch77a};
 sch78 	\citep{sch78a};
 sco85	\citep{sco85a};
 set05	\citep{set05a};
 sho73	\citep{sho73a};
 sho84	\citep{sho84a};
 sim02	\citep{sim02a};
 sof93	\citep{sof93a};
 sof96	\citep{sof96a};
 sof97	\citep{sof97a};
 sof09	\citep{sof09a};
 sor96	\citep{sor96a};
 spi92	\citep{spi92a};
 sta99	\citep{sta99a};
 sti02	\citep{sti02a};
 str77	\citep{str77a};
 swa97	\citep{swa97a};
 swa02	\citep{swa02a};
 tal79	\citep{tal79a};
 thi03	\citep{thi03a};
 til91	\citep{til91a};
 ton01	\citep{ton01a};
 tul74	\citep{tul74a};
 vad05	\citep{vad05a};
 vau58	\citep{vau58a};
 vau59	\citep{vau59a};
 vau64	\citep{vau64a};
 vau80	\citep{vau80a};
 vau82	\citep{vau82a};
 veg01	\citep{veg01a};
 ver00	\citep{ver00a};
 wal87	\citep{wal87a};
 wev86	\citep{wev86a};
 wil86	\citep{wil86a};
 woe93	\citep{woe93a};
 won04	\citep{won04a};
 woo07	\citep{woo07a};
 xue08	\citep{xue08a};
 zee02	\citep{zee02a}.


\end{TableNotes}

\setlength{\jot}{-0.0pt}

\begin{longtable}{l p{1.8cm} p{2.8cm} p{2.8cm} p{2.6cm} p{2.8cm}}

\caption{Sources of Observations\label{tbl_sources}}  
\\

\hline
\noalign{\smallskip}

Galaxy &
Velocity &
Orientation &
Rotation &
Photometry &
Distance
\\ 

\noalign{\smallskip}

(1) & 
(2) & 
(3) & 
(4) & 
(5) &
(6)
 \\

\noalign{\smallskip}
\hline
\noalign{\smallskip}
\endfirsthead

\caption{cont'd.} \\

\hline
\noalign{\smallskip}

Galaxy &
Velocity &
Orientation &
Rotation &
Photometry &
Distance
\\

\noalign{\smallskip}
 
(1) & 
(2) & 
(3) & 
(4) & 
(5) &
(6)
\\

\noalign{\smallskip}
\hline
\endhead

\noalign{\smallskip}
\hline

\endfoot
\hline
\noalign{\smallskip}
\insertTableNotes

\endlastfoot


\mbox{NGC 55\phantom{oo 1}}	&	\raggedright	puc91	&	\raggedright	kis88, pat03, puc91	&	\raggedright	puc91	&	\raggedright	pat03, fit90, jar03	&	\raggedright	gie08, set05	 \tabularnewline	\noalign{\smallskip}
\mbox{Andromeda\phantom{1}}	&	\raggedright	mar12, got70	&	\raggedright	sak00, pie92, vau58, got70	&	\raggedright	car06	&	\raggedright	wal87, jar03	&	\raggedright	fre01, cia89a, fer00, ton01, dur01, mcc05	 \tabularnewline	\noalign{\smallskip}
\mbox{NGC 247\phantom{o 1}}	&	\raggedright	car90	&	\raggedright	car85, car90, hla11	&	\raggedright	car90	&	\raggedright	car85, pat03, jar03	&	\raggedright	kar06, mou08	 \tabularnewline	\noalign{\smallskip}
\mbox{NGC 253\phantom{o 1}}	&	\raggedright	pen81, sco85	&	\raggedright	pen80, pen81	&	\raggedright	sof97	&	\raggedright	pen80, fit90, jar03	&	\raggedright	rek05, kar03c, mou05	 \tabularnewline	\noalign{\smallskip}
\mbox{NGC 300\phantom{o 1}}	&	\raggedright	puc90, rog79	&	\raggedright	car85, puc90	&	\raggedright	puc90	&	\raggedright	car85, pat03, jar03	&	\raggedright	fre01, sof96, riz07, sak04, but04	 \tabularnewline	\noalign{\smallskip}
\mbox{M33\phantom{ngeloo 1}}	&	\raggedright	new80a	&	\raggedright	sak00, pie92, vau59, new80a	&	\raggedright	new80a	&	\raggedright	pie92, vau59, jar03	&	\raggedright	fre01, cia04, riz07, kim02, gal04, mcc04, mcc05	 \tabularnewline	\noalign{\smallskip}
\mbox{M74\phantom{ngeloo 1}}	&	\raggedright	sho84, kam92	&	\raggedright	mol04, sho84, kam92	&	\raggedright	TF in $K_s$	&	\raggedright	pat03, mar01, jar03	&	\raggedright	her08	 \tabularnewline	\noalign{\smallskip}
\mbox{NGC 672\phantom{o 1}}	&	\raggedright	gar03	&	\raggedright	her96, gar03	&	\raggedright	gar03	&	\raggedright	pat03, her96, jar03	&	\raggedright	her96, gar03	 \tabularnewline	\noalign{\smallskip}
\mbox{NGC 891\phantom{o 1}}	&	\raggedright	oo07, swa97	&	\raggedright	sof93, mar01, oos07	&	\raggedright	san79, sof97	&	\raggedright	pat03, jar03	&	\raggedright	cia91, mou08	 \tabularnewline	\noalign{\smallskip}
\mbox{NGC 925\phantom{o 1}}	&	\raggedright	blo08	&	\raggedright	sak00, pis00, blo08	&	\raggedright	blo08	&	\raggedright	pat03, mac00, jar03	&	\raggedright	fre01	 \tabularnewline	\noalign{\smallskip}
\mbox{NGC 1023\phantom{ 1}}	&	\raggedright	noo08	&	\raggedright	noo08	&	\raggedright	noo08, agu03	&	\raggedright	bar75, pat03, jar03	&	\raggedright	cia91, ton01	 \tabularnewline	\noalign{\smallskip}
\mbox{Maffei 1\phantom{oo 1}}	&	\raggedright	fin03	&	\raggedright	but99	&		$\cdots$	&	\raggedright	but99, jar03	&	\raggedright	but99, fin03	 \tabularnewline	\noalign{\smallskip}
\mbox{Maffei 2\phantom{oo 1}}	&	\raggedright	hur96, but99	&	\raggedright	but99, hur96	&	\raggedright	hur96, fin07	&	\raggedright	but99, jar03	&	\raggedright	but99, hur96, fin07	 \tabularnewline	\noalign{\smallskip}
\mbox{Dwingeloo 1}	&	\raggedright	bur96, but99	&	\raggedright	but99, bur96	&	\raggedright	bur96	&	\raggedright	but99, jar03, twins N0598 and N0925	&	\raggedright	but99, bur96	 \tabularnewline	\noalign{\smallskip}
\mbox{NGC 1313\phantom{ 1}}	&	\raggedright	ryd95	&	\raggedright	ryd95, mar82	&	\raggedright	ryd95	&	\raggedright	pat03, ryd95, jar03	&	\raggedright	riz07	 \tabularnewline	\noalign{\smallskip}
\mbox{IC 342\phantom{loo 1}}	&	\raggedright	new80b	&	\raggedright	but99, new80b	&	\raggedright	sof97, new80b	&	\raggedright	but99, jar03	&	\raggedright	sah02, her08, fin07	 \tabularnewline	\noalign{\smallskip}
\mbox{NGC 1569\phantom{ 1}}	&	\raggedright	sti02	&	\raggedright	but99	&	\raggedright	sti02	&	\raggedright	hun06, but99, vad05	&	\raggedright	gro08	 \tabularnewline	\noalign{\smallskip}
\mbox{NGC 2403\phantom{ 1}}	&	\raggedright	fra02	&	\raggedright	sak00, pie92, pat03, fra02 	&	\raggedright	fra02	&	\raggedright	pie92, oka77, jar03	&	\raggedright	fre01, cia02	 \tabularnewline	\noalign{\smallskip}
\mbox{NGC 2683\phantom{ 1}}	&	\raggedright	cas91	&	\raggedright	her96, cas91	&	\raggedright	cas91	&	\raggedright	pat03, her96, jar03	&	\raggedright	ton01	 \tabularnewline	\noalign{\smallskip}
\mbox{NGC 2784\phantom{ 1}}	&	\raggedright	bla01	&	\raggedright	kir08	&	\raggedright	dre83	&	\raggedright	pat03, jar03	&	\raggedright	ton01	 \tabularnewline	\noalign{\smallskip}
\mbox{NGC 2787\phantom{ 1}}	&	\raggedright	ber95	&	\raggedright	nei99, erw03	&	\raggedright	erw03	&	\raggedright	pat03, fis08, jar03	&	\raggedright	ton01	 \tabularnewline	\noalign{\smallskip}
\mbox{NGC 2903\phantom{ 1}}	&	\raggedright	blo08	&	\raggedright	fis08, her96, her05	&	\raggedright	blo08	&	\raggedright	pat03, her96, fis08, jar03	&	\raggedright	her96, blo08	 \tabularnewline	\noalign{\smallskip}
\mbox{M81\phantom{ngeloo 1}}	&	\raggedright	blo08	&	\raggedright	but99, fis08, mol04, blo08	&	\raggedright	blo08, adl96	&	\raggedright	pat03, but99, fis08, jar03	&	\raggedright	fre01, jac89, ton01, riz07, sak04	 \tabularnewline	\noalign{\smallskip}
\mbox{M82\phantom{ngeloo 1}}	&	\raggedright	ach95	&	\raggedright	may05, ach95	&	\raggedright	ach95	&	\raggedright	pat03, ich95	&	\raggedright	sak99	 \tabularnewline	\noalign{\smallskip}
\mbox{NGC 3115\phantom{ 1}}	&	\raggedright	cap93, nor06	&	\raggedright	cap87, ems99	&	\raggedright	cap93	&	\raggedright	pat03, str77, jar03	&	\raggedright	cia02, ton01, els97	 \tabularnewline	\noalign{\smallskip}
\mbox{NGC 3344\phantom{ 1}}	&	\raggedright	ver00	&	\raggedright	ver00	&	\raggedright	ver00	&	\raggedright	pat03, jar03	&	\raggedright	pat03, ver00	 \tabularnewline	\noalign{\smallskip}
\mbox{M95\phantom{ngeloo 1}}	&	\raggedright	but88	&	\raggedright	sak00, her96, but88	&	\raggedright	but88	&	\raggedright	mac00, jar03	&	\raggedright	fre01, cia02, riz07, sak04	 \tabularnewline	\noalign{\smallskip}
\mbox{M96\phantom{ngeloo 1}}	&	\raggedright	her99	&	\raggedright	sak00, moi04	&	\raggedright	her99, veg01	&	\raggedright	mac00, jar03	&	\raggedright	fre01, fel97, ton01	 \tabularnewline	\noalign{\smallskip}
\mbox{NGC 3377\phantom{ 1}}	&	\raggedright	ems04	&	\raggedright	pat03, cap07, cop04	&	\raggedright	sim02, cop04	&	\raggedright	pat03, jar03	&	\raggedright	cia89b, ton01	 \tabularnewline	\noalign{\smallskip}
\mbox{M105\phantom{geloo 1}}	&	\raggedright	ems04	&	\raggedright	cap90, cap07, geb00	&	\raggedright	sta99	&	\raggedright	cap90, pat03, jar03	&	\raggedright	cia89b, ton01	 \tabularnewline	\noalign{\smallskip}
\mbox{NGC 3384\phantom{ 1}}	&	\raggedright	ems04	&	\raggedright	bus96, cap07	&	\raggedright	zee02	&	\raggedright	bus96, pat03, jar03	&	\raggedright	cia89b, ton01, mou09	 \tabularnewline	\noalign{\smallskip}
\mbox{NGC 3412\phantom{ 1}}	&	\raggedright	agu03	&	\raggedright	agu03	&	\raggedright	agu03	&	\raggedright	pat03, jar03	&	\raggedright	ton01	 \tabularnewline	\noalign{\smallskip}
\mbox{NGC 3489\phantom{ 1}}	&	\raggedright	ems04	&	\raggedright	nei99, cap07	&	\raggedright	cao00	&	\raggedright	pat03, jar03	&	\raggedright	ton01	 \tabularnewline	\noalign{\smallskip}
\mbox{NGC 3621\phantom{ 1}}	&	\raggedright	blo08	&	\raggedright	sak00, pat03, blo08	&	\raggedright	blo08	&	\raggedright	mac00, jar03	&	\raggedright	fre01, riz07, sak04	 \tabularnewline	\noalign{\smallskip}
\mbox{M66\phantom{ngeloo 1}}	&	\raggedright	blo08	&	\raggedright	sak00, fis08, pat03, blo08	&	\raggedright	blo08	&	\raggedright	mac00, jar03	&	\raggedright	fre01, cia02	 \tabularnewline	\noalign{\smallskip}
\mbox{NGC 4144\phantom{ 1}}	&	\raggedright	gar02	&	\raggedright	swa02, gar02	&	\raggedright	gar02	&	\raggedright	mak99, jar03	&	\raggedright	set05	 \tabularnewline	\noalign{\smallskip}
\mbox{NGC 4236\phantom{ 1}}	&	\raggedright	sho73	&	\raggedright	swa02, dai06	&	\raggedright	sho73	&	\raggedright	pie92, pat03, jar03	&	\raggedright	kar02a	 \tabularnewline	\noalign{\smallskip}
\mbox{NGC 4244\phantom{ 1}}	&	\raggedright	dah05	&	\raggedright	fry99, pat03, oll96, dah05 	&	\raggedright	oll96	&	\raggedright	pat03, jar03	&	\raggedright	kar03a, set05	 \tabularnewline	\noalign{\smallskip}
\mbox{M106\phantom{geloo 1}}	&	\raggedright	alb80, wev86	&	\raggedright	fis08, pat03, alb80, wev86		&	\raggedright	alb80	&	\raggedright	pat03, fis08, jar03	&	\raggedright	mac06, cia02, ton01, riz07, mac06	 \tabularnewline	\noalign{\smallskip}
\mbox{NGC 4449\phantom{ 1}}	&	\raggedright	hun98	&	\raggedright	hun99, hun98	&	\raggedright	hun98	&	\raggedright	mak99, jar03	&	\raggedright	ann08, kar03a	 \tabularnewline	\noalign{\smallskip}
\mbox{M104\phantom{geloo 1}}	&	\raggedright	baj84	&	\raggedright	bur86, hou61, sch78 	&	\raggedright	baj84, rub85	&	\raggedright	bur86, pat03, jar03	&	\raggedright	for96, ton01	 \tabularnewline	\noalign{\smallskip}
\mbox{NGC 4631\phantom{ 1}}	&	\raggedright	ran94	&	\raggedright	hum90, pat03, ran94	&	\raggedright	ran94	&	\raggedright	hum90, jar03	&	\raggedright	set05	 \tabularnewline	\noalign{\smallskip}
\mbox{M94\phantom{ngeloo 1}}	&	\raggedright	blo08	&	\raggedright	mol04, fis08, blo08	&	\raggedright	blo08	&	\raggedright	pat03, fis08, jar03	&	\raggedright	her08, kar03a	 \tabularnewline	\noalign{\smallskip}
\mbox{M64\phantom{ngeloo 1}}	&	\raggedright	blo08	&	\raggedright	fis08, her96, blo08	&	\raggedright	rix95	&	\raggedright	pat03, fis08, jar03	&	\raggedright	mou08	 \tabularnewline	\noalign{\smallskip}
\mbox{NGC 4945\phantom{ 1}}	&	\raggedright	dah93, ott01	&	\raggedright	vau64, dah93, ott01	&	\raggedright	sof97	&	\raggedright	pat03, jar03, twins N3877 and N4157	&	\raggedright	mou08, mou05	 \tabularnewline	\noalign{\smallskip}
\mbox{NGC 5023\phantom{ 1}}	&	\raggedright	gar02	&	\raggedright	swa02, bot86, gar02	&	\raggedright	gar02	&	\raggedright	pat03, jar03	&	\raggedright	set05	 \tabularnewline	\noalign{\smallskip}
\mbox{NGC 5068\phantom{ 1}}	&	\raggedright	kor04	&	\raggedright	ryd94, hel04	&	\raggedright	kor04	&	\raggedright	pat03, jar03	&	\raggedright	her08	 \tabularnewline	\noalign{\smallskip}
\mbox{NGC 5102\phantom{ 1}}	&	\raggedright	woe93	&	\raggedright	pat03, woe93	&	\raggedright	woe93	&	\raggedright	pat03, jar03	&	\raggedright	mcm94, kar02b	 \tabularnewline	\noalign{\smallskip}
\mbox{Centaurus A}	&	\raggedright	hui95	&	\raggedright	hui95, duf79, wil86, woo07 	&	\raggedright	woo07	&	\raggedright	duf79, jar03	&	\raggedright	fer07, hui93, riz07, rej05, har99, sor96	 \tabularnewline	\noalign{\smallskip}
\mbox{M51\phantom{ngeloo 1}}	&	\raggedright	tul74, til91	&	\raggedright	fis08, pat03, tul74, dai06, kun97	&	\raggedright	dai06, kun97	&	\raggedright	oka76, fis08, jar03	&	\raggedright	fel97	 \tabularnewline	\noalign{\smallskip}
\mbox{NGC 5195\phantom{ 1}}	&	\raggedright	sch77	&	\raggedright	bri01, sch77, spi92	&	\raggedright	sch77	&	\raggedright	oka76, jar03	&	\raggedright	ton01	 \tabularnewline	\noalign{\smallskip}
\mbox{M83\phantom{ngeloo 1}}	&	\raggedright	cro02	&	\raggedright	pat03, hel04, cro02	&	\raggedright	cro02	&	\raggedright	tal79, jar03	&	\raggedright	thi03, her08, kar07	 \tabularnewline	\noalign{\smallskip}
\mbox{M101\phantom{geloo 1}}	&	\raggedright	won04	&	\raggedright	won04, her05	&	\raggedright	won04	&	\raggedright	oka76, jar03	&	\raggedright	fre01, fel97, riz07, sak04	 \tabularnewline	\noalign{\smallskip}
\mbox{Circinus\phantom{oo 1}}	&	\raggedright	jon99, cur08	&	\raggedright	fre77, cur08, pat03	&	\raggedright	cur08, jon99	&	\raggedright	fre77, jar03	&	\raggedright	fre77, cur08, jon99	 \tabularnewline	\noalign{\smallskip}
\mbox{E274-G001\phantom{ 1}}	&	\raggedright	kor04	&	\raggedright	pat03	&	\raggedright	kor04	&	\raggedright	pat03, jar03	&	\raggedright	kar07	 \tabularnewline	\noalign{\smallskip}
\mbox{Milky Way\phantom{ 1}}	&	\raggedright	mar12	&	\raggedright	defined	&	\raggedright	car06, xue08, sof09, rei09, mcm11, mar12	&	\raggedright	TF in V, TF in $K_s$, twin N0891	&	\raggedright	mar12	 \tabularnewline	\noalign{\smallskip}
\mbox{NGC 6503\phantom{ 1}}	&	\raggedright	beg87	&	\raggedright	her96, beg87	&	\raggedright	beg87	&	\raggedright	mak99, vau82, jar03	&	\raggedright	kar03b	 \tabularnewline	\noalign{\smallskip}
\mbox{NGC 6946\phantom{ 1}}	&	\raggedright	blo08	&	\raggedright	abl71, blo08	&	\raggedright	blo08	&	\raggedright	mak99, fis08, jar03	&	\raggedright	her08	 \tabularnewline	\noalign{\smallskip}
\mbox{IC 5052\phantom{oo 1}}	&	\raggedright	kor04	&	\raggedright	kir08	&	\raggedright	kor04	&	\raggedright	pat03, jar03	&	\raggedright	set05	 \tabularnewline	\noalign{\smallskip}
\mbox{NGC 7793\phantom{ 1}}	&	\raggedright	blo08	&	\raggedright	car85, blo08	&	\raggedright	blo08	&	\raggedright	car85, vau80, jar03	&	\raggedright	kar03c	 \tabularnewline	

\noalign{\smallskip}

\newpage

\end{longtable}

\end{ThreePartTable}

\end{center}
\clearpage

\begin{center}
\begin{ThreePartTable}

\begin{TableNotes}

\item[]
(1) 
Name of galaxy, in order of right ascension.
(2) 
Tilt from photometry, in degrees.
(3)
Tilt from kinematics, in degrees.
(4)
Adopted tilt, in degrees.
(5) 
Position angle of line of nodes from photometry, in degrees measured eastward from north (epoch 1950 assumed).
(6)
Position angle of line of nodes from kinematics, in degrees measured eastward from north (epoch 1950 assumed).
(7)
Adopted position angle of line of nodes, in degrees measured eastward from north (epoch 1950 assumed).
(8)
Points of relevance.
\smallskip
\item[]
Notes:  
(1) highly inclined; (2) tilt from relative scale heights of different populations; (3) near face-on; (4) tilt derived from Tully-Fisher relation in $K_s$;  (5) intrinsic axis ratio uncertain; (6) heavy extinction; (7) complex velocity field; (8) disturbed or warped; (9) gas captured;  (10) non-circular motions and isophotes; (11) properties are for PNe in spheroidal component.


\end{TableNotes}

\setlength{\jot}{-0.0pt}

\begin{longtable}{lccccccl}

\caption{Orientational Data\label{tbl_orientations}}  
\\

\hline
\noalign{\smallskip}

Galaxy &
$i \, \mbox{(phot)}$ &
$i \, \mbox{(kin)}$ &
$i$ &
$PA \, \mbox{(phot)}$ &
$PA \, \mbox{(kin)}$ &
$PA$ &
Notes
\\ 

\noalign{\smallskip}

(1) & 
(2) & 
(3) & 
(4) & 
(5) & 
(6) &
(7) &
(8)
\\

\noalign{\smallskip}
\hline
\noalign{\smallskip}
\endfirsthead

\caption{cont'd.} \\

\hline
\noalign{\smallskip}

\mbox{Galaxy} &
$i \, \mbox{(phot)}$ &
$i \, \mbox{(kin)}$ &
$i$ &
$PA \, \mbox{(phot)}$ &
$PA \, \mbox{(kin)}$ &
$PA$ &
Notes
\\ 

\noalign{\smallskip}
 
(1) & 
(2) & 
(3) & 
(4) & 
(5) & 
(6) &
(7) &
(8)
\\

\noalign{\smallskip}
\hline
\noalign{\smallskip}
\endhead

\noalign{\smallskip}
\hline

\endfoot
\hline
\noalign{\smallskip}
\insertTableNotes

\endlastfoot


\mbox{NGC 55\phantom{oo 1}}	&	$	81.2	\pm	\phantom{1}	1.6	$	&	$		77.0	\pm	\phantom{1}	2.0	$	&	$		81.2	\pm	\phantom{1}	1.6	$	&	$		101.0	\pm	\phantom{1}	1.0	$	&	$		109.0	\pm	\phantom{1}	3.0	$	&	$		105.0	\pm	\phantom{1}	4.0	$	&	1, 2	\\
\mbox{Andromeda\phantom{1}}	&	$	78.0	\pm	\phantom{1}	2.6	$	&	$		78.0	\pm	\phantom{1}	1.0	$	&	$		78.0	\pm	\phantom{1}	0.5	$	&	$	\phantom{1}	37.7	\pm	\phantom{1}	0.2	$	&	$	\phantom{1}	38.0	\pm	\phantom{1}	1.0	$	&	$	\phantom{1}	37.9	\pm	\phantom{1}	0.5	$	&		\\
\mbox{NGC 247\phantom{o 1}}	&	$	75.5	\pm	\phantom{1}	1.6	$	&	$		74.0	\pm	\phantom{1}	1.0	$	&	$		74.8	\pm	\phantom{1}	0.8	$	&	$		171.1	\pm	\phantom{1}	0.1	$	&	$		170.0	\pm	\phantom{1}	1.0	$	&	$		170.6	\pm	\phantom{1}	0.5	$	&		\\
\mbox{NGC 253\phantom{o 1}}	&	$	76.9	\pm	\phantom{1}	1.7	$	&	$		75.0	\pm	\phantom{1}	0.5	$	&	$		75.9	\pm	\phantom{1}	0.9	$	&	$	\phantom{1}	51.0	\pm	\phantom{1}	0.1	$	&	$	\phantom{1}	51.2	\pm	\phantom{1}	0.8	$	&	$	\phantom{1}	51.1	\pm	\phantom{1}	0.5	$	&		\\
\mbox{NGC 300\phantom{o 1}}	&	$	42.7	\pm	\phantom{1}	6.1	$	&	$		50.0	\pm	\phantom{1}	3.0	$	&	$		46.4	\pm	\phantom{1}	3.6	$	&	$		105.6	\pm	\phantom{1}	1.8	$	&	$		108.0	\pm	\phantom{1}	4.0	$	&	$		106.8	\pm	\phantom{1}	1.2	$	&		\\
\mbox{M33\phantom{ngeloo 1}}	&	$	54.0	\pm	\phantom{1}	2.0	$	&	$		54.0	\pm	\phantom{1}	1.0	$	&	$		54.0	\pm	\phantom{1}	0.5	$	&	$	\phantom{1}	23.0	\pm	\phantom{1}	1.0	$	&	$	\phantom{1}	22.0	\pm	\phantom{1}	1.0	$	&	$	\phantom{1}	22.5	\pm	\phantom{1}	0.5	$	&		\\
\mbox{M74\phantom{ngeloo 1}}	&	$	14.2	\pm	\phantom{1}	2.4	$	&	$	\phantom{1}	9.3	\pm	\phantom{1}	0.9	$	&	$	\phantom{1}	9.3	\pm	\phantom{1}	0.9	$	&	$	\phantom{1}	71.1	\pm	\phantom{1}	1.0	$	&	$	\phantom{1}	25.0	\pm	\phantom{1}	5.0	$	&	$	\phantom{1}	25.0	\pm	\phantom{1}	5.0	$	&	3, 4	\\
\mbox{NGC 672\phantom{o 1}}	&	$	63.6	\pm	\phantom{1}	0.8	$	&	$		65.0	\pm	\phantom{1}	3.0	$	&	$		64.3	\pm	\phantom{1}	0.7	$	&	$	\phantom{1}	71.0	\pm	\phantom{1}	1.0	$	&	$	\phantom{1}	64.0	\pm	\phantom{1}	3.0	$	&	$	\phantom{1}	67.5	\pm	\phantom{1}	3.5	$	&		\\
\mbox{NGC 891\phantom{o 1}}	&	$	88.3	\pm	\phantom{1}	1.5	$	&	$		\cdots	$	&	$		88.3	\pm	\phantom{1}	1.5	$	&	$	\phantom{1}	22.0	\pm	\phantom{1}	0.5	$	&	$	\phantom{1}	23.0	\pm	\phantom{1}	1.0	$	&	$	\phantom{1}	22.5	\pm	\phantom{1}	0.5	$	&	1	\\
\mbox{NGC 925\phantom{o 1}}	&	$	56.0	\pm	\phantom{1}	1.1	$	&	$		66.0	\pm	\phantom{1}	1.0	$	&	$		61.0	\pm	\phantom{1}	5.0	$	&	$		112.0	\pm	\phantom{1}	4.0	$	&	$		106.6	\pm	\phantom{1}	1.0	$	&	$		109.3	\pm	\phantom{1}	2.7	$	&		\\
\mbox{NGC 1023\phantom{ 1}}	&	$	72.2	\pm	\phantom{1}	1.3	$	&	$		\cdots	$	&	$		72.2	\pm	\phantom{1}	1.3	$	&	$	\phantom{1}	85.0	\pm	\phantom{1}	1.0	$	&	$		\cdots	$	&	$	\phantom{1}	85.0	\pm	\phantom{1}	1.0	$	&		\\
\mbox{Maffei 1\phantom{oo 1}}	&	$	\cdots	$	&	$		\cdots	$	&	$		\cdots	$	&	$	\phantom{1}	83.9	\pm	\phantom{1}	0.7	$	&	$		\cdots	$	&	$	\phantom{1}	83.9	\pm	\phantom{1}	0.7	$	&	5, 6	\\
\mbox{Maffei 2\phantom{oo 1}}	&	$	66.2	\pm	\phantom{1}	0.6	$	&	$		67.0	\pm	\phantom{1}	1.0	$	&	$		67.0	\pm	\phantom{1}	1.0	$	&	$	\phantom{1}	23.0	\pm	\phantom{1}	0.7	$	&	$	\phantom{1}	26.0	\pm	\phantom{1}	1.0	$	&	$	\phantom{1}	24.5	\pm	\phantom{1}	1.5	$	&	6	\\
\mbox{Dwingeloo 1}	&	$	46.2	\pm	\phantom{1}	0.3	$	&	$		51.0	\pm	\phantom{1}	2.0	$	&	$		51.0	\pm	\phantom{1}	2.0	$	&	$		110.7	\pm	\phantom{1}	2.0	$	&	$		112.0	\pm	\phantom{1}	1.0	$	&	$		111.4	\pm	\phantom{1}	0.6	$	&	6	\\
\mbox{NGC 1313\phantom{ 1}}	&	$	42.7	\pm	\phantom{1}	0.9	$	&	$		48.0	\pm	\phantom{1}	3.0	$	&	$		45.4	\pm	\phantom{1}	2.6	$	&	$	\phantom{11}	4.0	\pm		10.0	$	&	$	\phantom{11}	1.0	\pm	\phantom{1}	3.0	$	&	$	\phantom{11}	2.5	\pm	\phantom{1}	1.5	$	&		\\
\mbox{IC 342\phantom{loo 1}}	&	$	29.5	\pm	\phantom{1}	0.5	$	&	$		25.0	\pm	\phantom{1}	3.0	$	&	$		25.0	\pm	\phantom{1}	3.0	$	&	$	\phantom{1}	86.5	\pm	\phantom{1}	1.6	$	&	$	\phantom{1}	39.0	\pm	\phantom{1}	3.0	$	&	$	\phantom{1}	39.0	\pm	\phantom{1}	3.0	$	&	6	\\
\mbox{NGC 1569\phantom{ 1}}	&	$	90.0	\pm	\phantom{1}	1.0	$	&	$		\cdots	$	&	$		90.0	\pm	\phantom{1}	1.0	$	&	$		119.3	\pm	\phantom{1}	1.1	$	&	$		\cdots	$	&	$		119.3	\pm	\phantom{1}	1.1	$	&	7	\\
\mbox{NGC 2403\phantom{ 1}}	&	$	58.0	\pm	\phantom{1}	2.0	$	&	$		62.9	\pm	\phantom{1}	2.1	$	&	$		60.5	\pm	\phantom{1}	2.5	$	&	$		126.0	\pm	\phantom{1}	1.0	$	&	$		124.5	\pm	\phantom{1}	0.6	$	&	$		125.3	\pm	\phantom{1}	0.8	$	&		\\
\mbox{NGC 2683\phantom{ 1}}	&	$	79.3	\pm	\phantom{1}	2.1	$	&	$		\cdots	$	&	$		79.3	\pm	\phantom{1}	2.1	$	&	$	\phantom{1}	44.0	\pm	\phantom{1}	1.0	$	&	$	\phantom{1}	41.5	\pm	\phantom{1}	1.0	$	&	$	\phantom{1}	42.8	\pm	\phantom{1}	1.3	$	&	1	\\
\mbox{NGC 2784\phantom{ 1}}	&	$	66.4	\pm	\phantom{1}	1.1	$	&	$		\cdots	$	&	$		66.4	\pm	\phantom{1}	1.1	$	&	$	\phantom{1}	73.0	\pm	\phantom{1}	1.0	$	&	$		\cdots	$	&	$	\phantom{1}	73.0	\pm	\phantom{1}	1.0	$	&	5	\\
\mbox{NGC 2787\phantom{ 1}}	&	$	55.5	\pm	\phantom{1}	1.6	$	&	$		\cdots	$	&	$		55.5	\pm	\phantom{1}	1.6	$	&	$		109.0	\pm	\phantom{1}	1.0	$	&	$		\cdots	$	&	$		109.0	\pm	\phantom{1}	1.0	$	&	5	\\
\mbox{NGC 2903\phantom{ 1}}	&	$	60.9	\pm	\phantom{1}	0.8	$	&	$		61.5	\pm	\phantom{1}	0.5	$	&	$		61.2	\pm	\phantom{1}	0.5	$	&	$	\phantom{1}	24.0	\pm	\phantom{1}	1.0	$	&	$	\phantom{1}	22.0	\pm	\phantom{1}	1.0	$	&	$	\phantom{1}	23.0	\pm	\phantom{1}	1.0	$	&		\\
\mbox{M81\phantom{ngeloo 1}}	&	$	55.5	\pm	\phantom{1}	0.9	$	&	$		59.0	\pm	\phantom{1}	1.0	$	&	$		57.2	\pm	\phantom{1}	1.8	$	&	$		152.3	\pm	\phantom{1}	1.0	$	&	$		150.2	\pm	\phantom{1}	1.0	$	&	$		151.3	\pm	\phantom{1}	1.1	$	&		\\
\mbox{M82\phantom{ngeloo 1}}	&	$	76.0	\pm	\phantom{1}	1.8	$	&	$		73.0	\pm	\phantom{1}	3.0	$	&	$		76.0	\pm	\phantom{1}	1.8	$	&	$	\phantom{1}	64.0	\pm	\phantom{1}	1.0	$	&	$	\phantom{1}	70.0	\pm	\phantom{1}	3.0	$	&	$	\phantom{1}	67.0	\pm	\phantom{1}	3.0	$	&	8	\\
\mbox{NGC 3115\phantom{ 1}}	&	$	86.0	\pm	\phantom{1}	5.2	$	&	$		86.0	\pm	\phantom{1}	1.0	$	&	$		86.0	\pm	\phantom{1}	0.5	$	&	$	\phantom{1}	43.5	\pm	\phantom{1}	1.0	$	&	$		\cdots	$	&	$	\phantom{1}	43.5	\pm	\phantom{1}	1.0	$	&		\\
\mbox{NGC 3344\phantom{ 1}}	&	$	25.3	\pm	\phantom{1}	0.3	$	&	$		25.5	\pm	\phantom{1}	0.5	$	&	$		25.5	\pm	\phantom{1}	0.5	$	&	$		159.8	\pm	\phantom{1}	1.6	$	&	$		156.1	\pm	\phantom{1}	0.7	$	&	$		156.1	\pm	\phantom{1}	0.7	$	&	3	\\
\mbox{M95\phantom{ngeloo 1}}	&	$	45.0	\pm	\phantom{1}	2.0	$	&	$		\cdots	$	&	$		45.0	\pm	\phantom{1}	2.0	$	&	$		179.0	\pm	\phantom{1}	1.0	$	&	$	\phantom{1}	13.0	\pm	\phantom{1}	1.0	$	&	$	\phantom{11}	6.0	\pm	\phantom{1}	7.0	$	&	7	\\
\mbox{M96\phantom{ngeloo 1}}	&	$	49.5	\pm	\phantom{1}	1.6	$	&	$		\cdots	$	&	$		49.5	\pm	\phantom{1}	1.6	$	&	$		135.0	\pm	\phantom{1}	5.0	$	&	$		\cdots	$	&	$		135.0	\pm	\phantom{1}	5.0	$	&	9	\\
\mbox{NGC 3377\phantom{ 1}}	&	$	\cdots	$	&	$		90.0	\pm		10.0	$	&	$		90.0	\pm		10.0	$	&	$	\phantom{1}	41.3	\pm	\phantom{1}	1.0	$	&	$	\phantom{1}	46.0	\pm	\phantom{1}	1.0	$	&	$	\phantom{1}	43.7	\pm	\phantom{1}	2.4	$	&	5	\\
\mbox{M105\phantom{geloo 1}}	&	$	\cdots	$	&	$		90.0	\pm		10.0	$	&	$		90.0	\pm		10.0	$	&	$	\phantom{1}	67.9	\pm	\phantom{1}	1.0	$	&	$	\phantom{1}	72.0	\pm	\phantom{1}	2.0	$	&	$	\phantom{1}	70.0	\pm	\phantom{1}	2.1	$	&	5	\\
\mbox{NGC 3384\phantom{ 1}}	&	$	62.8	\pm	\phantom{1}	1.6	$	&	$		\cdots	$	&	$		62.8	\pm	\phantom{1}	1.6	$	&	$	\phantom{1}	53.0	\pm	\phantom{1}	1.0	$	&	$	\phantom{1}	48.0	\pm	\phantom{1}	1.5	$	&	$	\phantom{1}	50.5	\pm	\phantom{1}	2.5	$	&		\\
\mbox{NGC 3412\phantom{ 1}}	&	$	57.0	\pm	\phantom{1}	1.6	$	&	$		\cdots	$	&	$		57.0	\pm	\phantom{1}	1.6	$	&	$		151.0	\pm	\phantom{1}	0.9	$	&	$		\cdots	$	&	$		151.0	\pm	\phantom{1}	0.9	$	&		\\
\mbox{NGC 3489\phantom{ 1}}	&	$	58.5	\pm	\phantom{1}	3.0	$	&	$		\cdots	$	&	$		58.5	\pm	\phantom{1}	3.0	$	&	$	\phantom{1}	71.2	\pm	\phantom{1}	1.0	$	&	$	\phantom{1}	73.0	\pm	\phantom{1}	1.0	$	&	$	\phantom{1}	72.1	\pm	\phantom{1}	0.9	$	&		\\
\mbox{NGC 3621\phantom{ 1}}	&	$	64.0	\pm	\phantom{1}	1.1	$	&	$		64.7	\pm	\phantom{1}	1.0	$	&	$		64.4	\pm	\phantom{1}	0.5	$	&	$		161.2	\pm	\phantom{1}	1.0	$	&	$		165.4	\pm	\phantom{1}	1.0	$	&	$		163.3	\pm	\phantom{1}	2.1	$	&		\\
\mbox{M66\phantom{ngeloo 1}}	&	$	65.0	\pm	\phantom{1}	1.3	$	&	$		61.8	\pm	\phantom{1}	1.0	$	&	$		63.4	\pm	\phantom{1}	1.6	$	&	$		173.0	\pm	\phantom{1}	1.0	$	&	$		173.0	\pm	\phantom{1}	1.0	$	&	$		173.0	\pm	\phantom{1}	0.5	$	&		\\
\mbox{NGC 4144\phantom{ 1}}	&	$	79.0	\pm	\phantom{1}	1.4	$	&	$		\cdots	$	&	$		79.0	\pm	\phantom{1}	1.4	$	&	$		102.0	\pm	\phantom{1}	1.0	$	&	$		102.0	\pm	\phantom{1}	1.0	$	&	$		102.0	\pm	\phantom{1}	0.5	$	&		\\
\mbox{NGC 4236\phantom{ 1}}	&	$	76.4	\pm	\phantom{1}	1.2	$	&	$		76.1	\pm	\phantom{1}	0.7	$	&	$		76.2	\pm	\phantom{1}	0.5	$	&	$		160.0	\pm	\phantom{1}	1.0	$	&	$		156.1	\pm	\phantom{1}	1.6	$	&	$		158.1	\pm	\phantom{1}	2.0	$	&		\\
\mbox{NGC 4244\phantom{ 1}}	&	$	88.1	\pm	\phantom{1}	6.9	$	&	$		84.5	\pm	\phantom{1}	0.5	$	&	$		84.5	\pm	\phantom{1}	0.5	$	&	$	\phantom{1}	42.2	\pm	\phantom{1}	1.0	$	&	$	\phantom{1}	45.0	\pm	\phantom{1}	2.0	$	&	$	\phantom{1}	43.6	\pm	\phantom{1}	1.4	$	&	1, 5	\\
\mbox{M106\phantom{geloo 1}}	&	$	66.9	\pm	\phantom{1}	0.9	$	&	$		72.0	\pm	\phantom{1}	1.0	$	&	$		66.9	\pm	\phantom{1}	0.9	$	&	$		150.0	\pm	\phantom{1}	1.0	$	&	$		150.0	\pm	\phantom{1}	1.0	$	&	$		150.0	\pm	\phantom{1}	0.5	$	&	8	\\
\mbox{NGC 4449\phantom{ 1}}	&	$	56.3	\pm	\phantom{1}	2.6	$	&	$		60.0	\pm	\phantom{1}	5.0	$	&	$		56.3	\pm	\phantom{1}	2.6	$	&	$	\phantom{1}	64.0	\pm	\phantom{1}	1.0	$	&	$	\phantom{1}	50.0	\pm		17.0	$	&	$	\phantom{1}	57.0	\pm	\phantom{1}	7.0	$	&	7	\\
\mbox{M104\phantom{geloo 1}}	&	$	84.8	\pm	\phantom{1}	0.6	$	&	$		\cdots	$	&	$		84.8	\pm	\phantom{1}	0.6	$	&	$	\phantom{1}	89.9	\pm	\phantom{1}	0.3	$	&	$		\cdots	$	&	$	\phantom{1}	89.9	\pm	\phantom{1}	0.3	$	&		\\
\mbox{NGC 4631\phantom{ 1}}	&	$	84.6	\pm	\phantom{1}	2.6	$	&	$		85.5	\pm	\phantom{1}	1.5	$	&	$		85.1	\pm	\phantom{1}	0.5	$	&	$	\phantom{1}	63.3	\pm	\phantom{1}	1.0	$	&	$	\phantom{1}	86.0	\pm	\phantom{1}	1.0	$	&	$	\phantom{1}	74.6	\pm		11.4	$	&	1, 5, 8	\\
\mbox{M94\phantom{ngeloo 1}}	&	$	39.7	\pm	\phantom{1}	3.8	$	&	$		41.4	\pm	\phantom{1}	1.0	$	&	$		40.5	\pm	\phantom{1}	0.9	$	&	$		106.5	\pm	\phantom{1}	1.0	$	&	$		116.1	\pm	\phantom{1}	1.0	$	&	$		111.3	\pm	\phantom{1}	4.8	$	&	10	\\
\mbox{M64\phantom{ngeloo 1}}	&	$	57.7	\pm	\phantom{1}	0.9	$	&	$		55.0	\pm	\phantom{1}	2.0	$	&	$		56.4	\pm	\phantom{1}	1.4	$	&	$		114.0	\pm	\phantom{1}	1.0	$	&	$		113.0	\pm	\phantom{1}	2.0	$	&	$		113.5	\pm	\phantom{1}	0.5	$	&	9	\\
\mbox{NGC 4945\phantom{ 1}}	&	$	81.7	\pm	\phantom{1}	2.2	$	&	$		78.0	\pm	\phantom{1}	1.0	$	&	$		81.7	\pm	\phantom{1}	2.2	$	&	$	\phantom{1}	42.0	\pm	\phantom{1}	1.0	$	&	$	\phantom{1}	45.0	\pm	\phantom{1}	2.0	$	&	$	\phantom{1}	43.5	\pm	\phantom{1}	1.5	$	&	1	\\
\mbox{NGC 5023\phantom{ 1}}	&	$	78.2	\pm	\phantom{1}	1.3	$	&	$		87.0	\pm	\phantom{1}	2.0	$	&	$		87.0	\pm	\phantom{1}	2.0	$	&	$	\phantom{1}	28.0	\pm	\phantom{1}	1.0	$	&	$	\phantom{1}	28.0	\pm	\phantom{1}	1.0	$	&	$	\phantom{1}	28.0	\pm	\phantom{1}	0.5	$	&	1, 5	\\
\mbox{NGC 5068\phantom{ 1}}	&	$	28.6	\pm	\phantom{1}	1.2	$	&	$		\cdots	$	&	$		28.6	\pm	\phantom{1}	1.2	$	&	$		104.0	\pm	\phantom{1}	1.0	$	&	$		\cdots	$	&	$		104.0	\pm	\phantom{1}	1.0	$	&	3	\\
\mbox{NGC 5102\phantom{ 1}}	&	$	70.6	\pm	\phantom{1}	2.7	$	&	$		70.0	\pm	\phantom{1}	2.0	$	&	$		70.0	\pm	\phantom{1}	2.0	$	&	$	\phantom{1}	49.2	\pm	\phantom{1}	1.0	$	&	$	\phantom{1}	43.0	\pm	\phantom{1}	3.0	$	&	$	\phantom{1}	46.1	\pm	\phantom{1}	3.1	$	&	5	\\
\mbox{Centaurus A}	&	$	\cdots	$	&	$		79.0	\pm		10.0	$	&	$		79.0	\pm		10.0	$	&	$	\phantom{1}	35.0	\pm	\phantom{1}	3.0	$	&	$	\phantom{1}	80.0	\pm	\phantom{1}	5.0	$	&	$	\phantom{1}	80.0	\pm	\phantom{1}	5.0	$	&	11	\\
\mbox{M51\phantom{ngeloo 1}}	&	$	39.1	\pm	\phantom{1}	3.7	$	&	$		20.0	\pm	\phantom{1}	5.0	$	&	$		20.0	\pm	\phantom{1}	5.0	$	&	$		163.0	\pm	\phantom{1}	1.0	$	&	$		169.0	\pm	\phantom{1}	4.2	$	&	$		169.0	\pm	\phantom{1}	4.2	$	&		\\
\mbox{NGC 5195\phantom{ 1}}	&	$	42.5	\pm	\phantom{1}	1.8	$	&	$		\cdots	$	&	$		42.5	\pm	\phantom{1}	1.8	$	&	$	\phantom{1}	91.0	\pm	\phantom{1}	5.0	$	&	$	\phantom{1}	98.0	\pm		25.0	$	&	$	\phantom{1}	94.5	\pm	\phantom{1}	3.5	$	&		\\
\mbox{M83\phantom{ngeloo 1}}	&	$	11.1	\pm		24.7	$	&	$		25.0	\pm	\phantom{1}	5.0	$	&	$		25.0	\pm	\phantom{1}	5.0	$	&	$	\phantom{1}	85.0	\pm	\phantom{1}	1.0	$	&	$	\phantom{1}	46.0	\pm	\phantom{1}	1.0	$	&	$	\phantom{1}	46.0	\pm	\phantom{1}	1.0	$	&	3	\\
\mbox{M101\phantom{geloo 1}}	&	$	35.3	\pm	\phantom{1}	3.0	$	&	$		21.0	\pm	\phantom{1}	3.0	$	&	$		21.0	\pm	\phantom{1}	3.0	$	&	$	\phantom{1}	43.0	\pm	\phantom{1}	6.0	$	&	$	\phantom{1}	42.0	\pm	\phantom{1}	2.0	$	&	$	\phantom{1}	42.0	\pm	\phantom{1}	2.0	$	&		\\
\mbox{Circinus\phantom{oo 1}}	&	$	69.9	\pm	\phantom{1}	2.7	$	&	$		66.0	\pm	\phantom{1}	5.0	$	&	$		69.9	\pm	\phantom{1}	2.7	$	&	$	\phantom{1}	36.1	\pm	\phantom{1}	1.0	$	&	$	\phantom{1}	24.0	\pm	\phantom{1}	3.0	$	&	$	\phantom{1}	30.1	\pm	\phantom{1}	6.1	$	&	6, 8	\\
\mbox{E274-G001\phantom{ 1}}	&	$	83.9	\pm	\phantom{1}	2.6	$	&	$		\cdots	$	&	$		83.9	\pm	\phantom{1}	2.6	$	&	$	\phantom{1}	43.1	\pm	\phantom{1}	1.0	$	&	$		\cdots	$	&	$	\phantom{1}	43.1	\pm	\phantom{1}	1.0	$	&		\\
\mbox{Milky Way\phantom{ 1}}	&	$	90.0	\pm	\phantom{1}	0.0	$	&	$		90.0	\pm	\phantom{1}	0.0	$	&	$		90.0	\pm	\phantom{1}	0.0	$	&	$	\phantom{1}	31.7	\pm	\phantom{1}	0.0	$	&	$	\phantom{1}	31.7	\pm	\phantom{1}	0.0	$	&	$	\phantom{1}	31.7	\pm	\phantom{1}	0.0	$	&		\\
\mbox{NGC 6503\phantom{ 1}}	&	$	71.5	\pm	\phantom{1}	0.9	$	&	$		73.8	\pm	\phantom{1}	1.2	$	&	$		72.7	\pm	\phantom{1}	1.1	$	&	$		121.0	\pm	\phantom{1}	1.0	$	&	$		120.6	\pm	\phantom{1}	0.9	$	&	$		120.8	\pm	\phantom{1}	0.5	$	&		\\
\mbox{NGC 6946\phantom{ 1}}	&	$	33.2	\pm	\phantom{1}	1.1	$	&	$		32.6	\pm	\phantom{1}	1.0	$	&	$		32.6	\pm	\phantom{1}	1.0	$	&	$	\phantom{1}	69.0	\pm	\phantom{1}	5.0	$	&	$	\phantom{1}	62.7	\pm	\phantom{1}	1.0	$	&	$	\phantom{1}	65.9	\pm	\phantom{1}	3.2	$	&	6	\\
\mbox{IC 5052\phantom{oo 1}}	&	$	85.7	\pm	\phantom{1}	3.2	$	&	$		\cdots	$	&	$		85.7	\pm	\phantom{1}	3.2	$	&	$		140.0	\pm	\phantom{1}	1.0	$	&	$		\cdots	$	&	$		140.0	\pm	\phantom{1}	1.0	$	&		\\
\mbox{NGC 7793\phantom{ 1}}	&	$	53.8	\pm	\phantom{1}	1.5	$	&	$		49.6	\pm	\phantom{1}	1.0	$	&	$		51.7	\pm	\phantom{1}	2.1	$	&	$	\phantom{1}	99.3	\pm	\phantom{1}	1.1	$	&	$		110.1	\pm	\phantom{1}	1.0	$	&	$		104.7	\pm	\phantom{1}	5.4	$	&		\\

\noalign{\smallskip}


\end{longtable}

\end{ThreePartTable}

\end{center}
\clearpage

\begin{center}
\begin{ThreePartTable}

\begin{TableNotes}

\item[]
(1)
Mass of Council relative to mass of Local Group;
(2)
Radius of boundary of Local Group, in Mpc;
(3)
Radius of outer boundary of Council, in Mpc;
(4)
Factor by which mass of Local Group judged from timing exceeds that derived by density matching.
(5)
Fraction of mass of galaxies which is stellar relative to cosmic fraction of matter which is baryonic as judged from density matching alone (without correcting for the overdensity).

\smallskip

\item{$^a$}
${\mathcal M}_\textit{stars} / {\mathcal L}_\textit{Ks}$ derived from $B-V$ using algorithm of \citet{por04a}; 
${\mathcal M}_\textit{total} / {\mathcal M}_\textit{stars}$ fixed; cylindrical geometry; Council boundaries equidistant from Council.

\item{$^b$}
${\mathcal M}_\textit{stars} / {\mathcal L}_\textit{Ks}$ set to baseline value for $B-V = 0.6$.

\item{$^c$}
Relative values of total mass constrained by trends in ${\mathcal M}_\textit{total} / {\mathcal M}_\textit{stars}$ with ${\mathcal M}_\textit{stars}$ as revealed by weak lensing \citep{vel14a}.

\item{$^d$}
Council boundaries defined by mass-matching. 


\end{TableNotes}

\setlength{\jot}{-0.0pt}

\begin{longtable}{lccccc}

\caption{Sensitivities of Derived Parameters to Input\label{tbl_sensitivities}}  
\\

\hline
\noalign{\smallskip}

\mbox{Input} &
${\mathcal M}_\textit{C} / {\mathcal M}_\textit{LG}$ &
$R_\textit{LG}$ &
$R_\textit{edge}$ &
$\Delta$ &
$\mbox{Efficiency} \times \Delta$
\\

&
(1) & 
(2) & 
(3) & 
(4) &
(5)
\\

\noalign{\smallskip}
\hline
\noalign{\smallskip}

\endfirsthead

\caption{cont'd.} \\

\hline
\noalign{\smallskip}

\mbox{Input} &
${\mathcal M}_\textit{C} / {\mathcal M}_\textit{LG}$ &
$R_\textit{LG}$ &
$R_\textit{edge}$ &
$\Delta$ &
$\mbox{Efficiency} \times \Delta$
\\

&
(1) & 
(2) & 
(3) & 
(4) &
(5)
\\

\noalign{\smallskip}
\hline
\endhead

\noalign{\smallskip}
\hline

\endfoot
\hline
\noalign{\smallskip}
\insertTableNotes

\endlastfoot


Baseline\,$^a$ & 
$2.72 \pm 0.54$ & $2.56 \pm 0.17$ & $4.94 \pm 0.27$ & $1.04 \pm 0.25$ & $0.365 \pm 0.060$ \\  

${\mathcal M}_\textit{stars} / {\mathcal L}_\textit{Ks}$ fixed\,$^b$ & 
$2.26 \pm 0.45$ & $2.67 \pm 0.18$ & $4.82 \pm 0.26$ & $0.92 \pm 0.21$ & $0.322 \pm 0.054$ \\  
${\mathcal M}_\textit{total} / {\mathcal M}_\textit{stars}$ variable\,$^c$ & 
$3.19 \pm 0.65$ & $2.46 \pm 0.17$ & $5.03 \pm 0.27$ & $1.17 \pm 0.29$ & $0.410 \pm 0.069$ \\  
Spherical Geometry & 
$2.72 \pm 0.54$ & $2.94 \pm 0.17$ & $4.55 \pm 0.24$ & $1.03 \pm 0.22$ & $0.361 \pm 0.065$ \\  
Equal-mass Shells\,$^d$ & 
$2.72 \pm 0.54$ & $2.44 \pm 0.18$ & $4.70 \pm 0.24$ & $1.21 \pm 0.31$ & $0.421 \pm 0.070$ \\  
\noalign{\smallskip}


\end{longtable}

\end{ThreePartTable}

\end{center}
\clearpage

\begin{center}
\begin{ThreePartTable}

\begin{TableNotes}

\item[]
(1) 
Number of members with 
$M_\textit{Ks} \le -22.5$.  
(2)
Supergalactic longitude and latitude of north pole of Local Sheet.
(3)
Inclination of plane of Local Sheet with respect to supergalactic plane.
(4)
Perpendicular offset of plane of Local Sheet from the Sun.
(5) 
Twice the standard deviation of giants about the plane.
(6) 
The diameter, as given both by the radius of the Council of Giants augmented by three times the standard deviation of the members and by the radius of the edge of the density-matched zone of the Council of Giants augmented by its uncertainty.
(7)
Supergalactic Cartesian coordinates of the centre of the Council of Giants.
(8)
Diameter of the Council of Giants.
(9)
Physical separation of Maffei~1 and Centaurus~A
(10)
Angular separation of Maffei~1 and Centaurus~A projected on to the Sheet plane, as seen from the centre of the Council of Giants.
(11)
Supergalactic longitude and latitude of pole of angular momentum vectors for Council giants.
(12)
Diameter of the zone of influence of the Local Group, as indicated by both density matching and the potential surface of the Sheet.
(13)
Tilt of the Andromeda--Milky Way axis to the plane of the Sheet.
(14)
Total luminosity of giants in $K_s$.
(15)
Stellar mass of giants, based upon luminosities in $K_s$ and mass-to-light ratios derived from $B-V$ colours using the algorithm of \citet{por04a}.
(16)
Total mass of giants, based upon the mass scale defined by the timing mass of the Local Group.
(17)
Ratio of total mass of giants to stellar mass of giants.
(18)
Fraction of the mass of the Sheet in the Local Group.
(19)
Fraction of the mass of the Sheet in giant ellipticals.
(20)
Smoothed mass of giants per unit area.
(21)
Radial dispersion of velocities of Council giants with respect to the Council centre.
(22)
Time to cross the thickness of the Sheet for a galaxy moving vertically with a velocity equal to the velocity dispersion radially.
(23)
Temperature expected for a virialized gas with an extent equal to that of the Sheet if the mass of galaxies were uniformly spread over that extent.
(24)
Vertical velocity required to escape the mid-plane based upon the surface mass density and extent of the Sheet.
(25) 
Factor by which mass of the Local Group judged from timing exceeds that derived from density-matching.
(26)
Fraction of the mass of galaxies which is stellar relative to the cosmic fraction of matter which is baryonic.

\medskip

\item{$^\dagger$}
Properties are founded upon a distance scale set by M106 (NGC 4258), the distance to which was adopted to be the geometric maser estimate 
$7.60 \, \rm Mpc$ \citep{hum13a}.  
The distance to the centre of the Milky Way was adopted to be 
$8.29 \, \rm kpc$ \citep{mar12a}.  
Velocity corrections were based upon a Hubble constant of $71.6 \, \rm km \, s^{-1} \, Mpc^{-1}$ \citep{rie11a, rie12a, hum13a}.
Uncertainties in tabulations are discussed in the text.


\end{TableNotes}

\setlength{\jot}{-0.0pt}

\begin{longtable}{llcllc}

\caption{The Local Sheet$^\dagger$\label{tbl_summary}}  
\\

\hline
\noalign{\smallskip}

Parameter &
Units &
Value &
Parameter &
Units &
Value
\\ 

\noalign{\smallskip}
\hline
\noalign{\smallskip}

\endfirsthead

\caption{cont'd.} \\

\hline
\noalign{\smallskip}

Parameter &
Units &
Value &
Parameter &
Units &
Value
\\ 

\noalign{\smallskip}
\hline
\endhead

\noalign{\smallskip}
\hline

\endfoot
\hline
\noalign{\smallskip}
\insertTableNotes

\endlastfoot


\phantom{1}1. \, Giant Membership & &
14  & 
14. \, Stellar Luminosity & $\rm \mathcal{L}_{\odot, \textit{Ks}}$ &
$1.53 \times 10^{12}$ \\

\phantom{1}2. \, $(L, B)$ of North Pole & deg & 
$(241.74, +82.05)$   & 
15. \, Stellar Mass & $\rm \mathcal{M}_\odot$ &
$8.7 \times 10^{11}$ \\  

\phantom{1}3. \, Tilt & deg &
$7.95$ &  
16. \, Total Mass & $\rm \mathcal{M}_\odot$ & 
$1.6 \times 10^{13}$ \\  

\phantom{1}4. \, Offset from Sun & Mpc & 
$0.129$ &  
17. \, Total Mass$/$Stellar Mass &  &
$18.5$ \\  

\phantom{1}5. \, Thickness & Mpc & 
$0.465$ & 
18. \, Local Group Fraction & & 
$0.27$ \\  

\phantom{1}6. \, Extent & Mpc & 
$10.4$ & 
19. \, Elliptical Fraction & &
$0.27$ \\  

\phantom{1}7. \, $(X, Y, Z)$ of Council Centre & Mpc &
$(-0.25, +0.77, -0.05)$ &  
20. \, Surface Mass Density & $\rm \mathcal{M}_\odot \, pc^{-2}$ & 
$0.21$ \\  

\phantom{1}8. \, Council Diameter & Mpc &
$7.49$ &  
21. \, Velocity Dispersion & $\rm km \, s^{-1}$ & 
$47$ \\  

\phantom{1}9. \, Separation of Ellipticals & Mpc &
$6.73$ &  
22. \, Crossing Time & $\rm Gyr$ &
$9.6$ \\  

10. \, Angle between Ellipticals & deg &
$175.0$ &  
23. \, Virial Temperature & K & 
$7.3 \times 10^5$ \\  

11. \, $(L, B)$ of Council Spin Pole & deg &
$(125.1, +41.5)$ &  
24. \, Escape Velocity & $\rm km \, s^{-1}$ & 
$244$ \\  

12. \, Local Group Realm & Mpc &
$5.11$ &  
25. \, Overdensity & & 
$1.04$ \\  

13. \, Tilt of Local Group & deg &
$11.3$ &  
26. \, Efficiency & &
$0.35$ \\  

\noalign{\smallskip}


\end{longtable}

\end{ThreePartTable}

\end{center}

\end{document}